\documentclass[prd,aps,floats,showpacs,twocolumn,twoside,preprintnumbers,
superscriptaddress,floatfix]{revtex4}

\usepackage{bm}
\usepackage[T1]{fontenc}
\usepackage[latin1]{inputenc}
\usepackage{amsmath}
\usepackage{amssymb}
\usepackage{mathrsfs}
\usepackage{epsfig}

\newcommand{\ie}{,~{\it i.e.},~}
\newcommand{\eg}{,~{\it e.g.},~}

\newcommand{\dslash}{{\partial\!\!\!/}}

\newcommand\expect[1]{\langle#1\rangle}
\renewcommand\exp[1]{e^{\,#1}}
\newcommand\eqn[1]{(\ref{#1})}
\newcommand\eqns[2]{(\ref{#1})-(\ref{#2})}
\newcommand\fig[1]{Fig.~\ref{#1}}
\newcommand\figs[2]{Figs.~\ref{#1}-\ref{#2}}

\begin{document}

\title{
The quark core of protoneutron stars in the phase diagram of quark matter
}

\author{F. Sandin}
\email{fredrik.sandin@ltu.se}
\affiliation{ 
Department of Physics, Lule{\aa} University of Technology,
SE-97187 Lule\aa , Sweden}

\author{D. Blaschke}
\email{blaschke@ift.uni.wroc.pl}
\affiliation{Instytut Fizyki Teoretycznej, Uniwersytet Wroc{\l}awski, 
PL-50-204 Wroc{\l}aw, Poland}
\affiliation{Bogoliubov Laboratory for Theoretical Physics, JINR  
Dubna, RU-141980 Dubna, Russia}
\affiliation{Institut f\"ur Physik, Universit\"at Rostock, D-18051 Rostock, 
Germany}

\begin{abstract}
We study the effect of neutrino trapping in new-born quark stars within a 
three-flavor Nambu--Jona-Lasinio (NJL) model with self-consistently calculated
quark masses. 
The phase diagrams and equations of state for charge neutral quark matter in 
$\beta$-equilibrium are presented, with and without trapped neutrinos. 
The compact star sequences for different neutrino untrapping scenarios are 
investigated and the energy release due to neutrino untrapping is found to be 
of the order of $10^{53}$~erg.
We find that hot quark stars characterized\eg by an entropy per baryon of
$1-2$ and a lepton fraction of $0.4$, as models for the cores of newborn
protoneutron stars, are in the two-flavor color superconducting (2SC) state.
High temperatures and/or neutrino chemical potentials disfavor 
configurations with a color-flavor-locked (CFL) phase. 
Stable quark star solutions with CFL cores exist only at low temperatures
and neutrino chemical potentials.
\end{abstract}

\pacs{12.38.Aw, 12.39.-x, 24.85.+p, 26.60.+c, 97.60.-s} 

\maketitle

%%%%%%%%%%%%%%%%%%%%%%%%%%%%%%%%%%%%%%%%%%%%%%%%%%%%%%%%%%%%%%%%%%%%%%%

\section{Introduction}

% 1 MeV = 1.6E-6 erg
% M_sun = 1.116E60 MeV

The engine of explosive phenomena in astrophysics, such as gamma-ray bursts
(GRBs) and type-II supernov{\ae} is presently not fully understood.
The generation and propagation of neutrino fluxes, as well as the neutrino
interactions in the hot and dense nuclear matter envelope play key roles
in models of the evolution of core-collapse supernov{\ae} \cite{Janka:2006}.
Detailed hydrodynamic simulations of the gravitational collapse of a massive 
star show that the possibility to obtain a successful explosion depends 
to a large extent on the properties of the protoneutron star (PNS) that
forms in the compressed core. About 99\% of the gravitational binding energy
is released by neutrino emission \cite{Lattimer:2006xb,Burrows:1986me,Prakash:1996xs}.
The supernova collapse and prompt neutrino production proceeds within
milliseconds and the shock-compressed matter is heated up to about
$30-50$~MeV. At such high temperature the neutrino mean free path is much
shorter than the radius of a PNS, $R\sim 10$~km, and neutrinos diffuse
on a time scale of $\sim 10$ seconds.

During this `neutrino trapping' regime the number of neutrinos is quasi
conserved and the neutrino chemical potential, $\mu_\nu$, is of the
order $200$~MeV \cite{Burrows:1986me,Prakash:1996xs}. This state lasts
until the temperature is low enough for the neutrino mean free path to 
become comparable to $R$. 
The behavior of the neutrino mean free path, which determines the timescale
for the untrapping transition and the onset of the 'second shock'
of the supernova, depends in a sensitive way on the microphysics of the
hot and dense PNS interior. In particular, if superfluid and superconducting
phases of hadronic and/or quark matter are created, the transport
properties are affected and the consequences for simulations of
core-collapse supernov{\ae} could be important.

In a scenario where color superconducting quark matter is the preferred
state of matter at high density there are several new aspects to consider,
for example:
\begin{enumerate}
        \item The kinetics of the phase transition, which eventually proceeds with
        the release of latent heat and requires a nucleation time scale.
        \item The possibility that color superconducting phases have large pairing
        gaps ($\sim 100$ MeV), which affect the transport properties in the PNS core.
        \item The possibility that color superconducting phases have a high critical
        temperature ($\propto$ the gaps) and therefore affect the formation and
        evolution of PNS.
\end{enumerate}
For recent reviews on dense color superconducting quark matter, see
\cite{Rajagopal:2000wf,Alford:2001dt,Schafer:2003vz,Rischke:2003mt,Casalbuoni:2003wh,Buballa:2003qv,Huang:2004ik,Shovkovy:2004me} and references therein. 

The energy released due to a phase transition to deconfined quark matter in
a PNS core can reach the order of 100 bethe (=$10^{53}$~erg)
\cite{Bombaci:2000cv,Berezhiani:2002ks,Aguilera:2002dh}, which is the correct
order of magnitude for the energy of GRBs. 
Moreover, the nucleation timescale for a quark matter phase transition could
explain the time delay statistics of GRB subpulse structure
\cite{Drago:2005qb,Drago:2006xa}. 
It has also been emphasized that in the presence of a strong magnetic field,
neutrinos propagating in hot superconducting quark matter can become collimated
(beaming) and asymmetric, thus explaining a resulting kick velocity for the PNS
\cite{Berdermann:2006rk}. 

In the present work we consider a microscopic, albeit schematic chiral 
quark model of NJL type, where the quark masses and pairing gaps are
calculated self-consistently at the mean-field level, see 
\cite{Abuki:2004zk,Ruster:2005jc,Blaschke:2005uj,Abuki:2005ms,Warringa:2005jh}.
We describe the phase structure of color superconducting three-flavor 
quark matter for two different strengths of the phenomenological diquark
pairing interaction, in systems with and without trapped neutrinos.    
The corresponding quark star solutions are considered as models for
PNS cores and their properties are described.
In particular, we are interested in the answers to the following questions:
\begin{enumerate}
        \item What influence has the neutrino chemical potential on the phase 
        diagram of quark matter?
        \item In which regions of the quark matter phase diagram can stable quark 
        stars be found?
        \item Can both two-flavor superconductivity (2SC) and three-flavor
        color-flavor-locking (CFL) phases be realized under PNS conditions?
        \item How much energy can be released in the cooling and untrapping
        evolution for these models of a PNS core? 
\end{enumerate}
The effect of neutrino trapping on the phase structure of color 
superconducting three-flavor quark matter has been investigated before in 
Refs. \cite{Steiner:2002gx,Ruster:2005ib},
for homogeneous phases, and in Ref. \cite{Laporta:2005be} for inhomogeneous
phases of the LOFF type \cite{LOFF}. In this paper we focus on the stability
of quark stars for different phase structures and estimates of the energy
release in the cooling and untrapping transitions.

The paper is organized as follows. In Section \ref{sec:model} we define the 
NJL-type model that we use to describe color superconducting quark matter 
with trapped neutrinos. Section \ref{sec:results} gives a summary of all 
results, which are discussed in detail in subsections according to the
main questions posed in the Introduction. The Conclusions summarize 
our main findings.

%%%%%%%%%%%%%%%%%%%%%%%%%%%%%%%%%%%%%%%%%%%%%%%%%%%%%%%%%%%%%%%%%%%%%%%

\section{Model of hot quark matter with trapped neutrinos}
\label{sec:model}

Due to the high density in compact stars, strange quarks could exist in
their interior. We therefore consider a grand canonical ensemble with
up, down, and strange quark degrees of freedom. Because strange quarks
have a relatively high mass, they should arise only at high density.
Therefore, at not too high densities, where matter is in a two-flavor
state, there should be an excess of down quarks for matter to be
charge neutral. Consequently, the difference between the chemical
potentials of the up and down quarks can be sufficiently large for muons
to be created by weak interactions, 
$d\leftrightarrow u + \mu^- + \bar{\nu}_\mu$.
It is therefore necessary to include both electron and muon lepton degrees
of freedom in the model. 
The $\tau$ lepton is, however, not included, because it
is too massive to play a significant role in compact stars. We neglect the
influence of neutrino oscillations and therefore omit the $\tau$ neutrino
as well.

The thermodynamics of the quark matter phase is described with an NJL-type
model. The path-integral representation of the quark partition
function is \cite{Blaschke:2005uj,Ruster:2005jc,Abuki:2005ms}
\begin{subequations}
\label{Z}
\begin{eqnarray}
        Z(T,\hat{\mu})&=&\int{\mathcal D}\bar{q}{\mathcal D}q\,\exp{
        \int d^4x (
          {\mathcal L}_f + {\mathcal L}_{\bar{q}q} + {\mathcal L}_{qq})}, \\
        {\mathcal L}_f &=& \bar{q}(i\dslash-\hat{m}+\hat{\mu}\gamma^0)q, \\
        {\mathcal L}_{\bar{q}q} &=& G_S\sum_{a=0}^8(\bar{q}\tau_aq)^2, \\
        {\mathcal L}_{qq} &=& G_D\!\!\!\!\sum_{A=2,5,7}\!\!\!
        (\bar{q}i\gamma_5\tau_A\lambda_AC\bar{q}^T) 
        (q^TiC\gamma_5\tau_A\lambda_Aq),\,\,\,\,\,\,\,\,\,\,\,\,
\end{eqnarray} 
\end{subequations}
where $\hat{\mu}$ and $\hat{m}$ are the diagonal chemical potential and
current quark mass matrices in color and flavor space.
For $a=0$, $\tau_0=(2/3)^{1/2}{\mathbf 1}_f$, otherwise $\tau_a$ and 
$\lambda_a$ are Gell-Mann matrices acting in flavor and color space,
respectively. $C=i\gamma^2\gamma^0$ is the charge conjugation operator
and $\bar{q}=q^\dagger\gamma^0$. The coupling constants, $G_S$ and $G_D$,
determine the coupling strengths in the $\bar{q}q$ and $qq$ channels,
which represent current-current interactions in the color singlet scalar
meson channel and the scalar color antitriplet diquark channel.
We follow the argument in \cite{Blaschke:2005uj} concerning the U$_A$(1)
symmetry breaking in the pseudoscalar isoscalar meson sector, which
essentially is that the symmetry breaking is dominated by quantum
fluctuations, and we therefore omit the 't~Hooft determinant interaction.
In the following we use the relative coupling strength
\begin{equation}
        \eta = G_D/G_S
\label{diquarkcoupling}
\end{equation}
to parametrize the coupling in the diquark channel. In the choice of the
four-fermion interaction channels we have omitted the pseudoscalar $\bar{q}q$
terms, which should be present in a chirally symmetric theory. These terms
do not contribute to the thermodynamic properties of the deconfined quark
matter phase at the mean-field (Hartree) level \cite{Warringa:2005jh}, to
which we restrict the discussion in the present paper.
 
After bosonization using Hubbard-Stratonovich transformations, we obtain
an exact transformation of the original partition function \eqn{Z}. The
transformed expression constitutes the starting point for approximations,
defined as truncations of the Taylor expanded action functional to different
orders in the collective boson fields.
In the following, we use the mean-field (MF) approximation. This means 
that the bosonic functional integrals are omitted and the collective fields 
are fixed at the extremum of the action. The corresponding mean-field 
thermodynamic potential, from which all thermodynamic quantities can
be derived, is given by
%\begin{widetext}
%\begin{eqnarray} 
%        &&\Omega_\text{MF}(T,\hat{\mu},\mu_{L_e},\mu_{L_\mu})
%        = -\frac{1}{\beta V}\ln Z_\text{MF}(T,\hat{\mu})
%        + \Omega_l(T,\mu_{L_e},\mu_{L_\mu}) \nonumber \\
%        &&
%        = \sum_{i=u,d,s}\!\!\!\frac{(M_i-m_i)^2}{8 G_S}
%        +\frac{\Delta_{ud}^2+\Delta_{us}^2+\Delta_{ds}^2}{4\eta G_S}
%        -\int\frac{d^3p}{(2\pi)^3}\sum_{a=1}^{18} 
%        \left[E_a(p)+2T\ln\left(1+e^{-\frac{E_a(p)}{T}}\right)\right]
%        +\,\Omega_l(T,\mu_{L_e},\mu_{L_\mu}) - \Omega_0.
%\label{potential} 
%\end{eqnarray} 
%\end{widetext}
%\begin{eqnarray} 
%        &&\Omega_\text{MF}(T,\hat{\mu},\mu_{L_e},\mu_{L_\mu}) \nonumber \\
%        &&\,\,\,\,\,\,\,\,\,
%        = -\frac{1}{\beta V}\ln Z_\text{MF}(T,\hat{\mu})
%        + \Omega_l(T,\mu_{L_e},\mu_{L_\mu}) \nonumber \\
%        &&\,\,\,\,\,\,\,\,\,
%        = \sum_{i=u,d,s}\frac{(M_i-m_i)^2}{8 G_S}
%        +\frac{\Delta_{ud}^2+\Delta_{us}^2+\Delta_{ds}^2}{4\eta G_S} \nonumber 
%\\
%        &&\,\,\,\,\,\,\,\,\,
%        -\int\frac{d^3p}{(2\pi)^3}\sum_{a=1}^{18} 
%        \left[E_a(p)+2T\ln\left(1+e^{-E_a(p)/T}\right)\right] \nonumber \\
%        &&\,\,\,\,\,\,\,\,\,
%        +\,\Omega_l(T,\mu_{L_e},\mu_{L_\mu}) - \Omega_0.
%\label{potential} 
%\end{eqnarray} 
%
% The following equation was divided into two pieces in
% order to avoid a half-empty page before the figures
\begin{eqnarray} 
        &&\Omega_\text{MF}(T,\hat{\mu},\mu_{L_e},\mu_{L_\mu}) \nonumber \\
        &&\,\,\,\,\,\,\,\,\,
        = -\frac{1}{\beta V}\ln Z_\text{MF}(T,\hat{\mu})
        + \Omega_l(T,\mu_{L_e},\mu_{L_\mu}) \nonumber \\
        &&\,\,\,\,\,\,\,\,\,
		= \sum_{i=u,d,s}\frac{(M_i-m_i)^2}{8 G_S}
        +\frac{\Delta_{ud}^2+\Delta_{us}^2+\Delta_{ds}^2}{4\eta G_S} \nonumber \\
        &&\,\,\,\,\,\,\,\,\,
        -\int\frac{d^3p}{(2\pi)^3}\sum_{a=1}^{18} 
        \left[E_a(p)+2T\ln\left(1+e^{-E_a(p)/T}\right)\right] \nonumber
\end{eqnarray} 

\pagebreak

\begin{eqnarray} 
        \!\!\!\!\!\!\!\!\!\!\!\!\!\!\!\!\!\!\!\!\!\!\!\!\!\!\!\!\!\!\!\!\!
        +\,\Omega_l(T,\mu_{L_e},\mu_{L_\mu}) - \Omega_0.
\label{potential} 
\end{eqnarray} 
Here, $M_i=m_i+\phi_i$ are the renormalized quark masses, $m_i$ are the
current quark masses, $E_a(p)$ are the eighteen independent quark quasiparticle
dispersion relations, and $\phi_i$ ($\Delta_{ij}$) are chiral (diquark) gaps,
see \cite{Blaschke:2005uj} for details.
The gaps, $\phi_i$ and $\Delta_{ij}$, emerge from the auxiliary
boson fields introduced by the Hubbard-Stratonovich transformations and
represent collective modes generated by $\bar{q}q$ and $qq$, respectively.
$\Omega_l(T,\mu_{L_e},\mu_{L_\mu})$ is the thermodynamic
potential for an ideal gas of neutrinos, electrons ($m_e\simeq 0.511$~MeV),
and muons ($m_\mu\simeq 105.66$~MeV)
\begin{eqnarray}
        \Omega_l(T,\mu_{L_e},\mu_{L_\mu}) = -\sum_{l=e,\,\mu}\Bigg[
        \frac{\mu^4_{\nu_l}}{24\pi^2} + \frac{\mu^2_{\nu_l}T^2}{12}
        +\frac{7\pi^2T^4}{360} \nonumber \\
        +\frac{T}{\pi^2}\sum_\pm\int_0^\infty \text{d}k
        k^2\ln\left(1+\exp{-\frac{E_l\pm\mu_l}{T}}\right)\Bigg],
\end{eqnarray}
where $E_l = \sqrt{p^2+m^2_l}$.
$\Omega_0$ is a divergent term that is subtracted in order to get zero
pressure and energy density in vacuum ({\it i.e.}, at $T=\mu=0$)
\begin{equation}
        \Omega_0 = \sum_{i=u,d,s}\frac{(\phi^0_i)^2}{8 G_S}
        - 6\int\frac{d^3p}{(2\pi)^3}\sqrt{p^2+(m_i+\phi^0_i)^2}.
        \label{omega0}
\end{equation}

The model has six conserved charge densities and
associated chemical potentials.
The $U(3)_f\times SU(3)_c\times U(1)_Q$ symmetry of the quarks is broken in
the presence of diquarks. Therefore, there are only four mutually commuting
conserved charge densities\eg the quark number density
\begin{equation}
        n = \expect{q^\dagger q} = -\frac{\partial\Omega_\text{MF}}{\partial\mu},
\end{equation}
two color charge densities
\begin{subequations}
\label{colorcharge}
\begin{eqnarray}
        n_3 = \expect{q^\dagger \lambda_3 q} = -\frac{\partial\Omega_\text{MF}}{\partial\mu_3}, \\
        n_8 = \expect{q^\dagger \lambda_8 q} = -\frac{\partial\Omega_\text{MF}}{\partial\mu_8},
\end{eqnarray}
\end{subequations}
and the electric charge density
\begin{equation}
        n_Q = \expect{q^\dagger Qq} - n_e - n_\mu = -\frac{\partial\Omega_\text{MF}}{\partial\mu_Q}.
        \label{nQ}
\end{equation}
Here, $Q=\text{diag}_f(\frac{2}{3},-\frac{1}{3},-\frac{1}{3})$ and $n_e$
($n_\mu$) is the number density of electrons (muons). Consequently, the quark
chemical potential matrix, $\hat{\mu}$, is
\begin{equation}
        \hat{\mu} = \mu + Q\mu_Q + \lambda_3\mu_3 + \lambda_8\mu_8,
        \label{qchempot}
\end{equation}
where $\mu$ is the quark number chemical potential, $\mu_Q$ the (positive) 
electric charge chemical potential, and $\mu_3$ and $\mu_8$ are color charge 
chemical potentials. 
Note the discussion of color neutrality
in Ref. \cite{Buballa:2005bv}.

The quark number density is related to the baryon number density by
%\begin{equation}
%n_B = \frac{n}{3}.
%\end{equation}
\begin{equation}
n_B = n/3.
\end{equation}
The remaining two charges are the number densities of the lepton families
\begin{subequations}
\begin{eqnarray}
        n_{L_e} = n_e + n_{\nu_e} = -\frac{\partial\Omega_\text{MF}}{\partial\mu_{L_e}}, \\
        n_{L_\mu} = n_\mu + n_{\nu_\mu} = -\frac{\partial\Omega_\text{MF}}{\partial\mu_{L_\mu}},
\end{eqnarray}
\end{subequations}
which are conserved when the neutrinos are trapped in the system and oscillations
are neglected.
Electrons and muons have both electric charge and lepton number, while
neutrinos have lepton number only. Therefore
\begin{subequations}
\begin{eqnarray}
        \mu_e &=& \mu_{L_e} - \mu_Q, \\
        \mu_\mu &=& \mu_{L_\mu} - \mu_Q, \\
        \mu_{\nu_e} &=& \mu_{L_e}, \\
        \mu_{\nu_\mu} &=& \mu_{L_\mu}.
\end{eqnarray}
\end{subequations}
The lepton fractions, which represent the relative number of leptons and
baryons, are defined as
\begin{subequations}
\label{leptonfrac}
\begin{eqnarray}
        Y_{L_e} \equiv \frac{n_{L_e}}{n_B}, \\
        Y_{L_\mu} \equiv \frac{n_{L_\mu}}{n_B}.
\end{eqnarray}
\end{subequations}

Bulk matter in compact stars should be charge neutral.
In NJL models there are no gauge fields that neutralize the color charge
dynamically, because the gluons are replaced by effective point-like
$\bar{q}q$ and $qq$ interactions. Color neutrality must therefore be
enforced by solving for the charge chemical potentials, $\mu_3$ and $\mu_8$,
such that the corresponding charge densities, 
$n_a=\expect{q^\dagger\lambda_aq}$, are zero.
%~\cite{Buballa:2005bv}. 
In addition, matter in compact stars should be electrically neutral and in
$\beta$-equilibrium with respect to weak interactions. The chemical
potentials, $\mu_Q$, $\mu_3$, and $\mu_8$ are therefore determined 
such that the charge densities \eqns{colorcharge}{nQ} vanish
\begin{equation}
        \frac{\partial\Omega_\text{MF}}{\partial\mu_Q}=
        \frac{\partial\Omega_\text{MF}}{\partial\mu_3}=
        \frac{\partial\Omega_\text{MF}}{\partial\mu_8}=0.
        \label{neutrality}
\end{equation}
Observe that the definition of the quark chemical potential \eqn{qchempot} implies
that matter is in $\beta$-equilibrium. This holds true also when neutrinos are
trapped and $\mu_{L_l}>0$, because\eg
$\mu_e+\mu_{\bar{\nu}_e} = (\mu_{L_e}-\mu_Q)+(-\mu_{L_e})=-\mu_Q$.
The gaps, $\phi_i$ and $\Delta_{ij}$, are order parameters that are determined
by minimization of the mean-field thermodynamic potential \eqn{potential}
\begin{subequations}
\label{gapeq}
\begin{eqnarray}
        \frac{\partial\Omega_\text{MF}}{\partial\phi_u}=
        \frac{\partial\Omega_\text{MF}}{\partial\phi_d}=
        \frac{\partial\Omega_\text{MF}}{\partial\phi_s}=0, \\
        \frac{\partial\Omega_\text{MF}}{\partial\Delta_{ud}}=
        \frac{\partial\Omega_\text{MF}}{\partial\Delta_{us}}=
        \frac{\partial\Omega_\text{MF}}{\partial\Delta_{ds}}=0.
\end{eqnarray}
\end{subequations}
Local minima of the thermodynamic potential define competing phases and the
global minimum is the physical solution. 

\pagebreak

At the charge neutral global minimum
of the thermodynamic potential the pressure, entropy density, number densities,
and energy density are
\begin{subequations}
\label{eoseq}
\begin{eqnarray}
        P(T,\mu,\mu_{L_e},\mu_{L_\mu}) &=& -\Omega_\text{MF}, \\
        s(T,\mu,\mu_{L_e},\mu_{L_\mu}) &=& -\frac{\partial\Omega_\text{MF}}{\partial T}, \\
        n_a(T,\mu,\mu_{L_e},\mu_{L_\mu}) &=& -\frac{\partial\Omega_\text{MF}}{\partial \mu_a}, \\
        \epsilon(T,\mu,\mu_{L_e},\mu_{L_\mu}) &=& -P + Ts +
                \!\!\!\!\!\!\!\!\sum_{a=B,L_e,L_{\mu}}\!\!\!\!\!\!\!\!\mu_an_a.
\end{eqnarray}
\end{subequations}
The sum in the expression for the energy density should account for all 
conserved charges in the system. 
However, because we require that $n_Q=n_3=n_8=0$, only three
terms are included.
Once the current quark masses, coupling constants, and momentum regularization
method has been fixed, equations \eqns{diquarkcoupling}{omega0}, \eqn{qchempot}, and
\eqns{neutrality}{eoseq} define a self-consistent set of equations for
the quark matter model.

\subsection{Numerical methods}

The quark quasiparticle dispersion relations, $E_a(p)$, in \eqn{potential}
are eigenvalues of Hermitian matrices, see \cite{Blaschke:2005uj} for details.
LAPACK is used to calculate the eigenvalues and the corresponding eigenvectors
of these matrices. For given values of all parameters,
%($T$, $\hat{\mu}$, $\mu_{L_e}$, $\mu_{L_\mu}$, $\mu_Q$, $\mu_3$, $\mu_8$,
% $\phi_u$, $\phi_d$, $\phi_s$, $\Delta_{ud}$, $\Delta_{us}$, $\Delta_{ds}$)
the mean-field thermodynamic potential \eqn{potential} is calculated with
a Gaussian integration quadrature. Observe that the integrands depend only
on the magnitude of the three-momentum, so $d^3p \rightarrow 4\pi p^2 dp$.

The derivatives of the thermodynamic potential are explicitly calculated
using the eigenvectors of the Hermitian matrices. 
The derivatives are of the form
\begin{equation}
        \frac{\partial\Omega_\text{MF}}{\partial x} =
        F_1(x)-\int\frac{d^3p}{(2\pi)^3}\sum_{a=1}^{18}F_2\left(
        \frac{\partial E_a(p)}{\partial x}
        \right),
\end{equation}
for some simple functions $F_1$ and $F_2$ that depend on the choice
of the variable $x$. The numerical problem therefore is to calculate
the derivatives of the quasiparticle dispersion relations, $E_a(p)$\ie
the derivatives of the eigenvalues of the Hermitian matrices.
It is easy to show that
\begin{equation}
        \frac{\partial E_a}{\partial x} = 
        \frac{X_a^\dagger \frac{\partial H}{\partial x} X_a}{X_a^\dagger X_a},
\end{equation}
where $X_a$ is the eigenvector associated to an eigenvalue $E_a$ of
a Hermitian matrix $H$. The derivatives of the Hermitian matrices are
sparse and all derivatives of the thermodynamic potential are
therefore obtained practically at the cost of computing the eigenvectors.
The integrands of all derivatives and the thermodynamic potential are
calculated in parallel and the momentum integrals are evaluated simultaneously
to reduce computational redundancy.
The gap and charge neutrality equations \eqns{neutrality}{gapeq} are solved
with a modified Newton method in multidimensions\ie essentially a steepest
descent method, for different starting points in parameter space. The
solution with the highest pressure is used.
The results thereby obtained, and the consequences for the properties
of quark stars are discussed in the next Section.

%%%%%%%%%%%%%%%%%%%%%%%%%%%%%%%%%%%%%%%%%%%%%%%%%%%%%%%%%%%%%%%%%%%%%%%%%%%%%

\section{Results}
\label{sec:results}

We use the same parametrization of the quark matter model
as in \cite{Blaschke:2005uj}
\begin{subequations}
\label{parameters}
\begin{eqnarray}
m_{u,d}&=& 5.5~\text{MeV},\\
m_s &=& 112.0~\text{MeV},\\
G_S \Lambda^2 &=& 2.319,\\
\Lambda &=& 602.3~\text{MeV}.
\end{eqnarray}
\end{subequations}
The relative diquark coupling strength, $\eta$, is considered as a free
parameter of the model. Here we present results for $\eta=0.75$ (intermediate
coupling), which is the vacuum result obtained by a Fierz transformation,
and $\eta=1.0$ (strong coupling), which is motivated by the phenomenology
of compact stars and heavy ion collisions \cite{Klahn:2006iw}.
Quark matter phases are characterized by the order parameters, $\phi_i$
and $\Delta_{ij}$. In particular, the following phases have been
identified in the numerical investigation of the model
\label{phases}
\begin{eqnarray*}
        &&\text{2SC phase: $\Delta_{us}=\Delta_{ds}=0$ and $\Delta_{ud}\neq 0$}, \\
        &&\text{uSC phase: $\Delta_{ds}=0$ and $\Delta_{ud}\neq 0$, $\Delta_{us}\neq 0$}, \\
        &&\text{dSC phase: $\Delta_{us}=0$ and $\Delta_{ud}\neq 0$, $\Delta_{ds}\neq 0$}, \\
        &&\text{CFL phase: $\Delta_{ud}\neq 0$, $\Delta_{us}\neq 0$, and $\Delta_{ds}\neq 0$}.
        \,\,\,\,\,\,\,\,\,\,
\end{eqnarray*}
In addition, gapless phases exist, which are characterized by the presence
of one or more quasiparticle dispersion relations that have no forbidden 
energy band above the Fermi surface. 
Such excitations exist when the differences between the Fermi momenta and/or 
the renormalized masses of the paired quarks are sufficiently large
\cite{Sandin:2005um}. We denote gapless phases with a leading ``g''\eg g2SC 
for the gapless 2SC phase.

In the following we present results for two different neutrino untrapping
scenarios. In one scenario the initial state is a hot, $T=40$~MeV, quark core
with trapped neutrinos, $\mu_{\nu_e}=200$~MeV. The core cools rapidly by
neutrino emission and a cold, $T\sim 1$~MeV, configuration with trapped
neutrinos, $\mu_\nu=200$~MeV, forms. The mass defect due to cooling is
obtained by comparing the masses of these two states for configurations
of equal baryon number.
At low temperature, $T\sim 1$~MeV, the mean-free path of neutrinos increases
and becomes comparable to the size of the core/star. The neutrinos therefore
escape and the final state is a cold, $T\sim 1$~MeV, star with $\mu_\nu=0$.
The mass defect due to neutrino untrapping is obtained
by comparing the masses of $\mu_\nu=200$~MeV, $T\sim 1$~MeV configurations
with the masses of $\mu_\nu=0$, $T\sim 1$~MeV configurations of equal baryon
number. The ``final'' state will continue to cool for millions of years
with practically no effect on its structure, because temperatures below 
$1$ MeV have negligible effects on the quark matter equation of state. Phase
diagrams for intermediate and strong coupling are provided for matter with
trapped and untrapped neutrinos.
The second scenario has a more ``conventional'' initial state, characterized
by a fixed lepton number, $Y_{L_e}=0.4$, and entropy per baryon, $s/n_B=1-2$.
In this case the temperature and neutrino chemical potential varies over
the radius of the star. The properties of the quark star solutions and the
effect of neutrino untrapping are similar in these two scenarios. While the
initial state of the second scenario is motivated by detailed core-collapse
simulations, it is more complicated and to some extent it provides less
transparent results.

%%%%%%%%%%%%%%%%%%%%%%%%%%%%%%%%%%%%%%%%%%%%%%%%%%%%%%%%%%%%%%%%%%%%%%%%%%%%%

\subsection{Order parameters at fixed $\mu$}

\begin{figure*}
%\vspace{-5mm}
\centering
\begin{tabular}{ccc}
      \includegraphics[width=.32\textwidth,clip=]{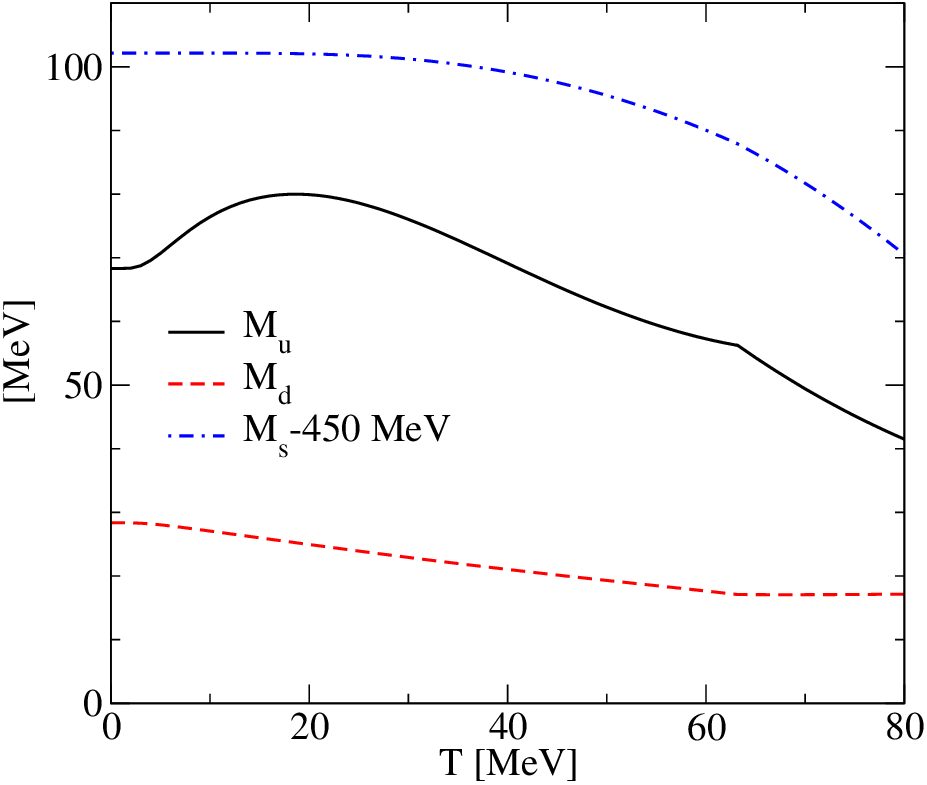} &
      \includegraphics[width=.32\textwidth,clip=]{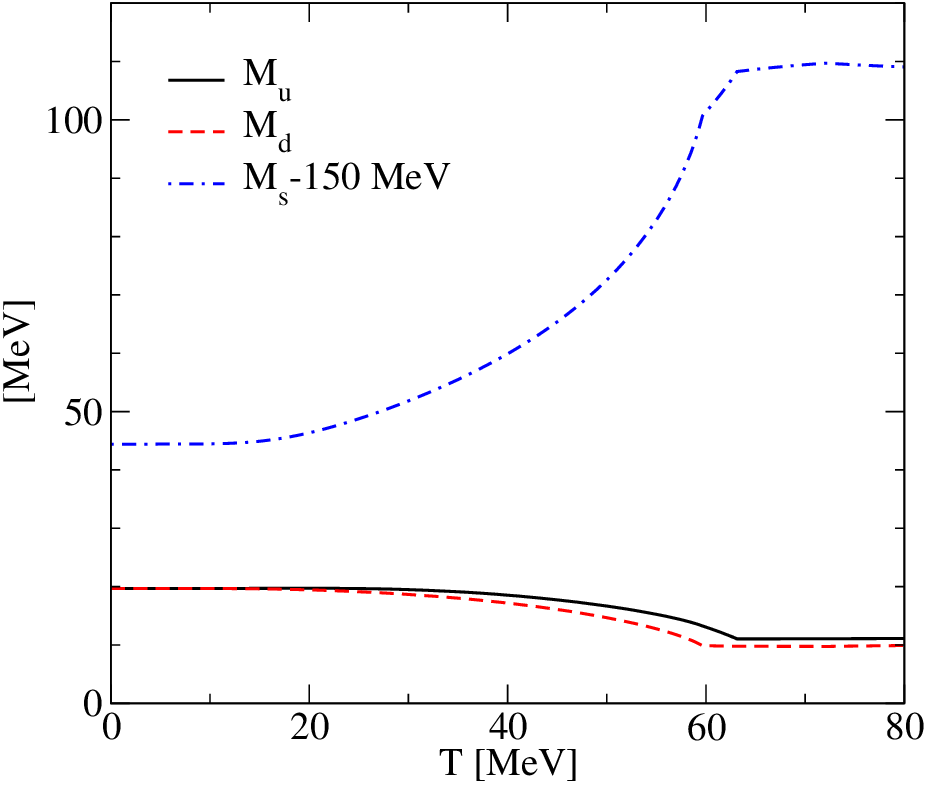} &
      \includegraphics[width=.32\textwidth,clip=]{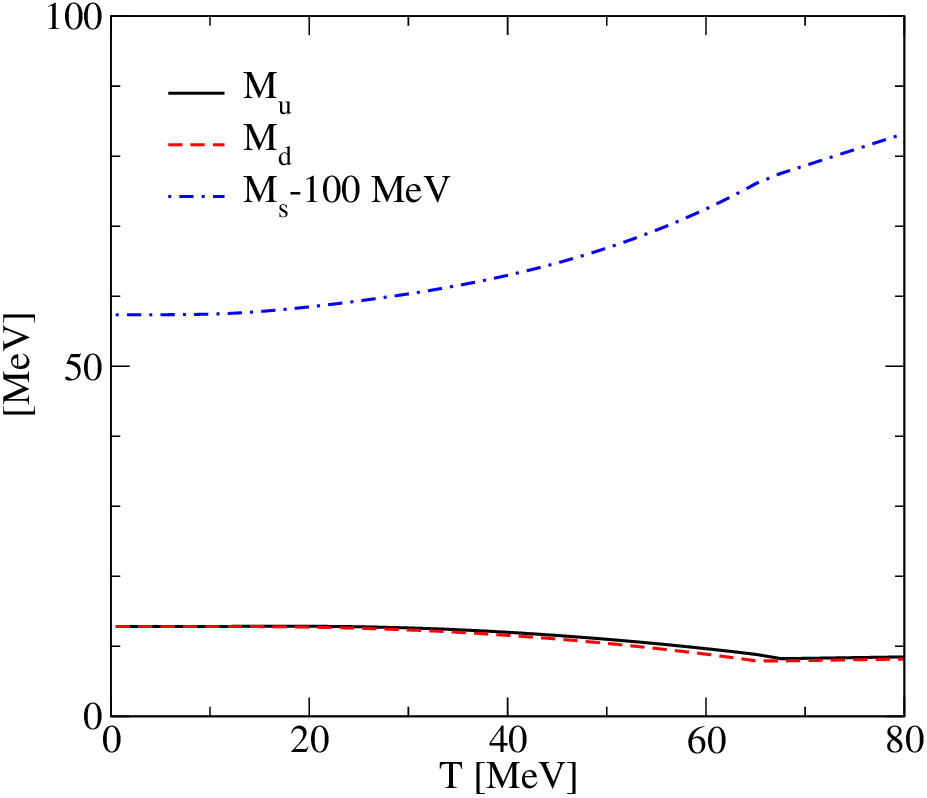} \\
      \includegraphics[width=.32\textwidth,clip=]{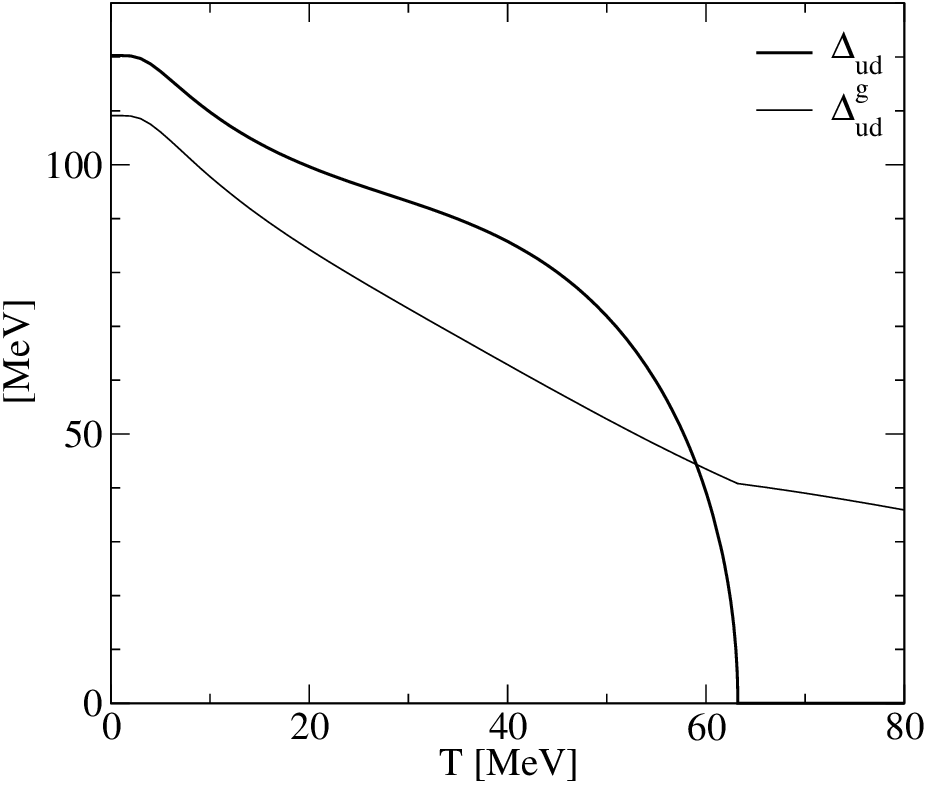} &
      \includegraphics[width=.32\textwidth,clip=]{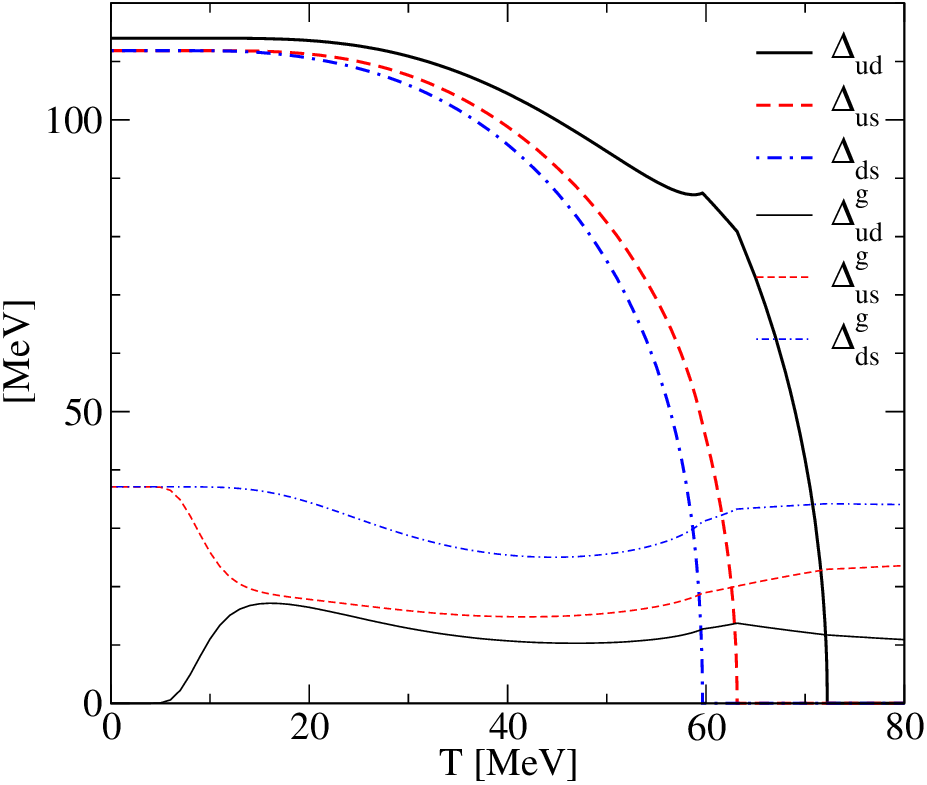} &
      \includegraphics[width=.32\textwidth,clip=]{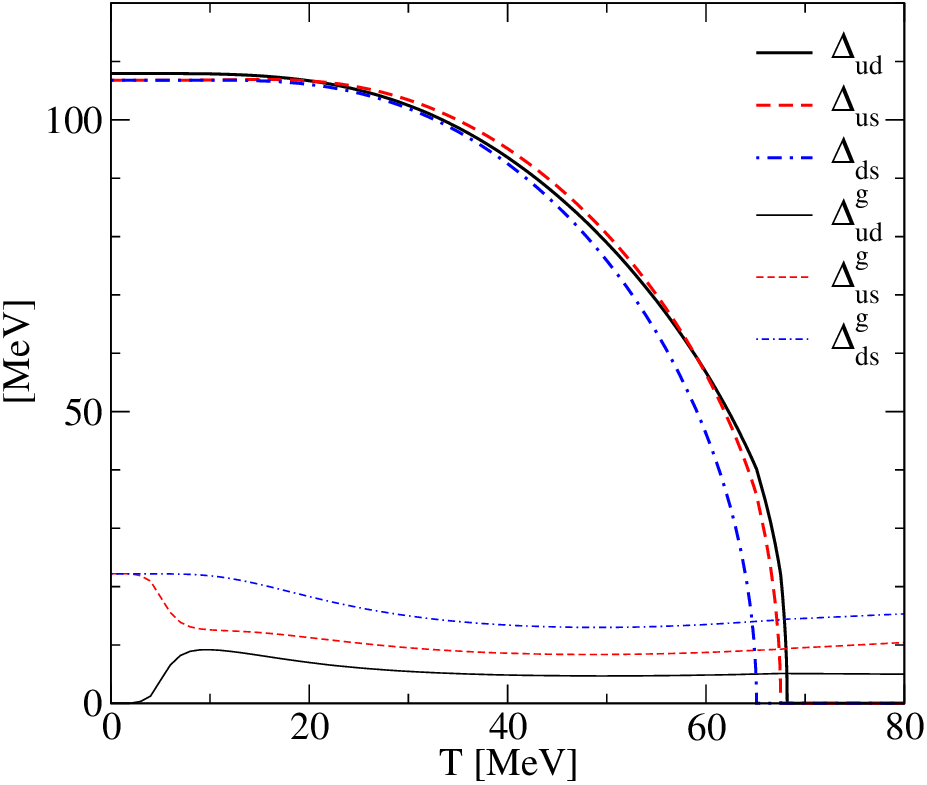} \\
      \includegraphics[width=.32\textwidth,clip=]{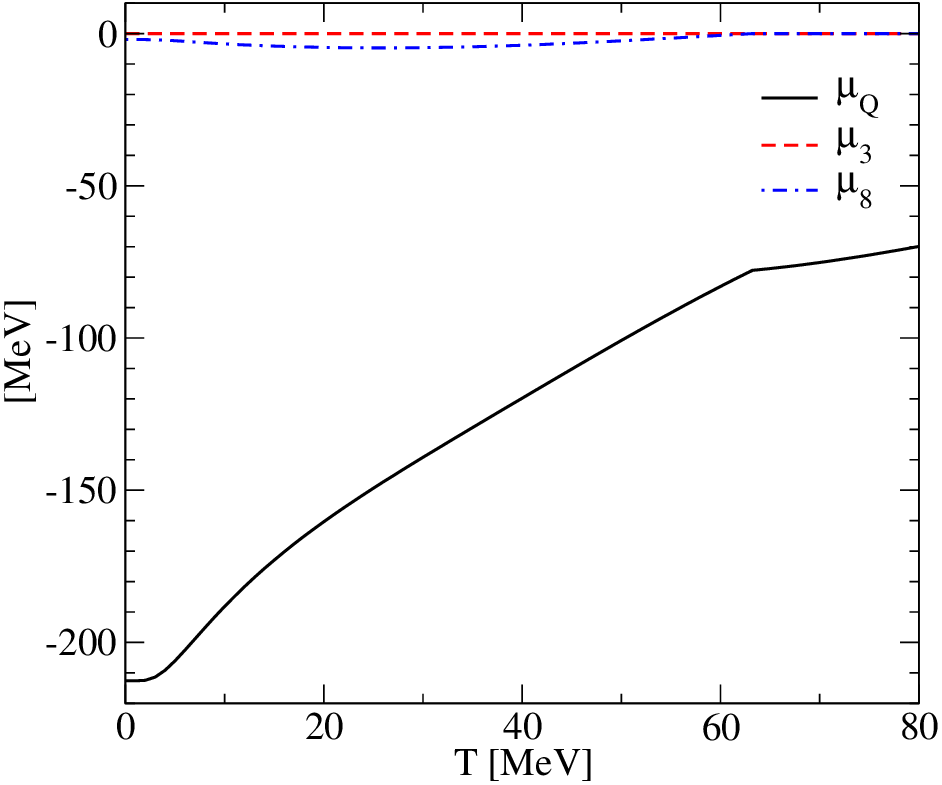} &
      \includegraphics[width=.32\textwidth,clip=]{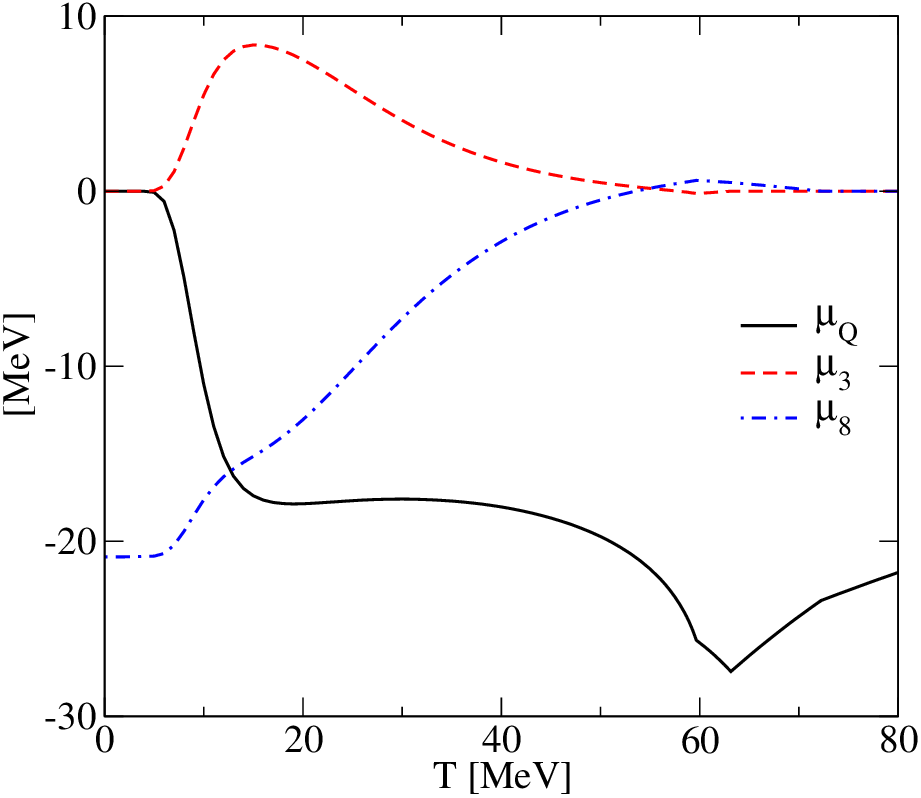} &
      \includegraphics[width=.32\textwidth,clip=]{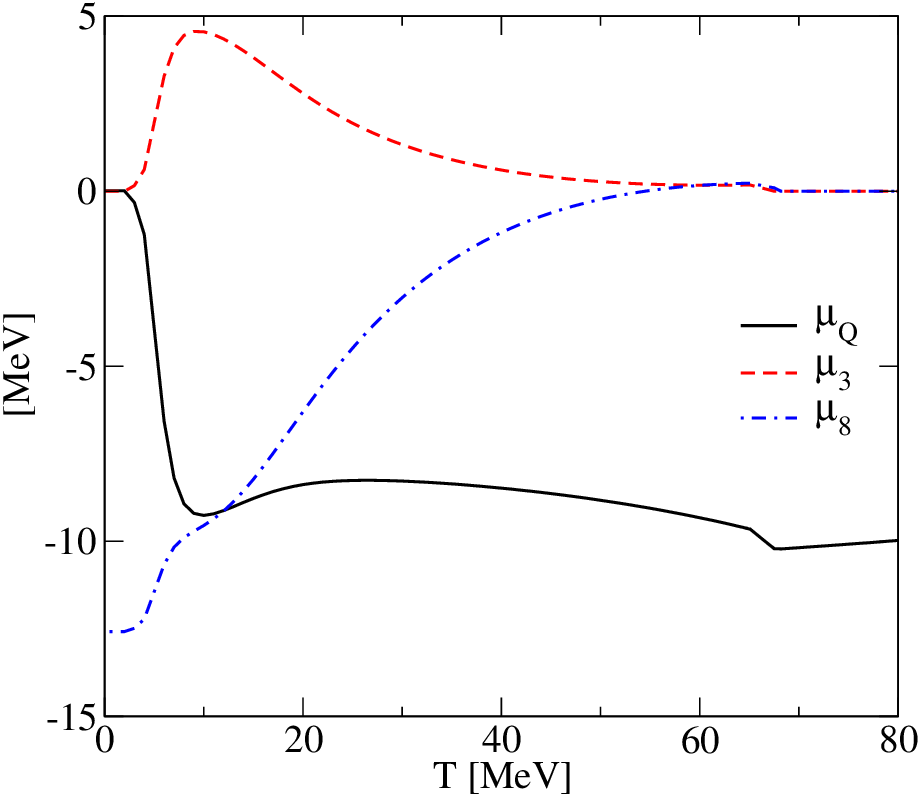} \\
      \includegraphics[width=.32\textwidth,clip=]{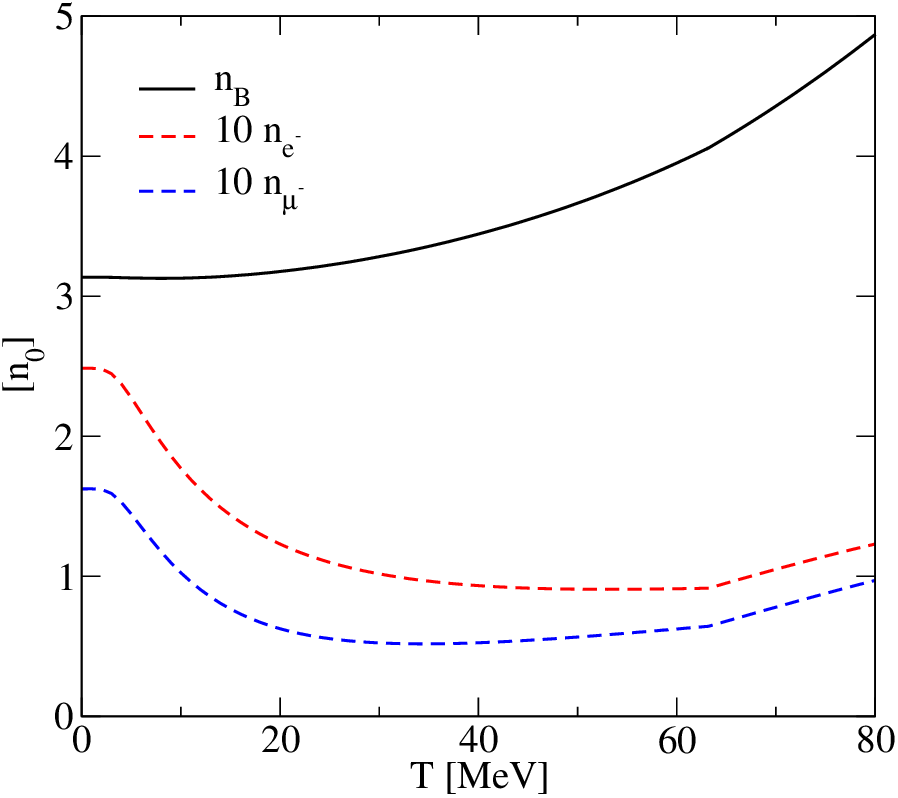} &
      \includegraphics[width=.32\textwidth,clip=]{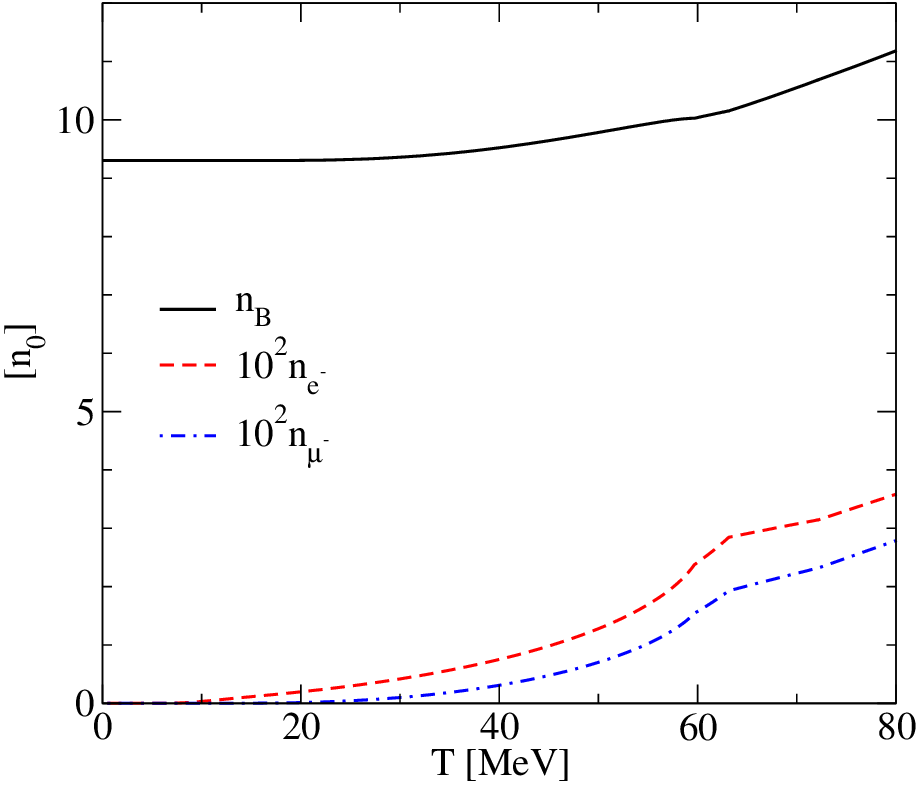} &
      \includegraphics[width=.32\textwidth,clip=]{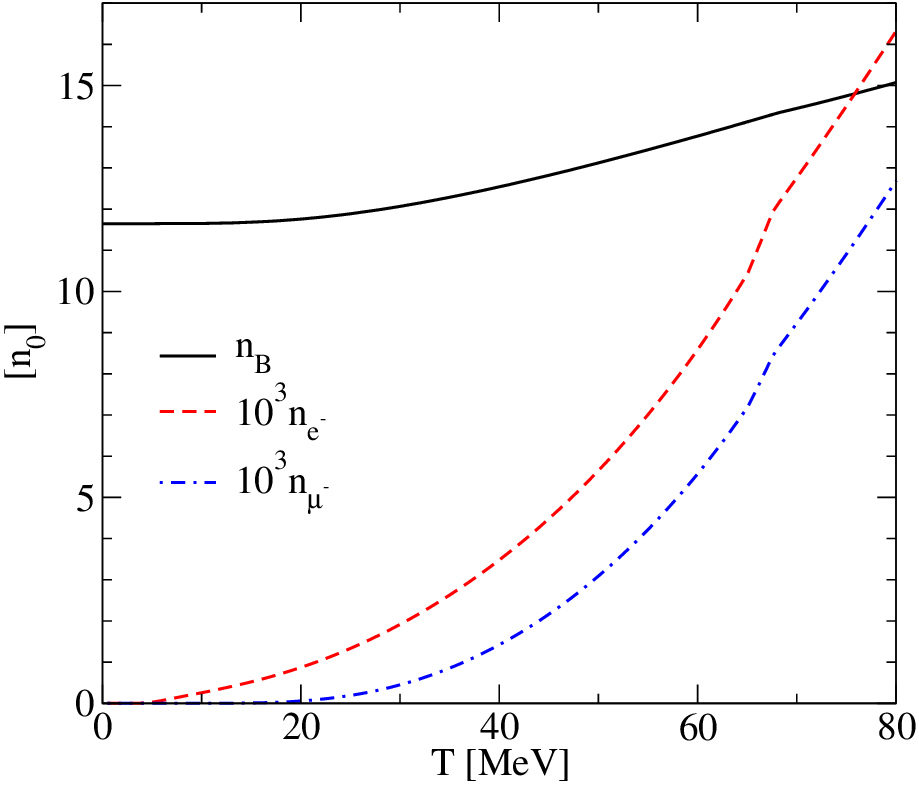}
\end{tabular}
\caption{(Color online)
The constituent quark masses, gaps, charge chemical potentials,
and number densities of charge neutral quark matter in $\beta$-equilibrium
with untrapped neutrinos, $\mu_\nu=0$,
at intermediate coupling, $\eta=0.75$. The three columns (from left to right)
represent solutions for $\mu=400$, $500$, and $550$~MeV. $\Delta^g_{ij}$
are thresholds for the existence of gapless excitations \cite{Sandin:2005um}\ie
gapless excitations exist if $\Delta_{ij} \leq \Delta^g_{ij}$.
%$n_0\sim0.17$~fm$^{-3}$ is the nuclear saturation density.
}
\label{figDetailLe0Eta075}
\end{figure*}

For the calculation of quark matter phase diagrams in the plane of
temperature and quark (or baryon) number chemical potential, the gap equations
for the order parameters of the model (the masses, $M_i$, and diquark gaps,
$\Delta_{ij}$) are solved self-consistently. The values of the order parameters
characterize the different quark matter phases, as described above.
In addition to solving the gap equations, the color and electric charge
neutrality conditions are enforced by solving for the corresponding chemical
potentials ($\mu_Q$, $\mu_3$, $\mu_8$), for given values of the quark number
chemical potential, $\mu$, and temperature, $T$. 
In \figs{figDetailLe0Eta075}{figDetailLe200Eta1}, the 
quark masses (upper row), diquark gaps (second row), chemical potentials
(third row) and densities of baryon number, electrons and muons (bottom row)
are plotted vs. the temperature for three different values of the quark number
chemical potential (columns).
The four Figures represent solutions without ($\mu_\nu=0$) and with 
($\mu_\nu=200$ MeV) trapped neutrinos for intermediate ($\eta=0.75$) and
strong  ($\eta=1.0$) coupling, respectively. Here we use the more compact
notation
\begin{equation}
        \mu_\nu\equiv\mu_{\nu_e}=\mu_{L_e},
\end{equation}
for the electron neutrino chemical potential, because the muon neutrino 
chemical potential is zero as neutrino oscillations are neglected.
Note the gapless constraints, $\Delta^g_{ij}$, that are included
in the second row of these Figures. If $\Delta_{ij}\le \Delta^g_{ij}$, the
corresponding quasiparticle has
a gapless dispersion relation\ie when this condition is met there is no
forbidden energy band above the Fermi surface.
Typically, the gapless conditions are fulfilled only near the critical
temperature ($60 - 70$~MeV for intermediate coupling and $100 - 110$~MeV for
strong coupling), where the gaps go to zero in a second order phase transition
to a non-superconducting state.
Since these temperatures are well above the maximum temperature relevant for 
PNS, we can neglect gapless phases in our discussion.

%%%%%%%%%%%%%%%%%%%%%%%%%%%%%%%%%%%%%%%%%%%%%%%%%%%%%%%%%%%%%%%%%%%%%%%%%%%%%

\subsection{Order parameters at fixed $Y_{L_e}$ and $s/n_B$}

\begin{figure*}
%\vspace{-5mm}
\centering
\begin{tabular}{ccc}
        \includegraphics[width=.3\textwidth,clip=]{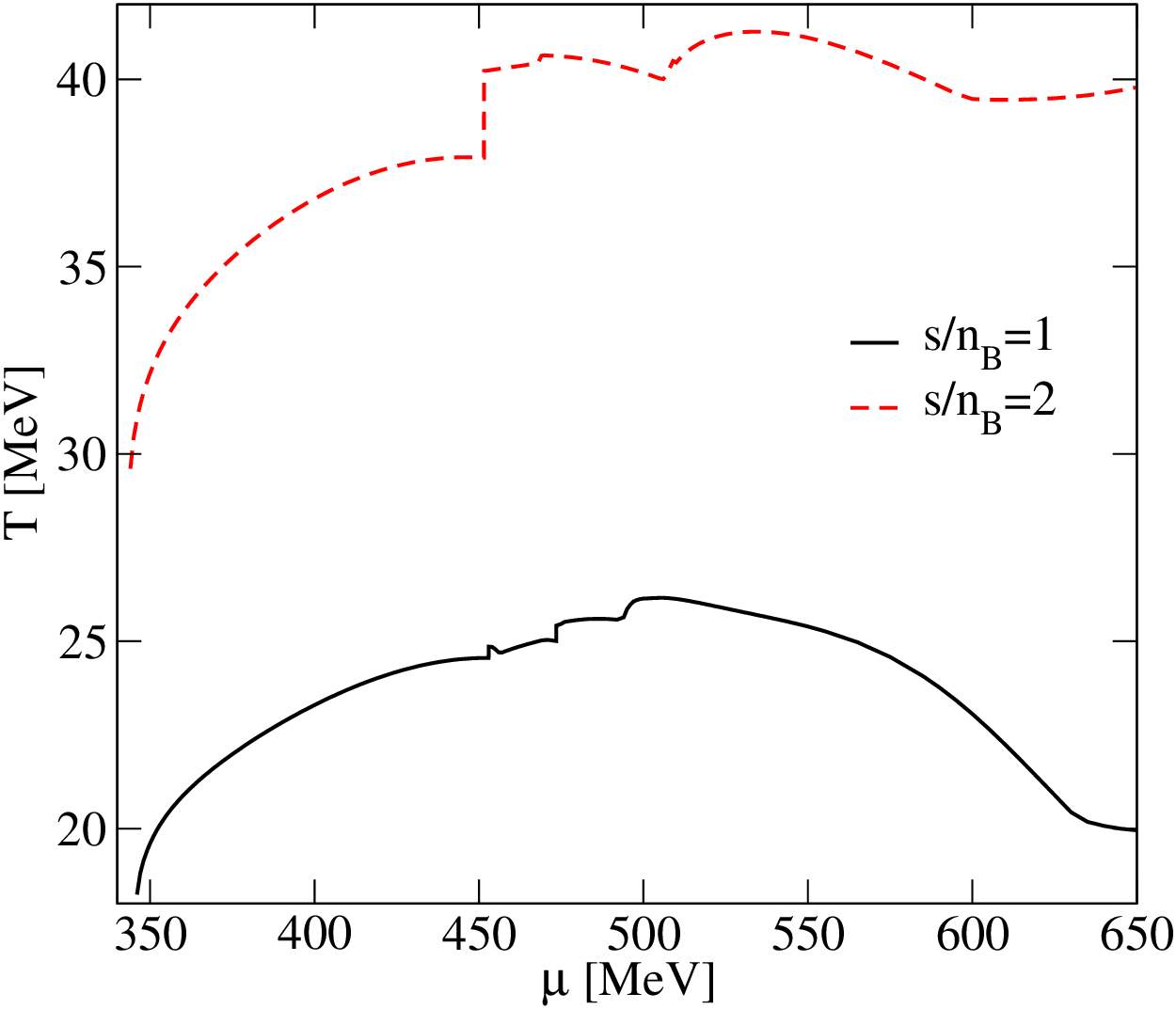} &
        \includegraphics[width=.3\textwidth,clip=]{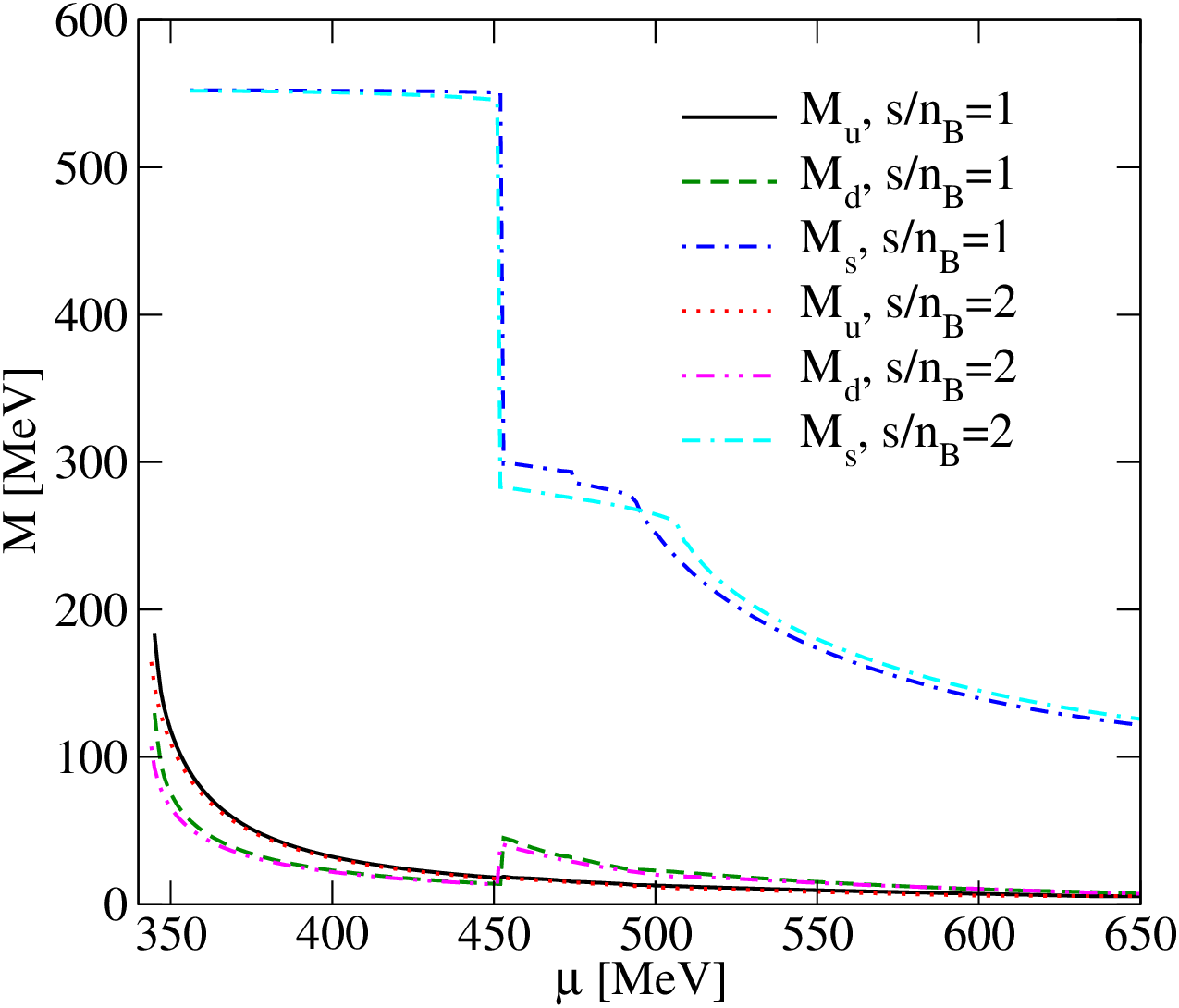} &
        \includegraphics[width=.3\textwidth,clip=]{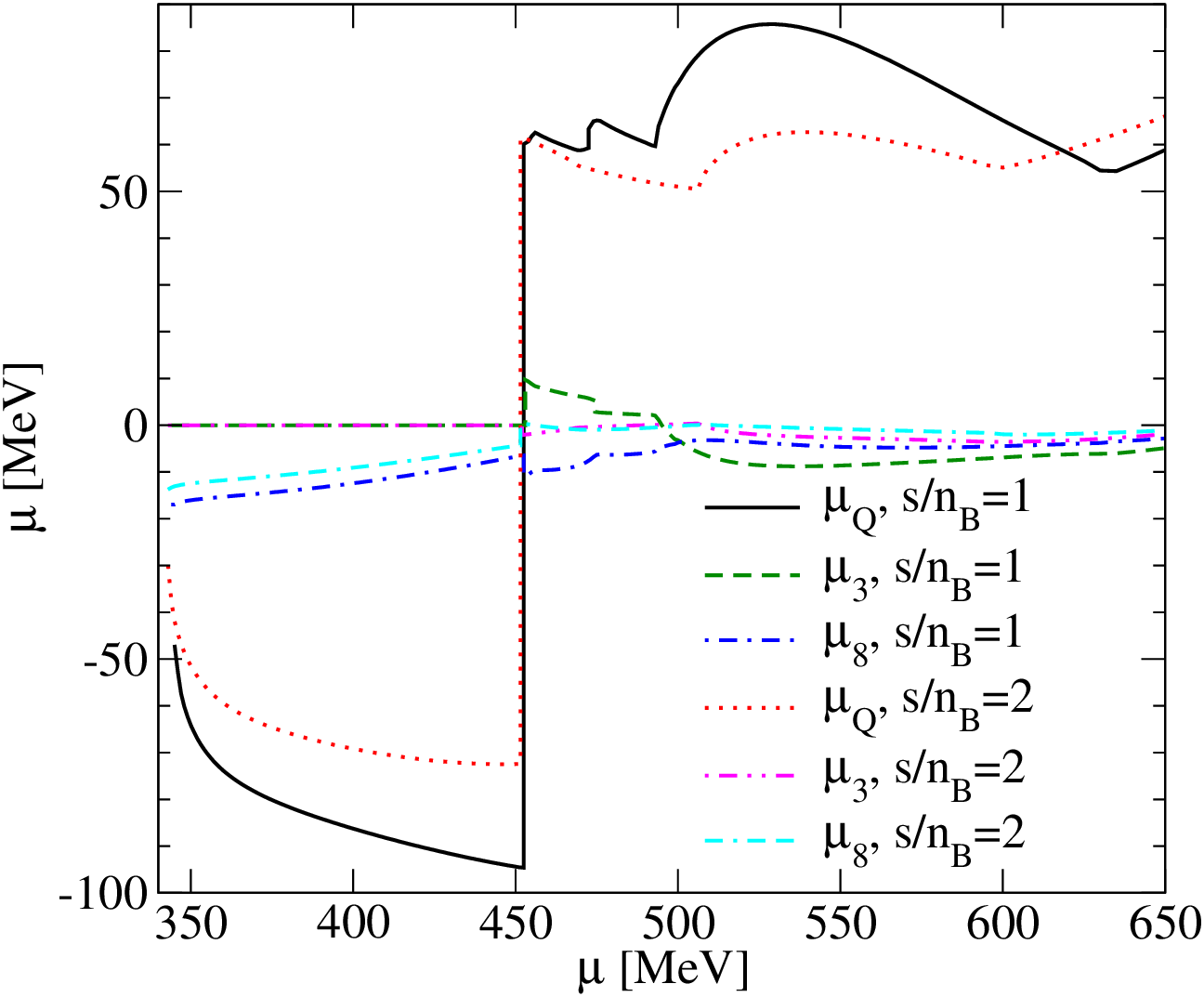} \\
        \includegraphics[width=.3\textwidth,clip=]{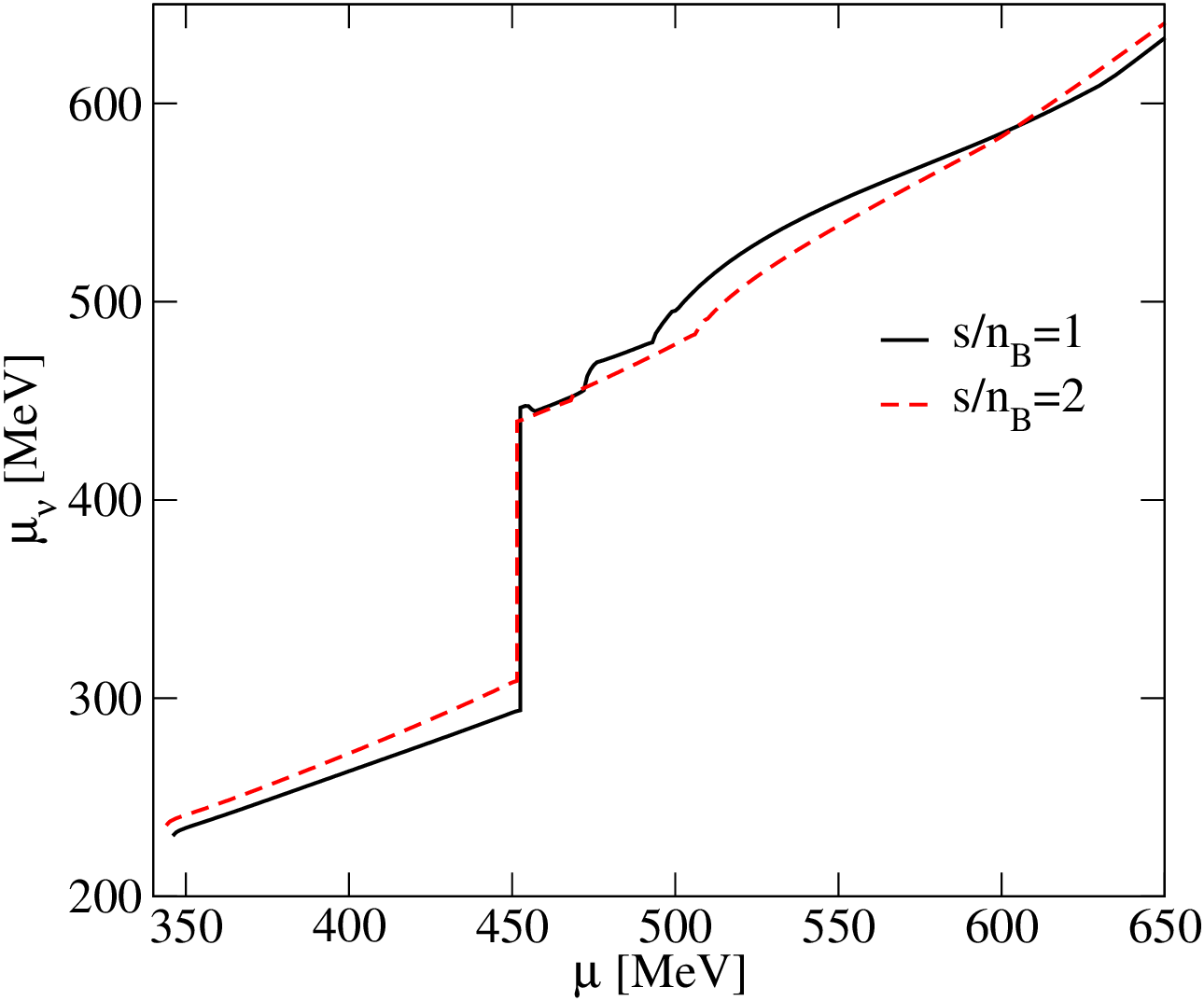} &
        \includegraphics[width=.3\textwidth,clip=]{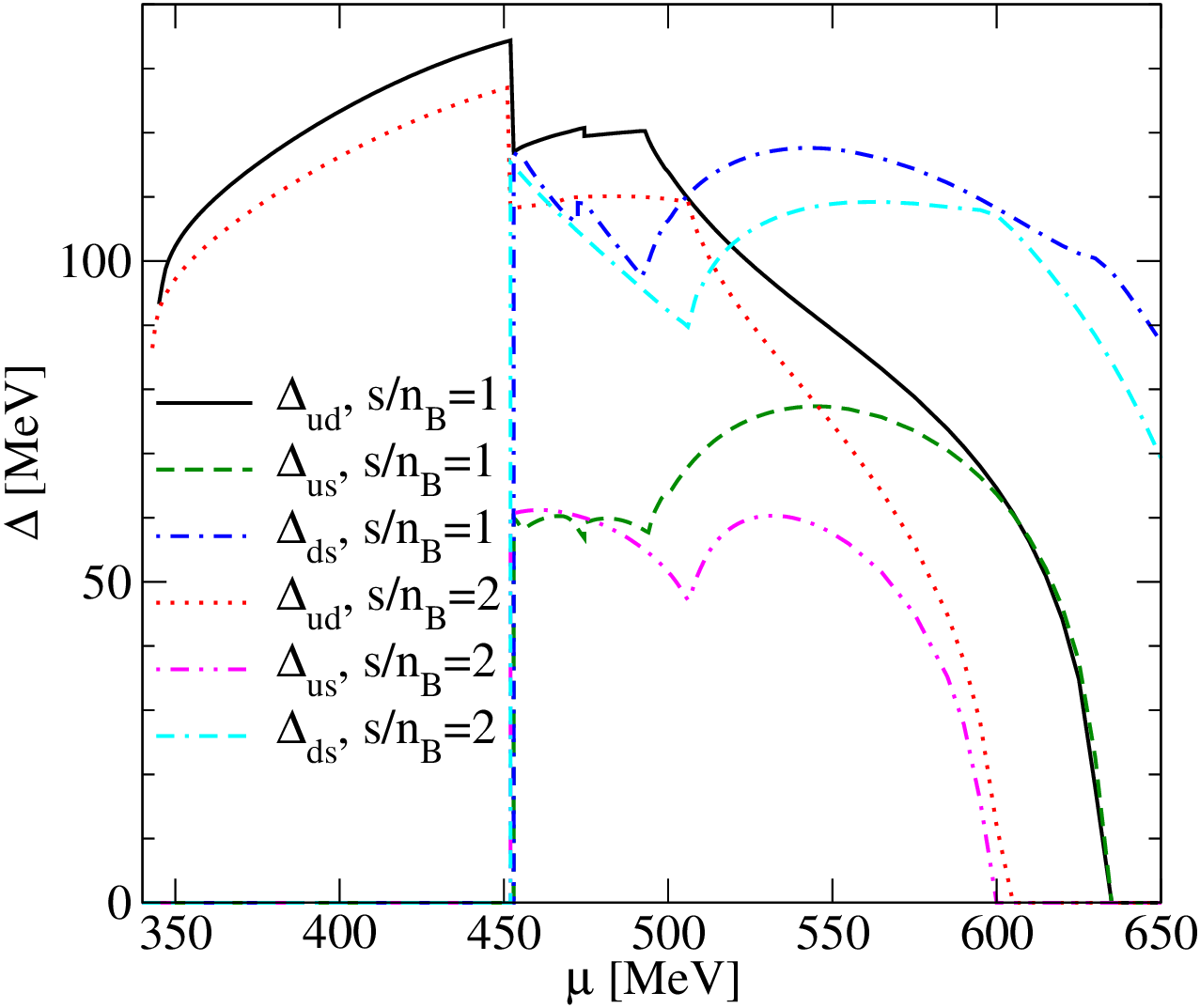} &
        \includegraphics[width=.3\textwidth,clip=]{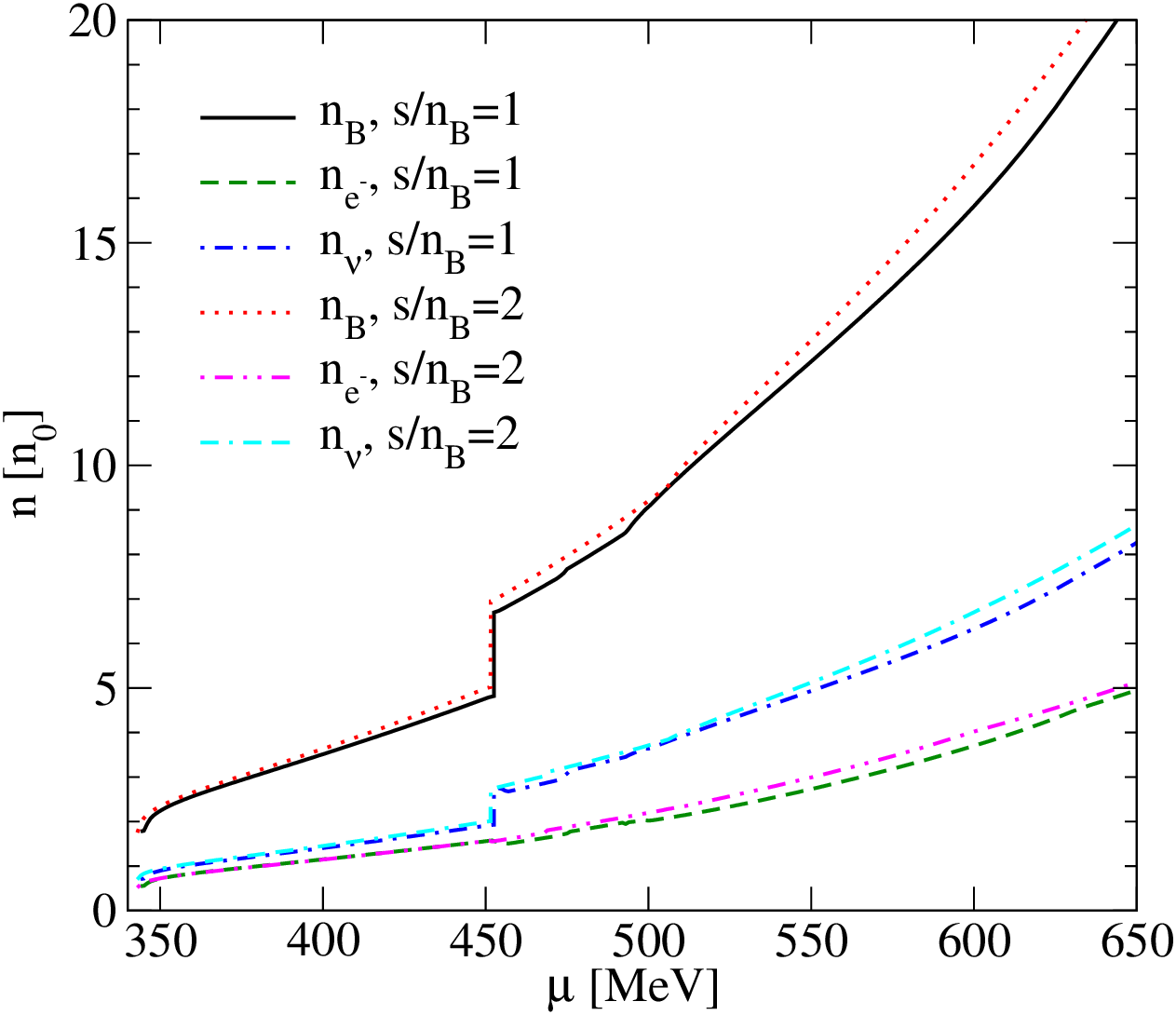}
\end{tabular}
\caption{(Color online)
The temperature, constituent quark masses, charge chemical potentials,
neutrino chemical potential, gaps, and number densities of charge neutral
quark matter in $\beta$-equilibrium at intermediate coupling ($\eta=0.75$)
for fixed lepton fraction $Y_{L_e}(T,\mu_\nu)=0.4$, and two values of the 
entropy per baryon, $s(T,\mu_\nu)/n_B(T,\mu_\nu) = 1, ~2$.}
\label{figYLe04eta075}
\end{figure*}

\begin{figure*}
%\vspace{-5mm}
\centering
\begin{tabular}{ccc}
        \includegraphics[width=.3\textwidth,clip=]{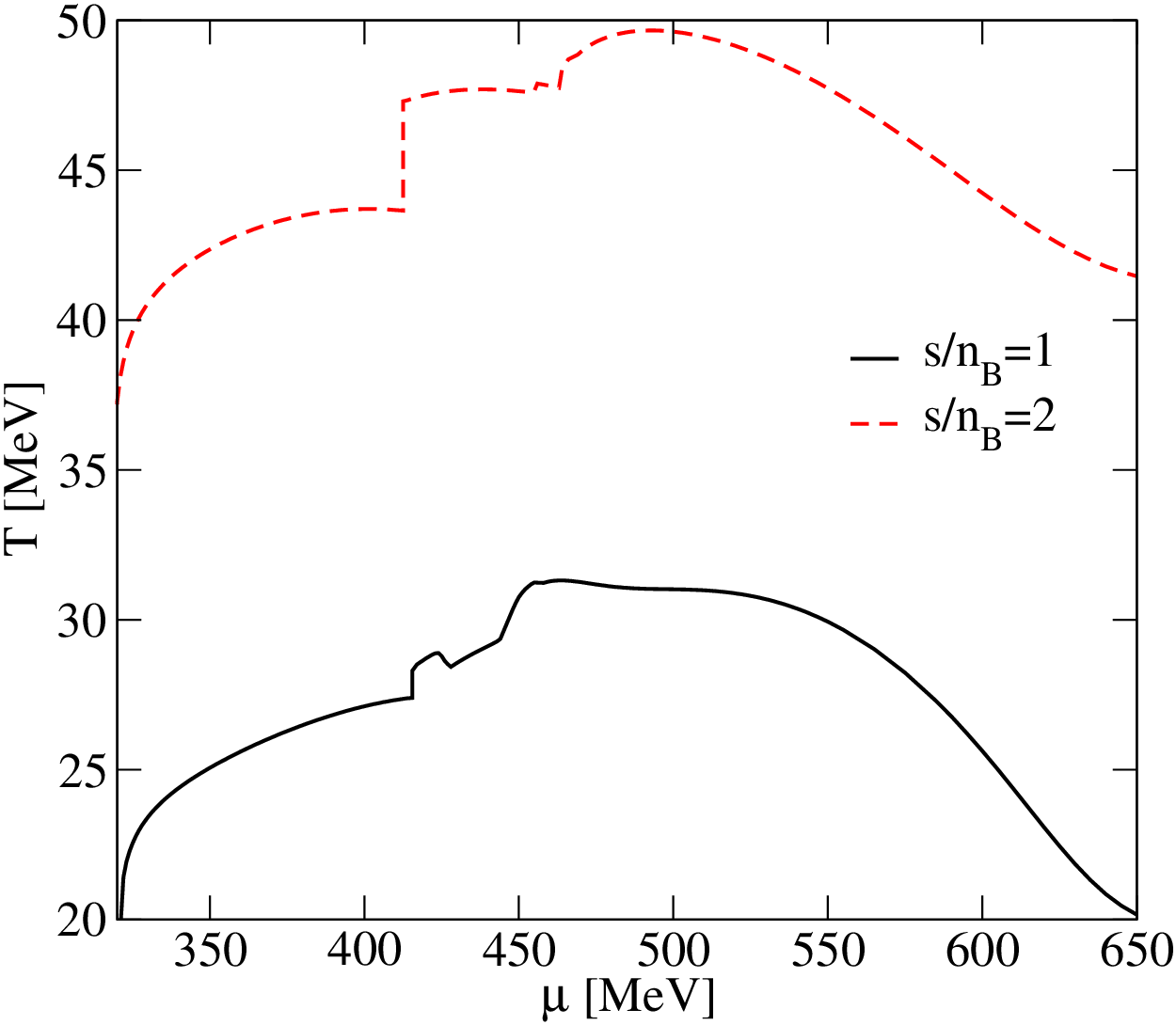} &
        \includegraphics[width=.3\textwidth,clip=]{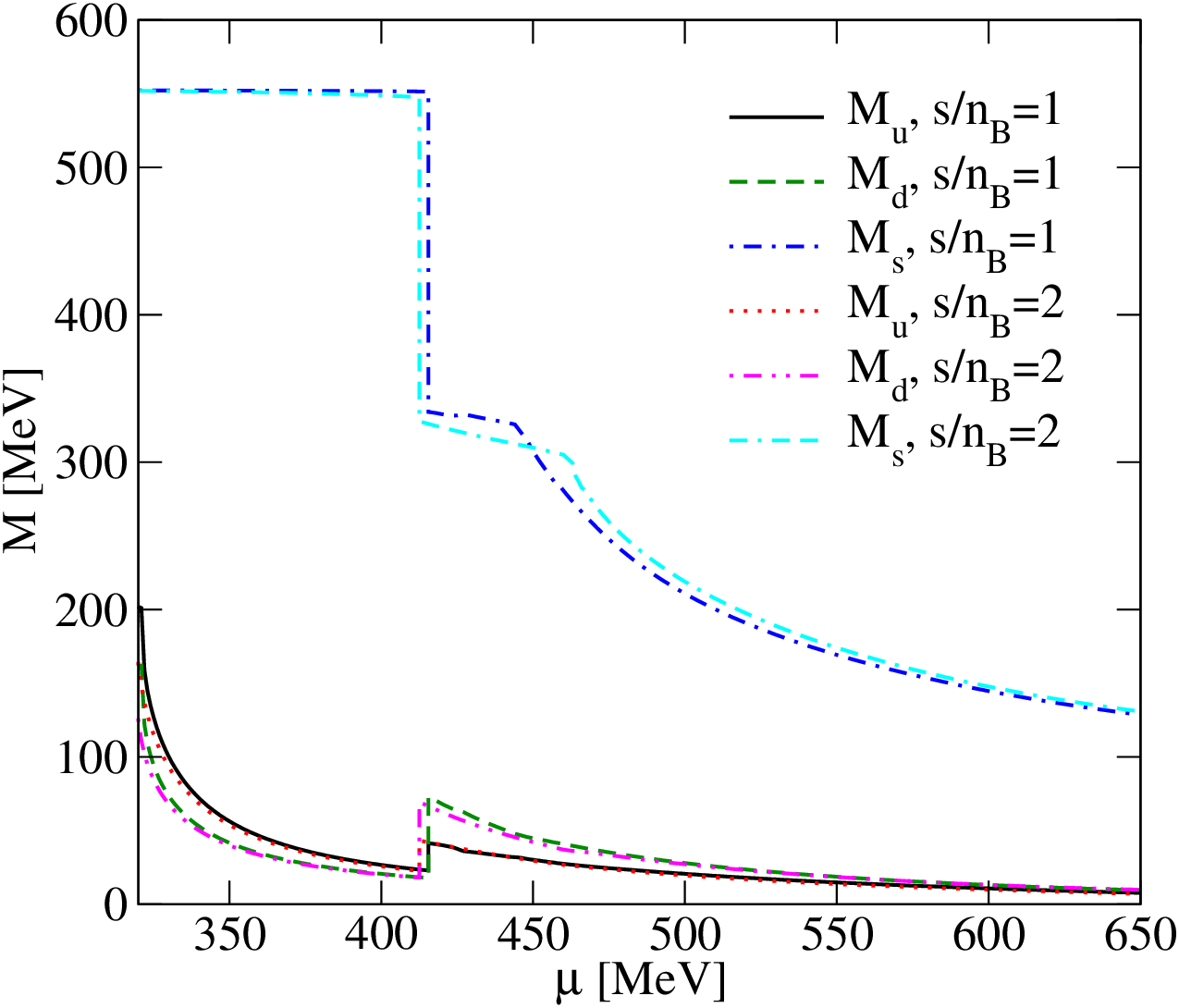} &
        \includegraphics[width=.3\textwidth,clip=]{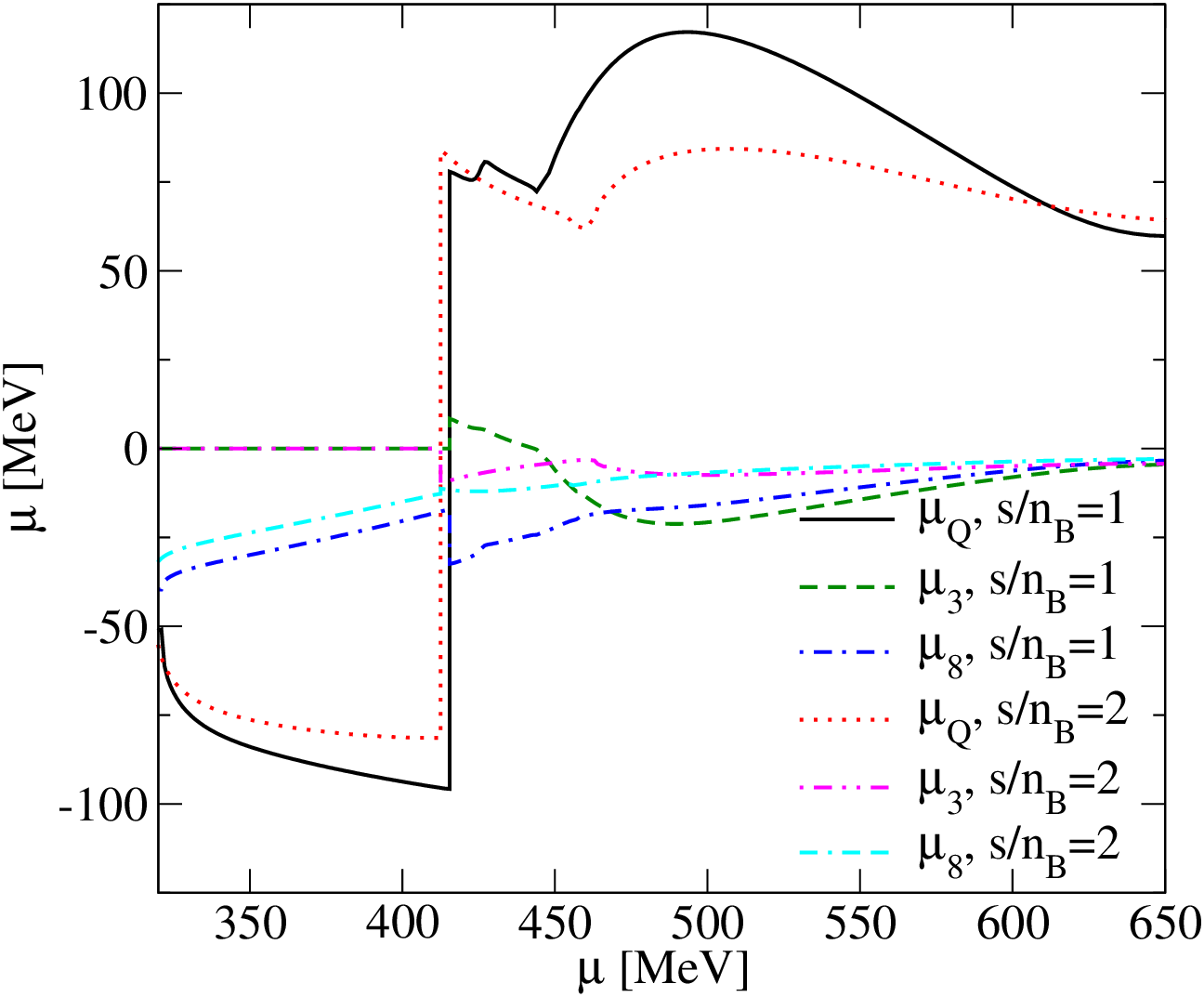} \\
        \includegraphics[width=.3\textwidth,clip=]{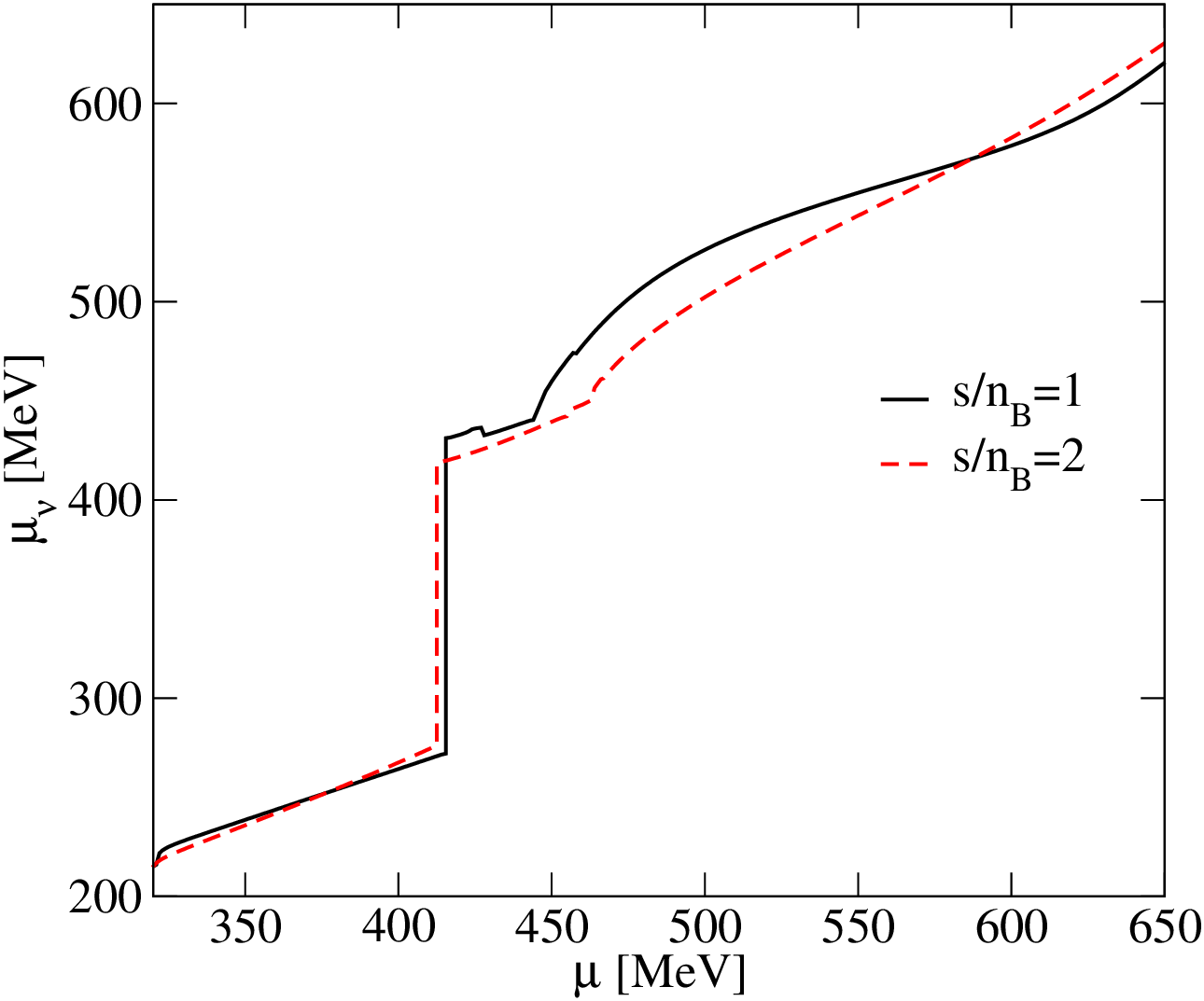} &
        \includegraphics[width=.3\textwidth,clip=]{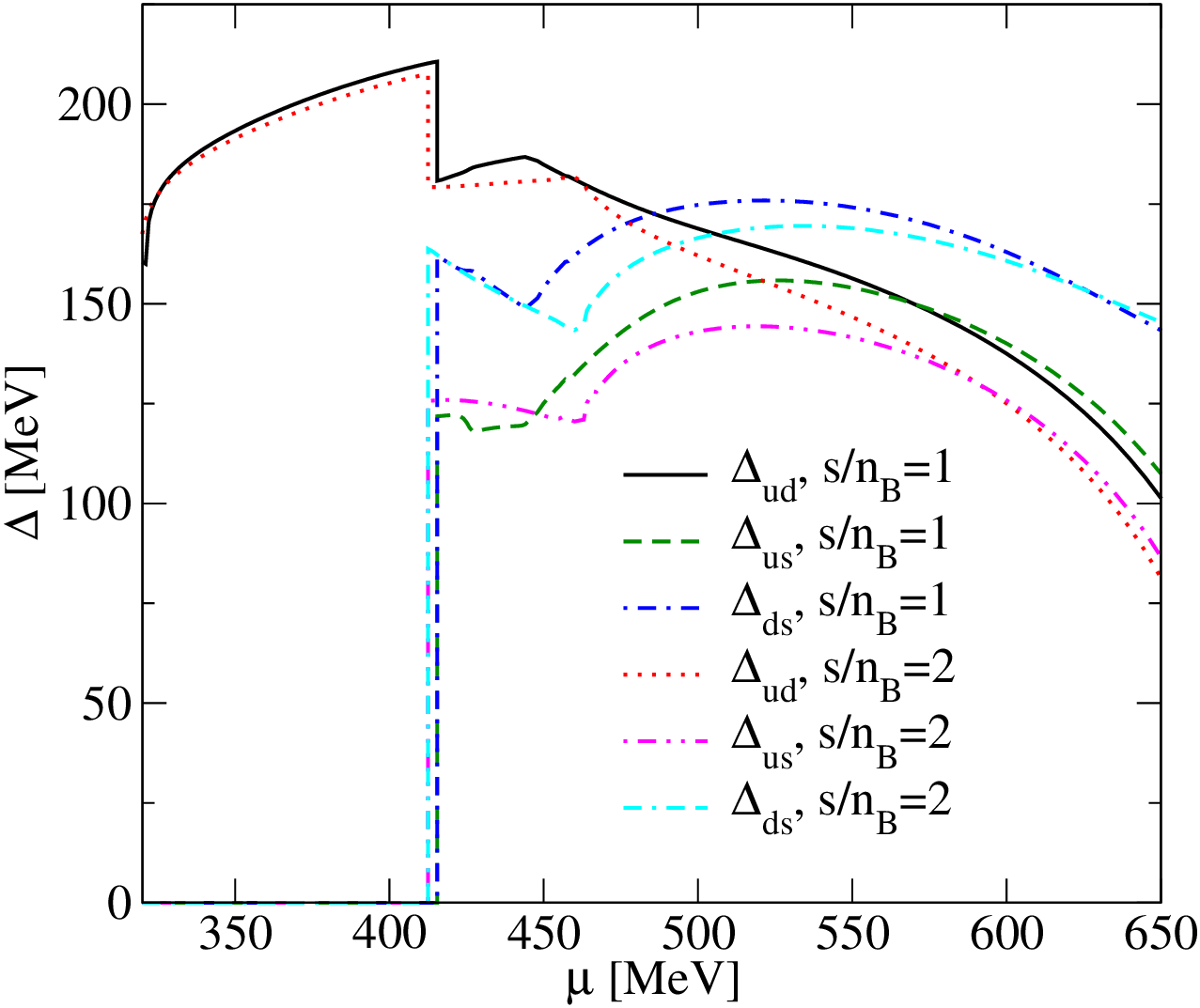} &
        \includegraphics[width=.3\textwidth,clip=]{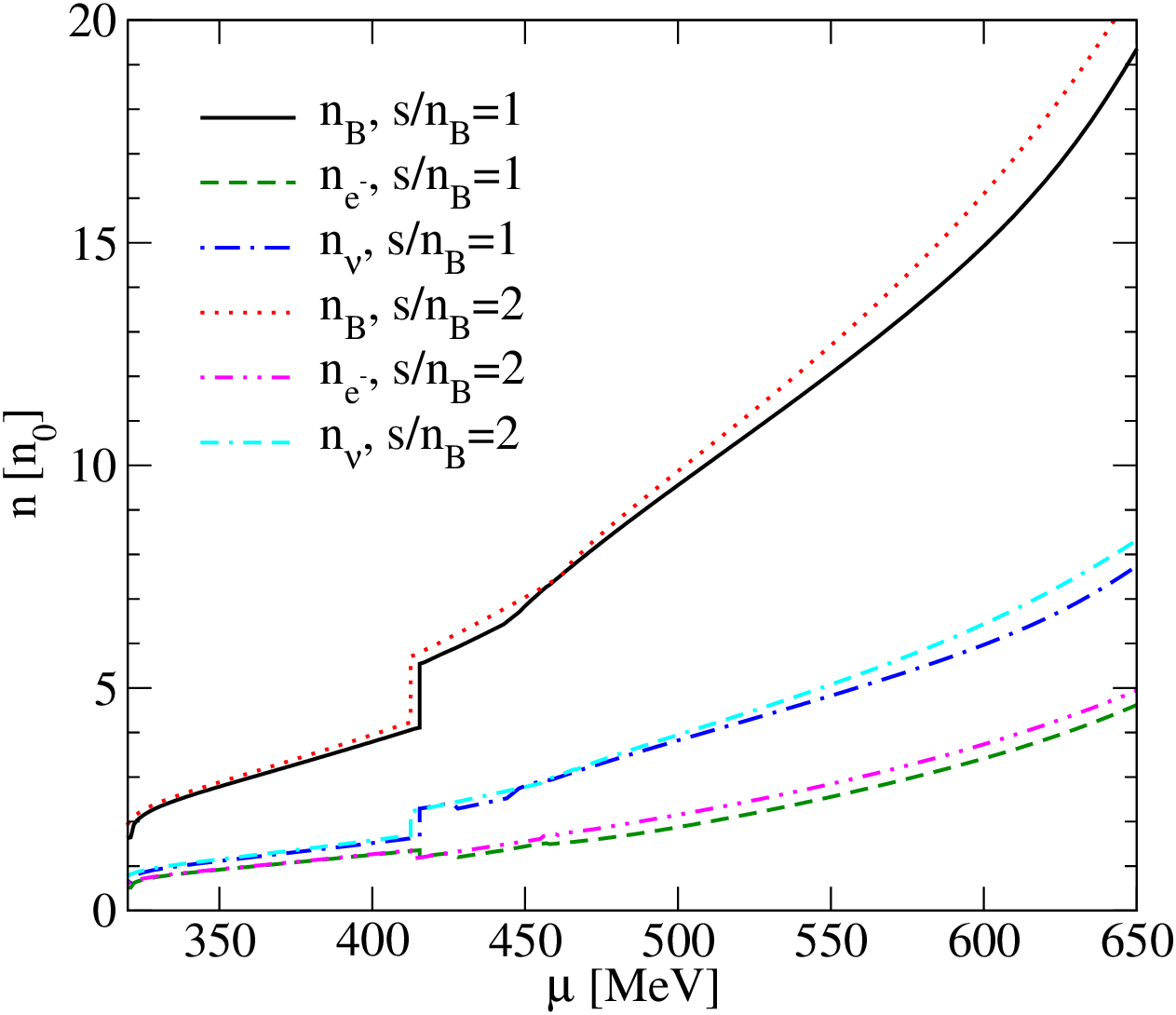}
\end{tabular}
\caption{(Color online)
The same as Fig. \ref{figYLe04eta075} but at strong coupling, $\eta=1$.}
\label{figYLe04eta1}
\end{figure*}

Next we consider a system with fixed values of the lepton fraction,
$Y_{L_e}(T,\mu_\nu)=0.4$, and the entropy per baryon, 
$s(T,\mu_\nu)/n_B(T,\mu_\nu)=1,~2$. Consequently, for a given value
of the quark number chemical potential, $\mu$, the temperature
and electron neutrino chemical potential are determined such that
$Y_{L_e}=0.4$ and $s/n_B=1$ (or $2$). These two equations are solved
in parallel with the gap and charge neutrality equations.
In \fig{figYLe04eta075} 
the temperature, constituent quark masses, charge chemical potentials,
neutrino chemical potential, gaps, and number densities are plotted vs.
the quark number chemical potential for
charge neutral quark matter in $\beta$-equilibrium at
intermediate coupling, $\eta=0.75$. \fig{figYLe04eta1} shows the same
relationships for strong coupling, $\eta=1$.

%%%%%%%%%%%%%%%%%%%%%%%%%%%%%%%%%%%%%%%%%%%%%%%%%%%%%%%%%%%%%%%%%%%%%%%%%%%%%

\subsection{Equations of state}

\begin{figure}
%\vspace{-5mm}
\includegraphics[width=0.45\textwidth,angle=0]{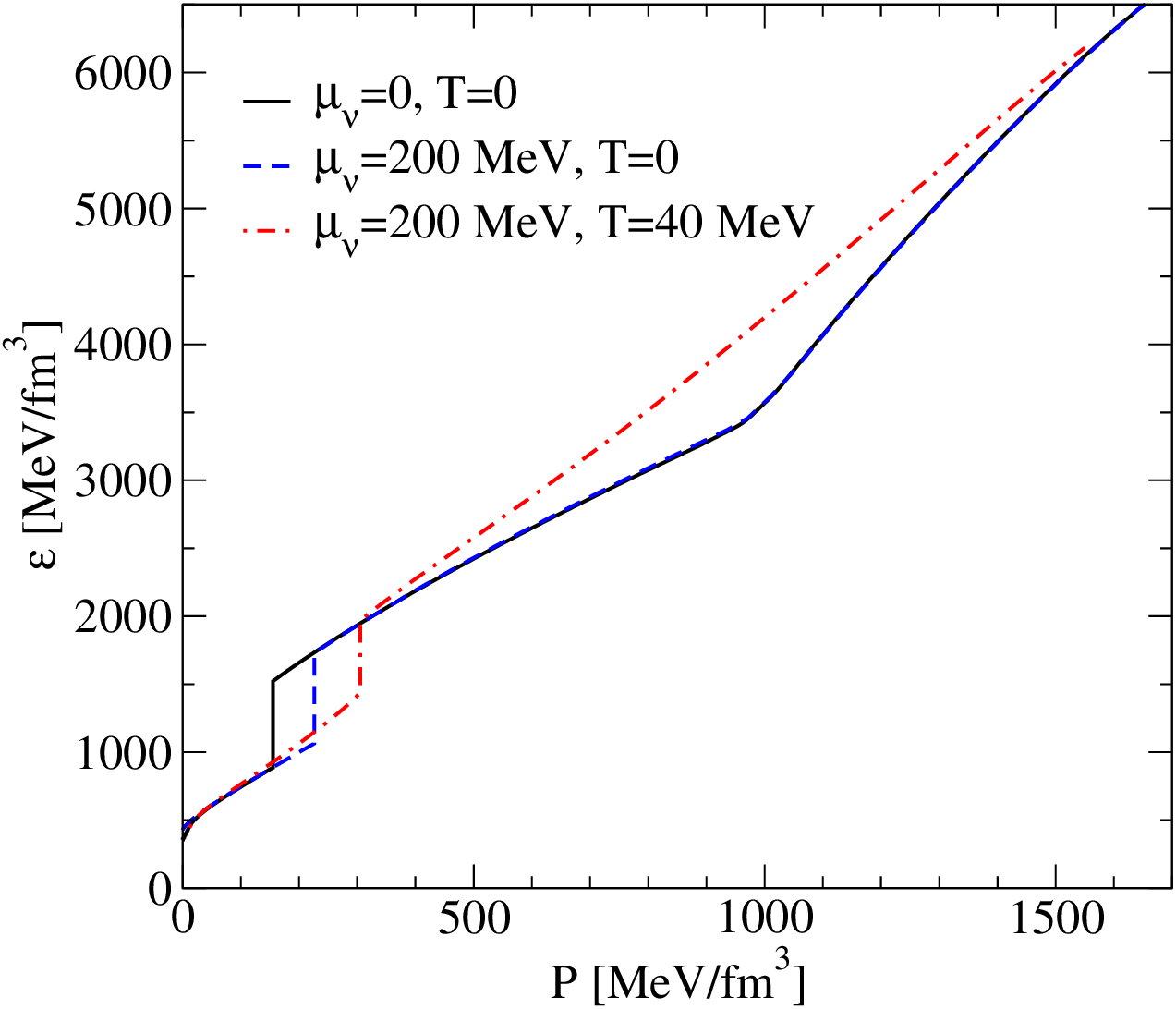}
\caption{(Color online)
        The equation of state for charge neutral quark matter in
        $\beta$-equilibrium at intermediate coupling, $\eta=0.75$, and
        fixed values of the temperature and electron neutrino chemical
        potential. The discontinuities appear at the 2SC-CFL phase
        transition due to the Maxwell construction, see text.}
\label{figEosEta075}
\end{figure}

The equation of state (EoS), $\epsilon(P)$, can be calculated using 
\eqn{eoseq}\eg for different values of the temperature, $T$, and lepton number 
chemical potential, $\mu_{L_e}=\mu_\nu$, or for fixed values of lepton 
fraction, $Y_{L_e}(T,\mu_{L_e})$, and the entropy per baryon, 
$s(T,\mu_{L_e})/n_B(T,\mu_{L_e})$, depending on the application.
In \figs{figEosEta075}{figEosEta1} the EoS for three different
pairs of $(\mu_\nu,T)$ values are plotted for intermediate, $\eta=0.75$,
and strong, $\eta=1.0$, coupling, respectively.
%
%The 2SC-CFL phase transition is first order, signalled by a jump in the energy 
%density. This behavior is triggered by the discontinuity of the strange quark 
%mass obtained from the selfconsistent solution of the system of gap equations 
%\eqn{gapeq}.
%
The energy density is discontinuous at the 2SC-CFL phase transition
due to the discontinuity of the strange quark mass, which is obtained
from the self-consistent solution of the system of gap equations \eqn{gapeq}.
The phase transition is first order.
When increasing $T$ and/or $\mu_\nu$, the pressure at the transition increases.
\fig{figEosYLe04} shows the EoS for a fixed lepton fraction,
$Y_{L_e}(T,\mu_\nu)=0.4$, and two values of entropy per baryon, 
$s(T,\mu_\nu)/n_B(T,\mu_\nu)=1$ and $2$, for intermediate and strong coupling.
When increasing the entropy per baryon, the pressure at the transition 
increases, whereas an increase of the coupling strength lowers it.

%%%%%%%%%%%%%%%%%%%%%%%%%%%%%%%%%%%%%%%%%%%%%%%%%%%%%%%%%%%%%%%%%%%%%%%%%%%%%

\subsection{Quark star sequences}

Given an EoS described in the previous Section,
the corresponding compact star sequence is calculated by solving
the Tolman-Oppenheimer-Volkoff equations for a static spherically
symmetric object
\begin{subequations}
\label{toveq}
\begin{eqnarray}
        \frac{dP(r)}{dr} &=& -\frac{[\epsilon(r)+P(r)][m(r)+4\pi r^{3}P(r)]}{r[r-2m(r)]}, \\
        m(r) &=& 4\pi \int_{0}^{r}\varepsilon(r')r'^{2}dr',
\end{eqnarray}
\end{subequations}
for different values of the central pressure.
Due to the thermal pressure, the high-temperature EsoS do not extend to
zero pressure.
%
% OBS: The following sentence is divided into two pieces in
% order to avoid a nearly empty page before the figures
The surface of hot configurations
is therefore defined to be the point where the (approxi-
% mate) chiral symmetry is broken by
% a first order phase transition into the $\chi$SB phase.

\begin{figure*}
%\vspace{-5mm}
\centering
\begin{tabular}{ccc}
     \includegraphics[width=.3\textwidth,clip=]{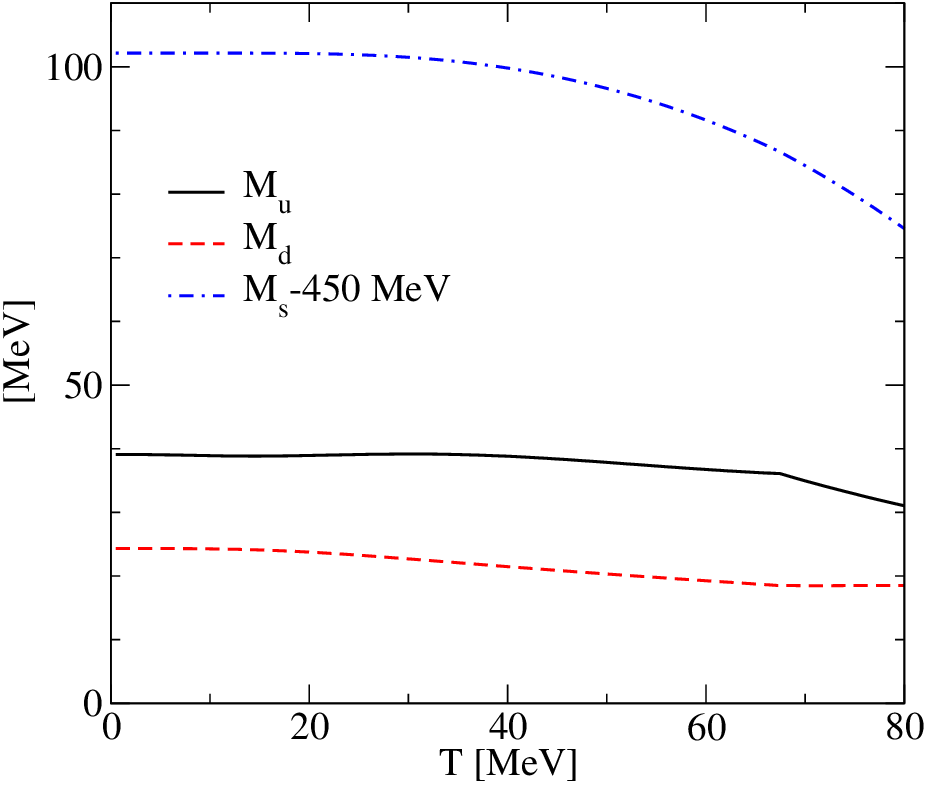} &
     \includegraphics[width=.3\textwidth,clip=]{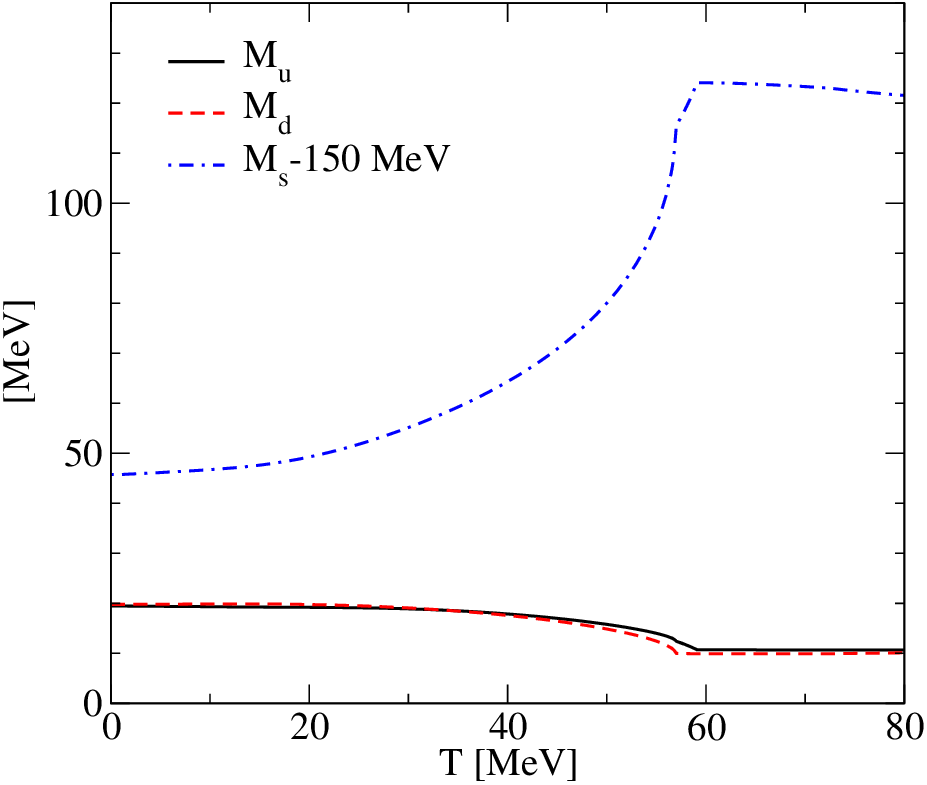} &
     \includegraphics[width=.3\textwidth,clip=]{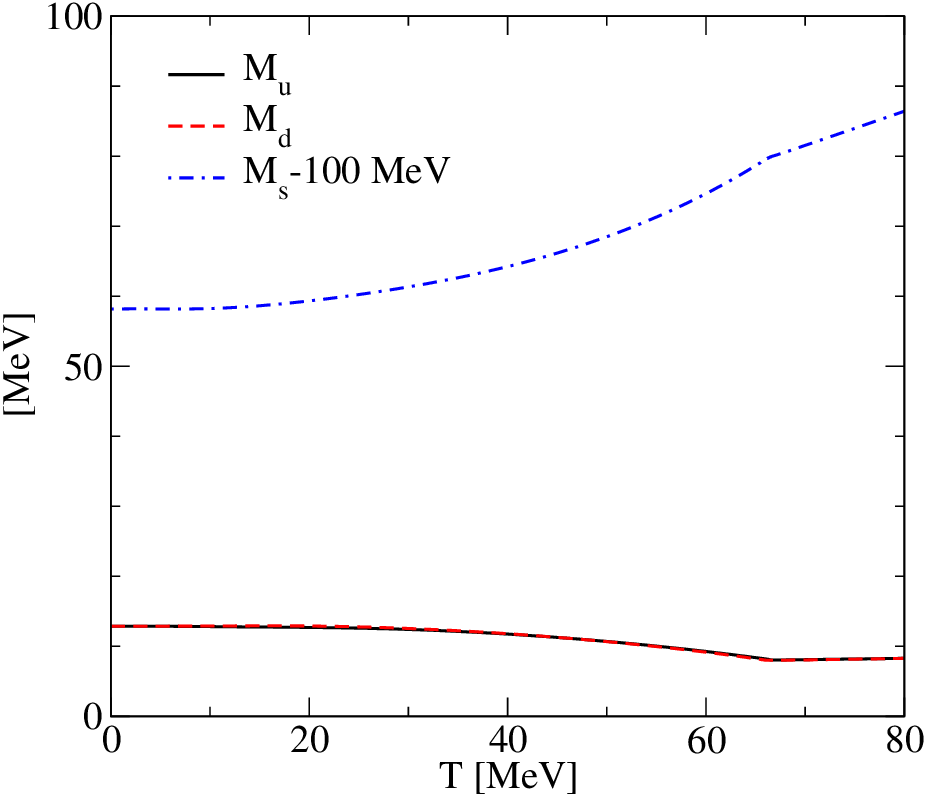} \\
     \includegraphics[width=.3\textwidth,clip=]{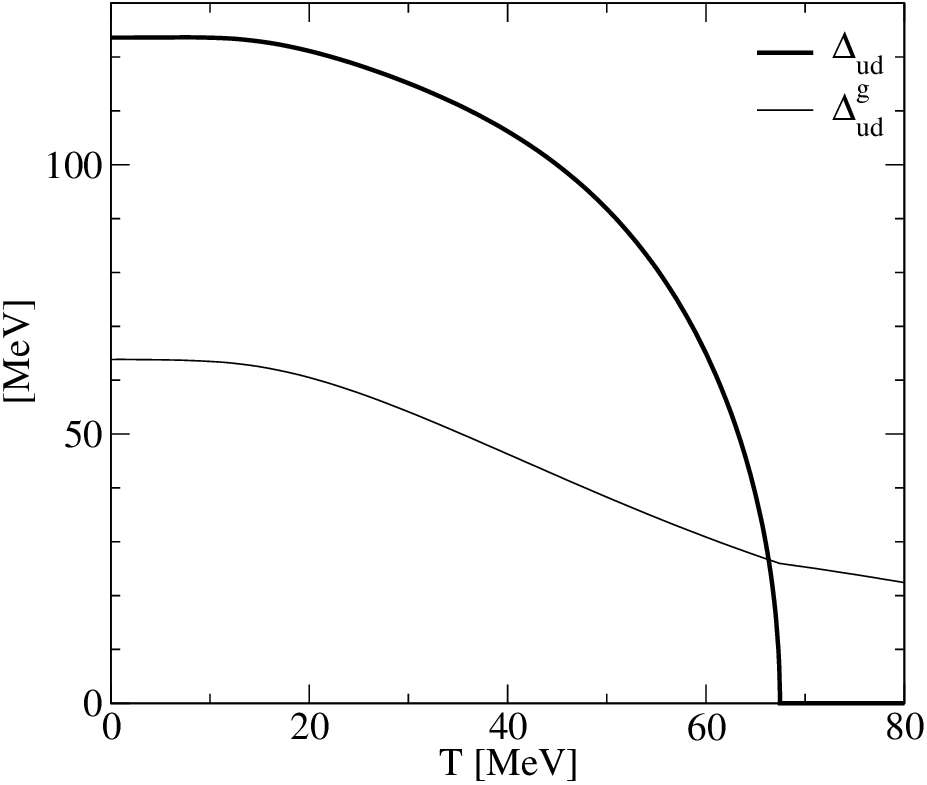} &
     \includegraphics[width=.3\textwidth,clip=]{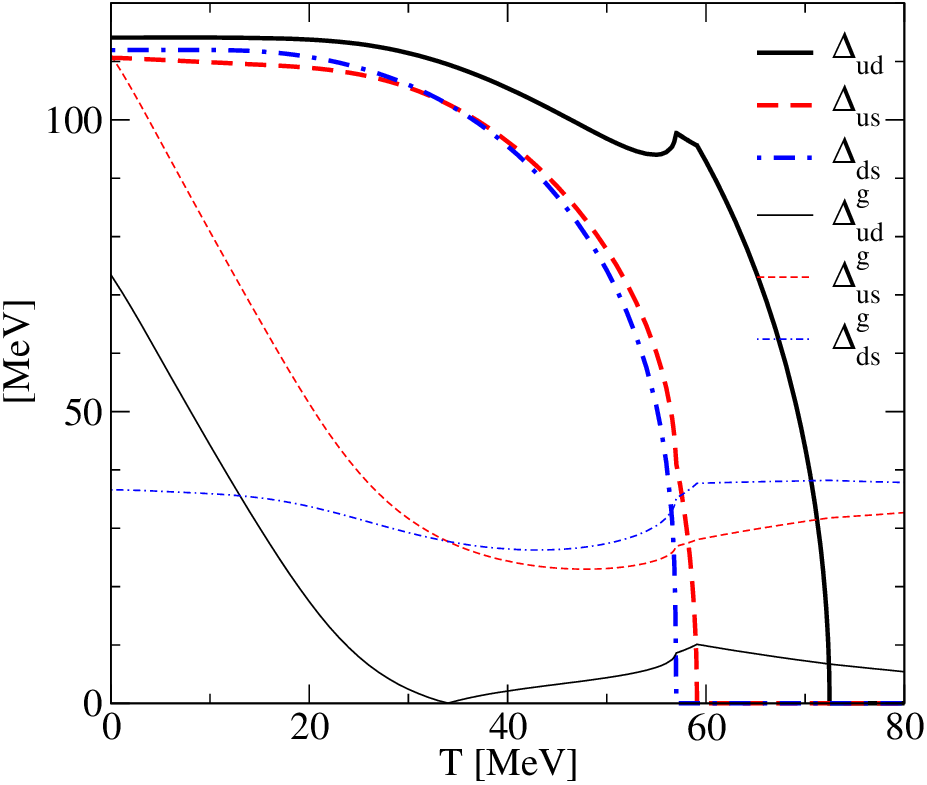} &
     \includegraphics[width=.3\textwidth,clip=]{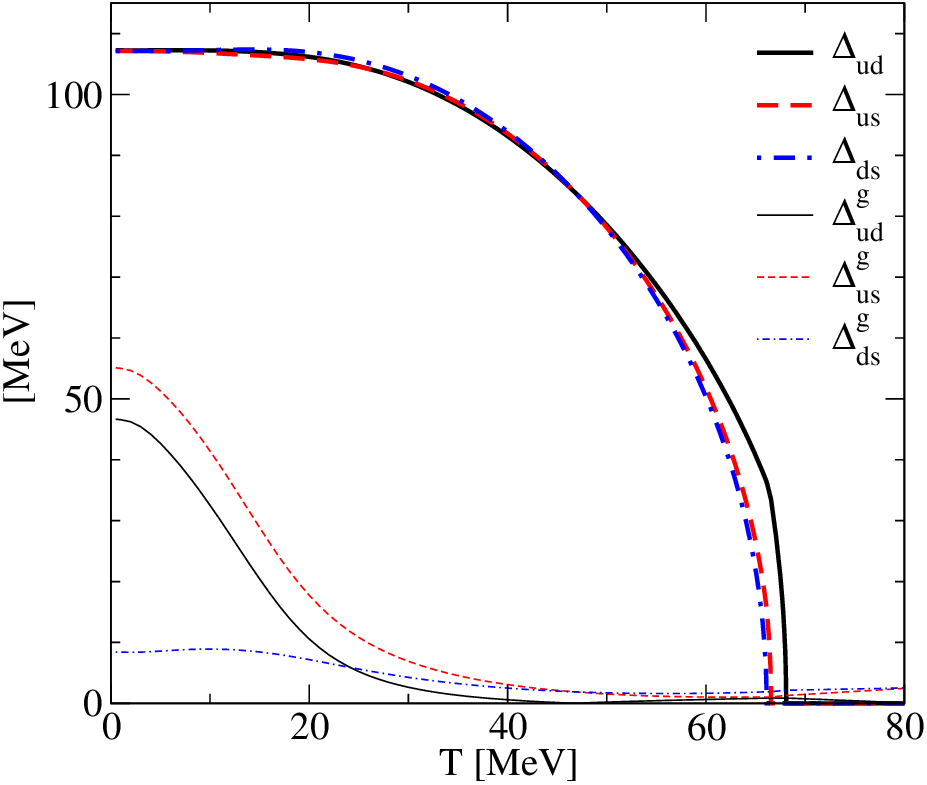} \\
     \includegraphics[width=.3\textwidth,clip=]{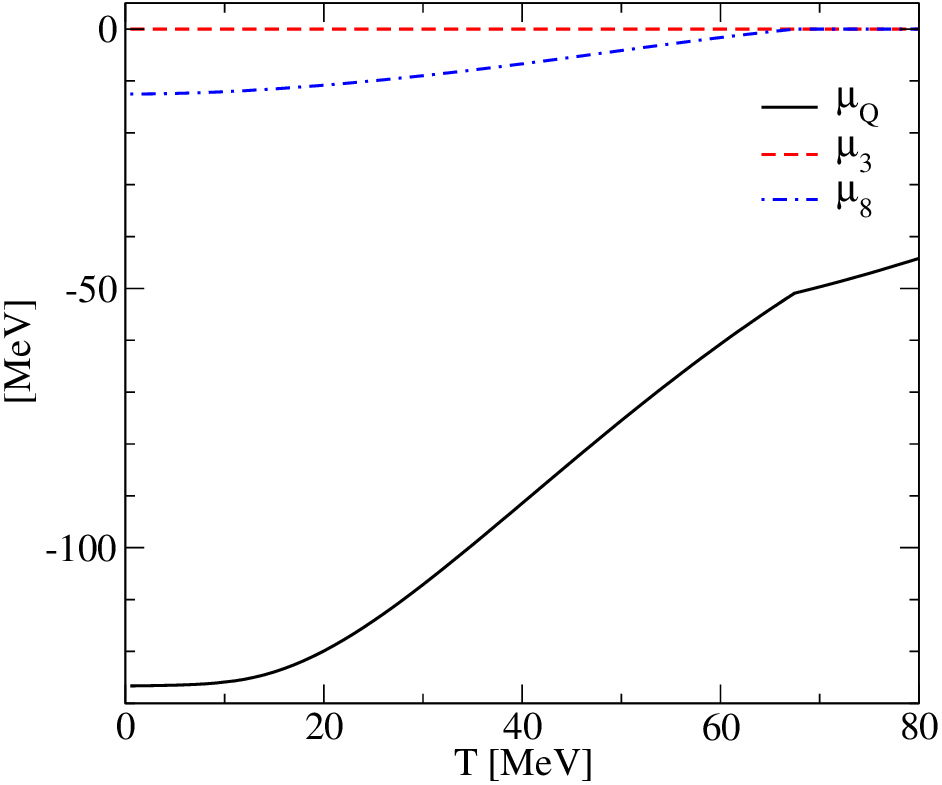} &
     \includegraphics[width=.3\textwidth,clip=]{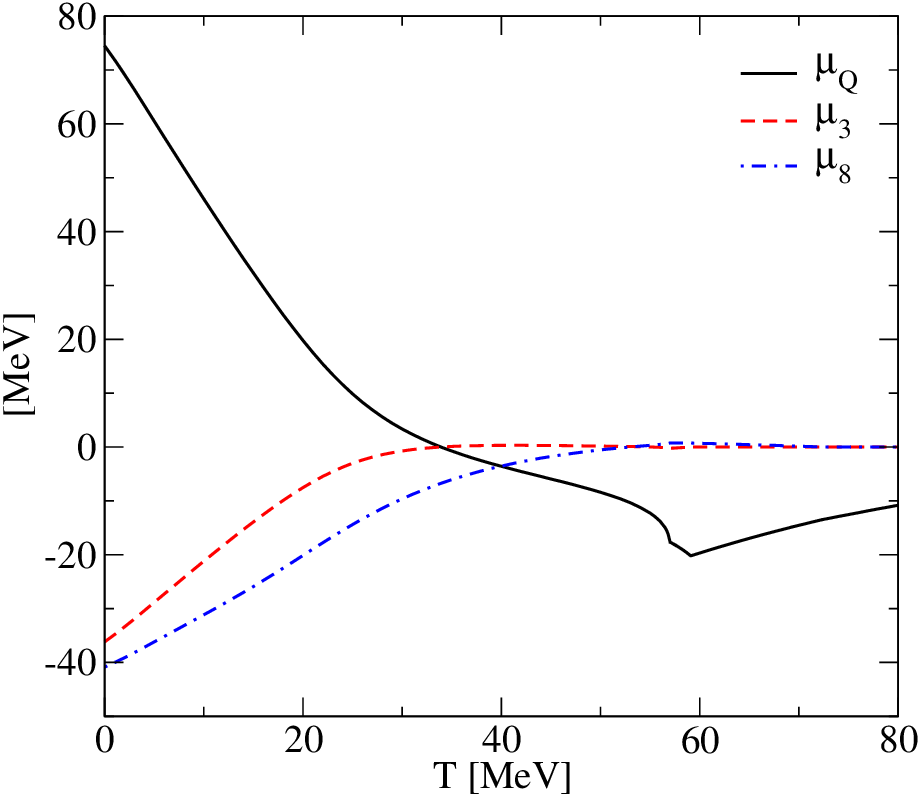} &
     \includegraphics[width=.3\textwidth,clip=]{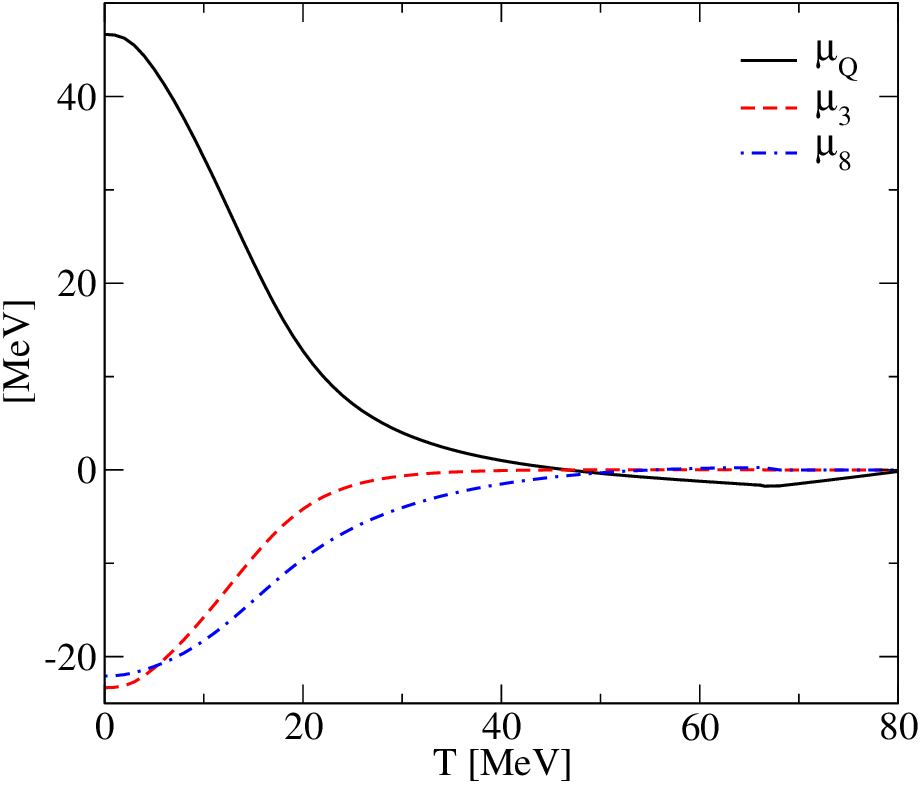} \\
     \includegraphics[width=.3\textwidth,clip=]{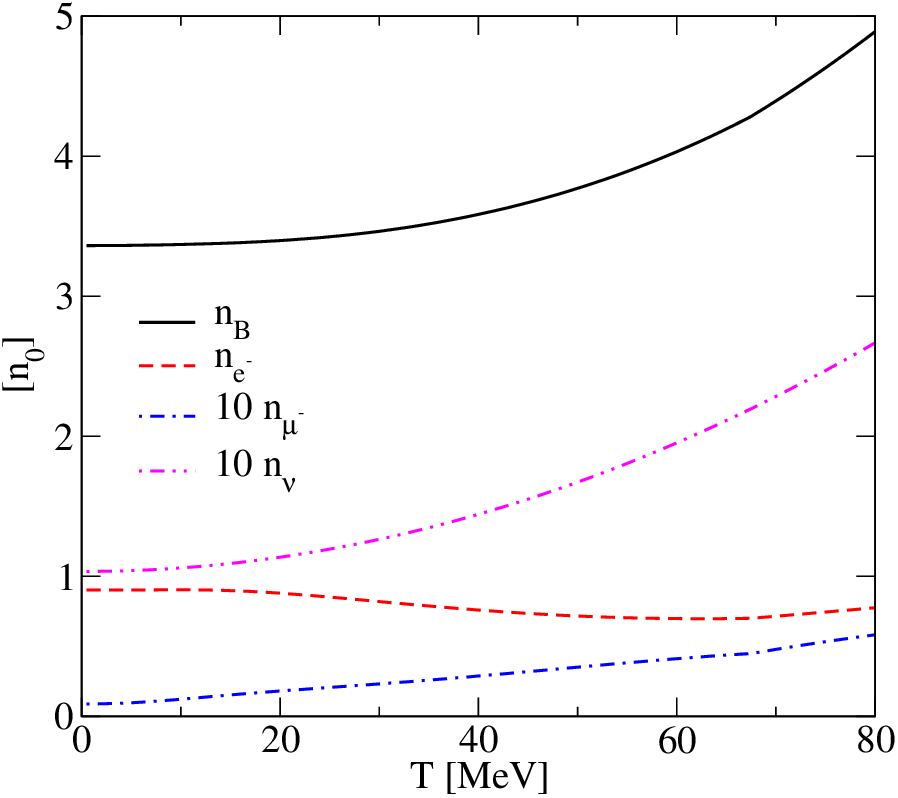} &
     \includegraphics[width=.3\textwidth,clip=]{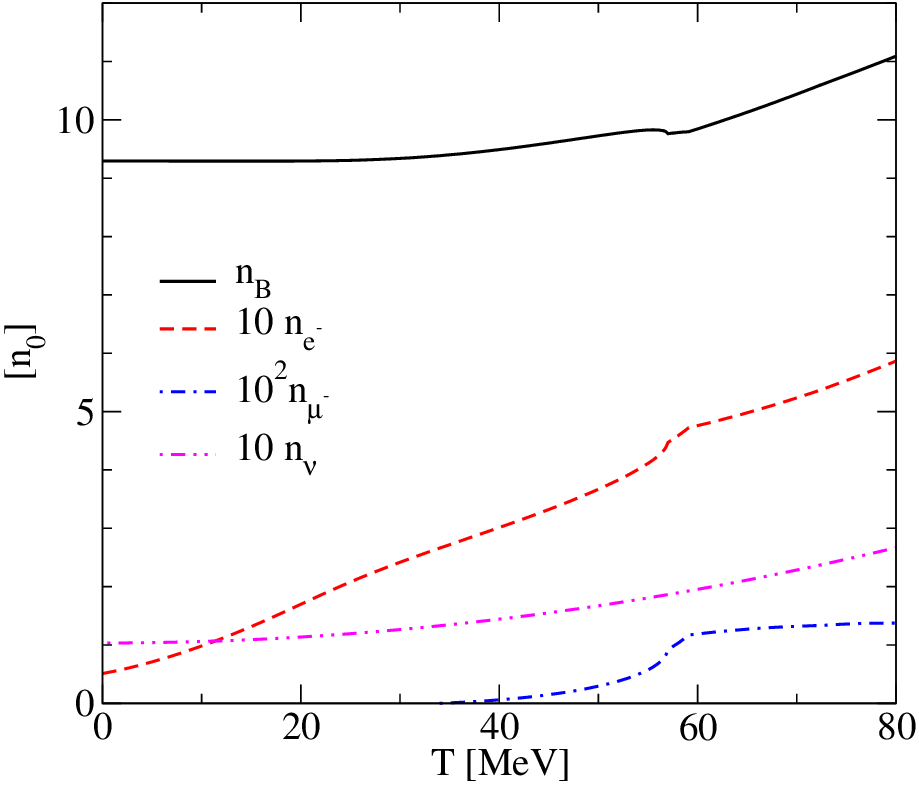} &
     \includegraphics[width=.3\textwidth,clip=]{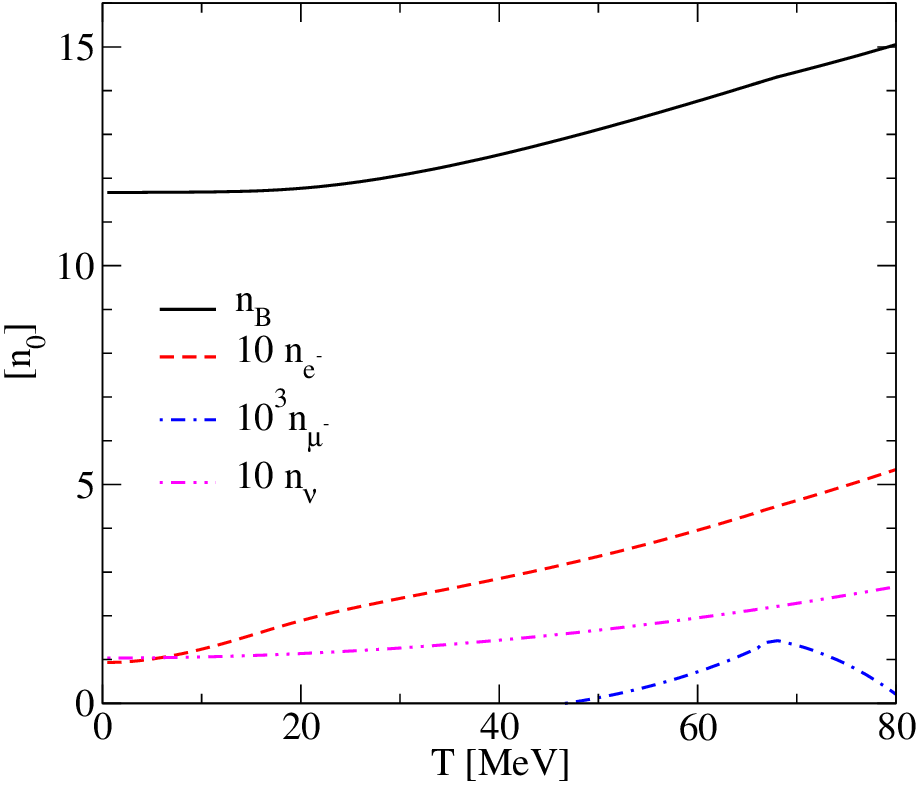}
\end{tabular}
\caption{(Color online)
The constituent quark masses, gaps, charge chemical potentials,
and number densities of charge neutral quark matter in $\beta$-equilibrium
with trapped neutrinos, $\mu_\nu=200$~MeV, at intermediate coupling, 
$\eta=0.75$. Line styles as in \fig{figDetailLe0Eta075}.}
\label{figDetailLe200Eta075}
\vspace{2cm} % Remove few lines of text at bottom of page
\end{figure*}

\begin{figure*}
%\vspace{-5mm}
\centering
\begin{tabular}{ccc}
        \includegraphics[width=.3\textwidth,clip=]{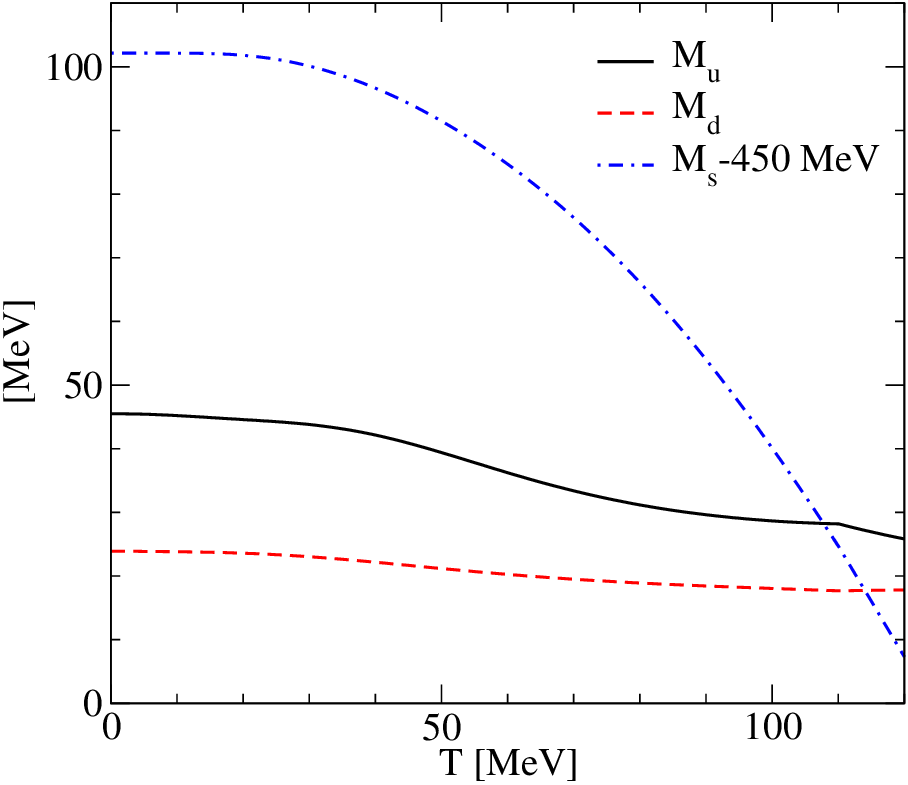} &
        \includegraphics[width=.3\textwidth,clip=]{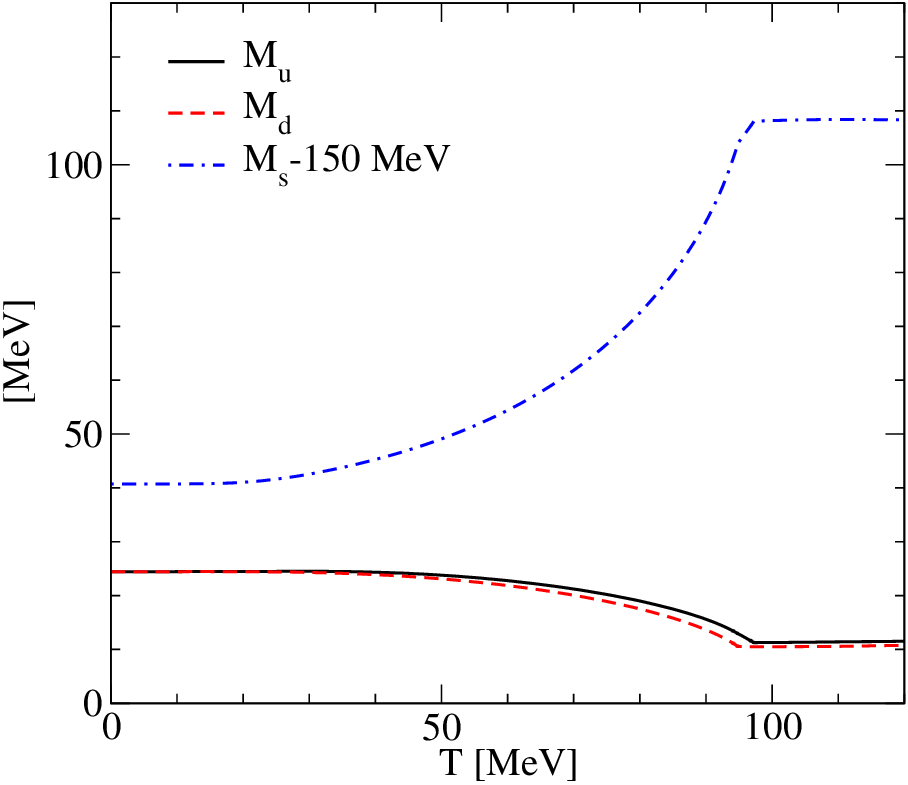} &
        \includegraphics[width=.3\textwidth,clip=]{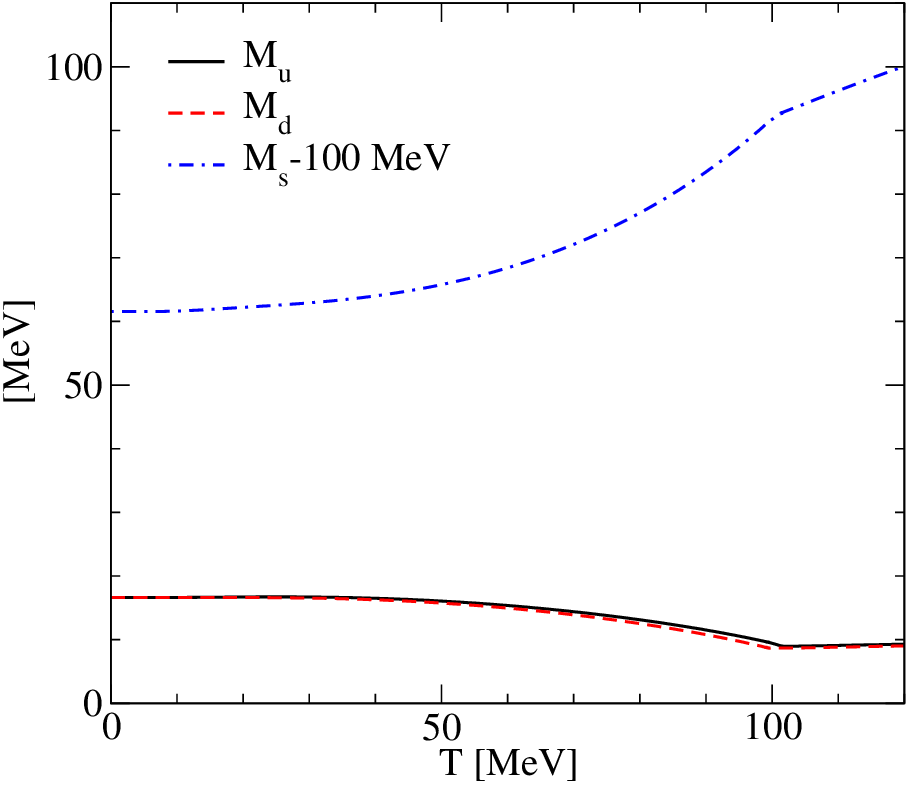} \\
        \includegraphics[width=.3\textwidth,clip=]{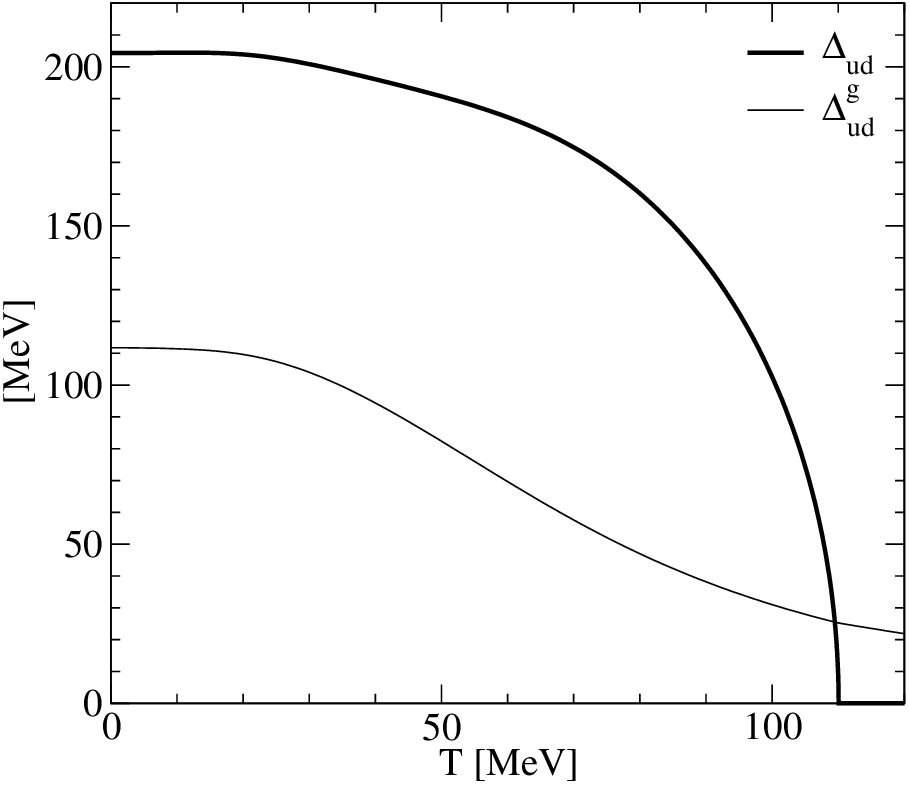} &
        \includegraphics[width=.3\textwidth,clip=]{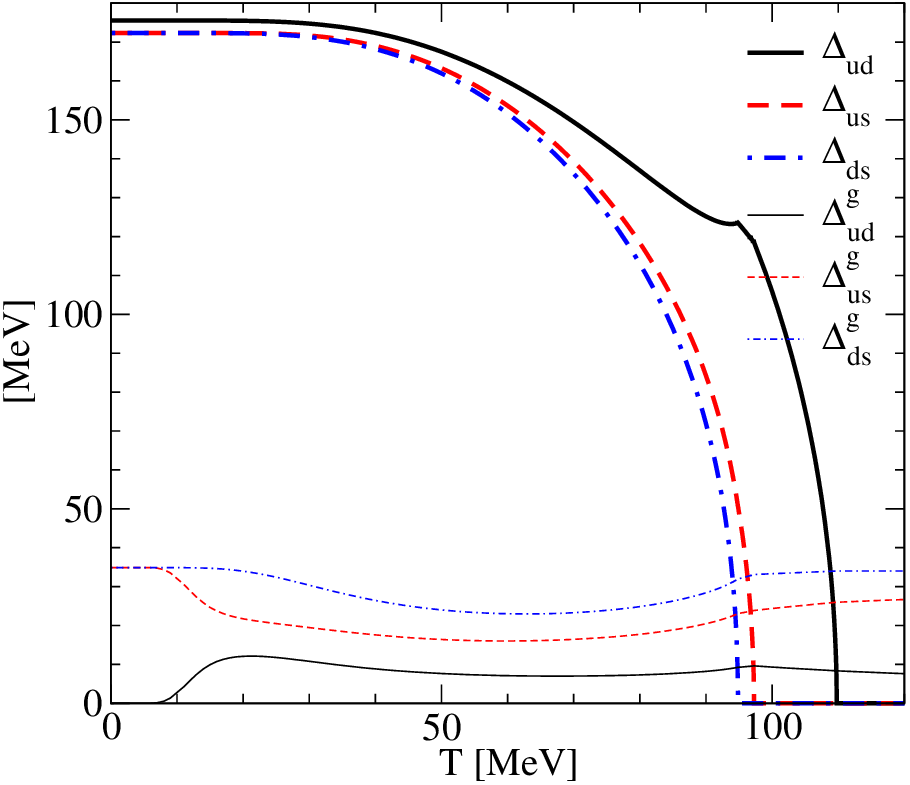} &
        \includegraphics[width=.3\textwidth,clip=]{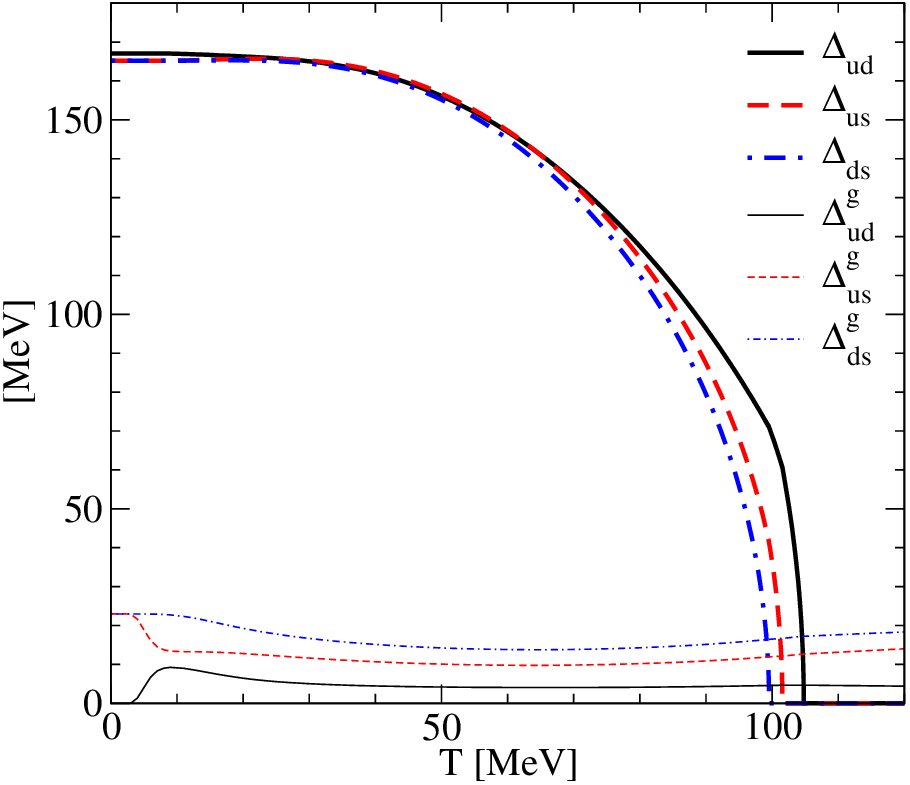} \\
        \includegraphics[width=.3\textwidth,clip=]{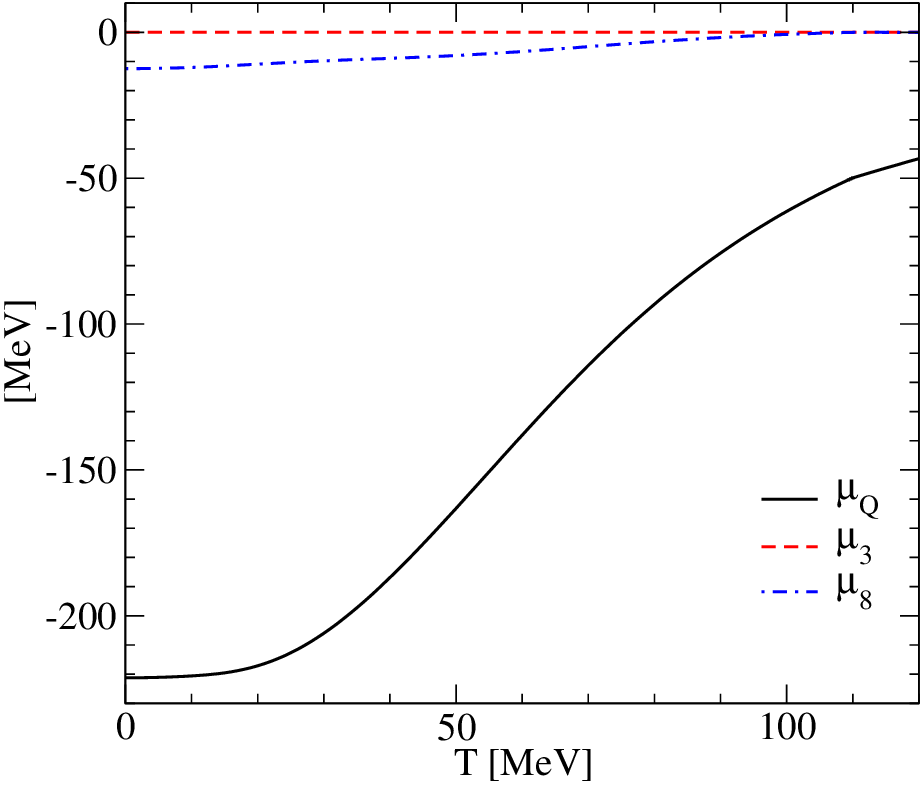} &
        \includegraphics[width=.3\textwidth,clip=]{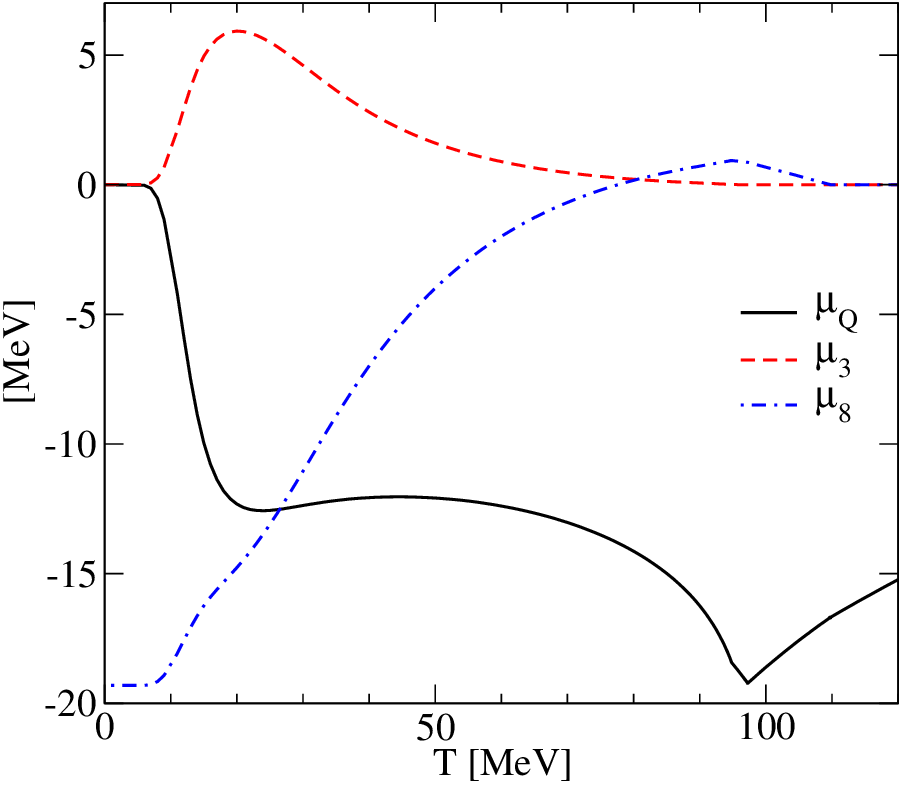} &
        \includegraphics[width=.3\textwidth,clip=]{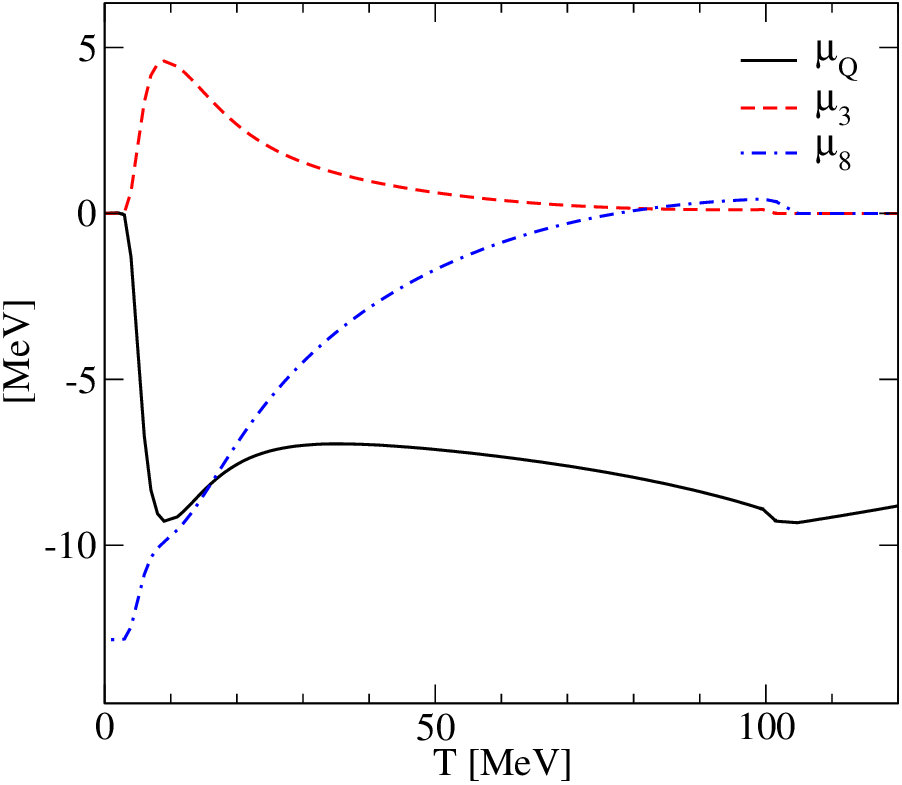} \\
        \includegraphics[width=.3\textwidth,clip=]{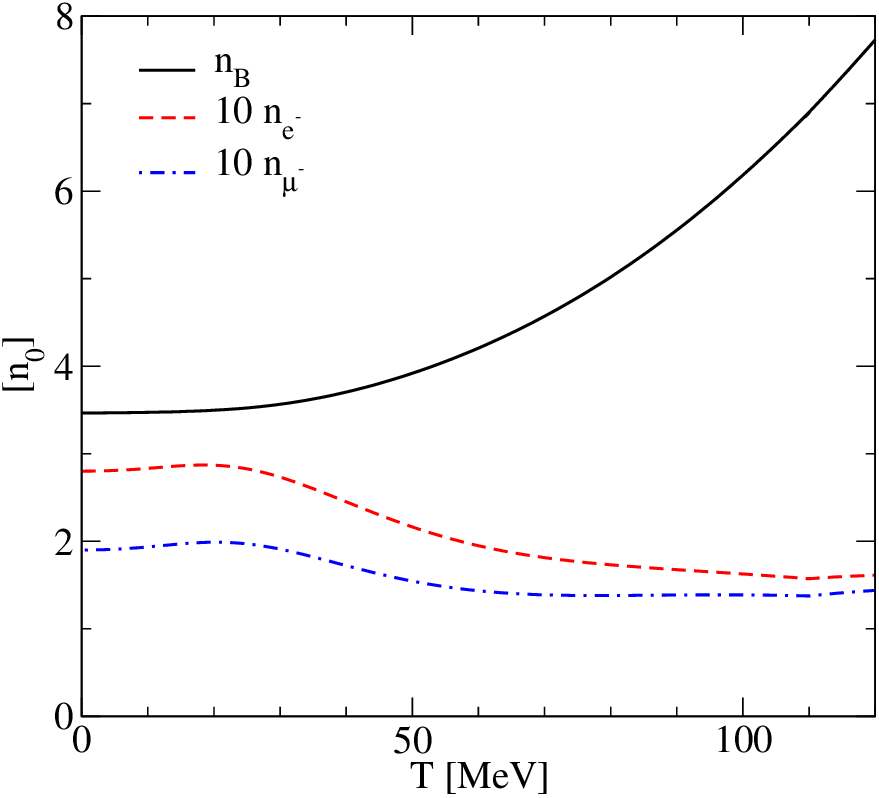} &
        \includegraphics[width=.3\textwidth,clip=]{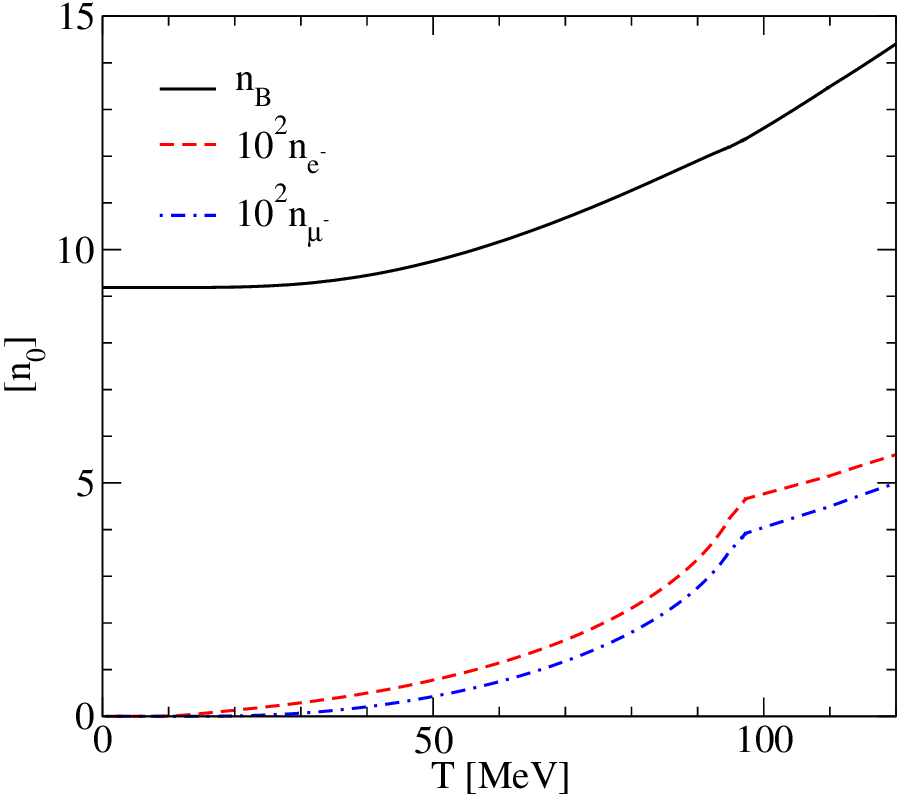} &
        \includegraphics[width=.3\textwidth,clip=]{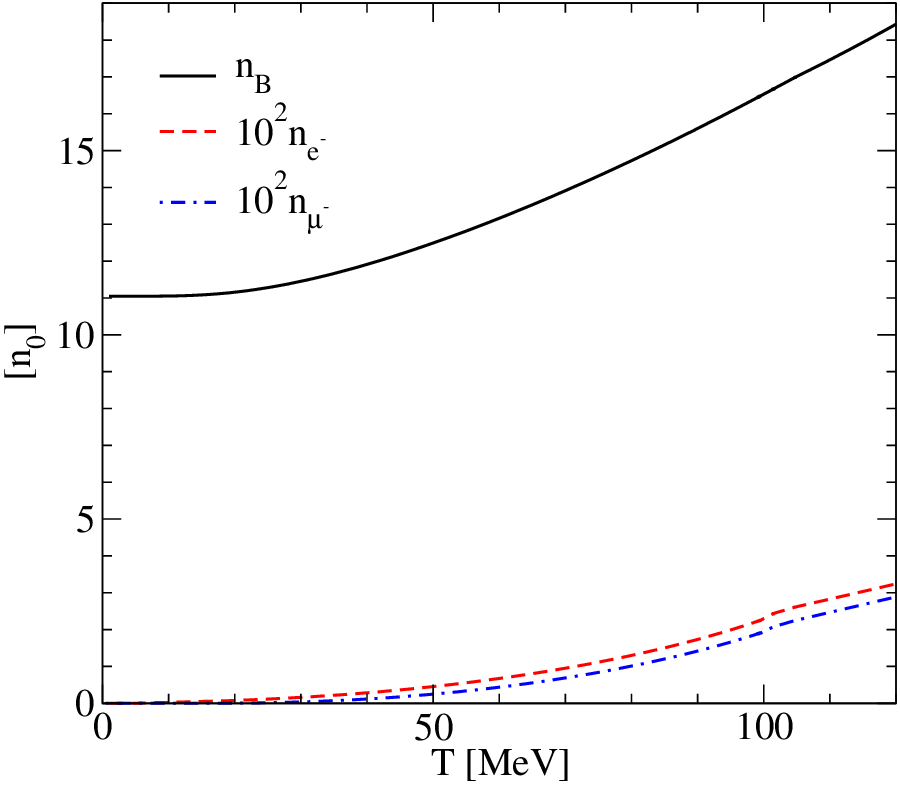}
\end{tabular}
\caption{(Color online)
The constituent quark masses, gaps, charge chemical potentials,
and number densities of charge neutral quark matter in $\beta$-equilibrium
with untrapped neutrinos, $\mu_\nu=0$,
at strong coupling, $\eta=1.0$. Line styles as in \fig{figDetailLe0Eta075}.}
\label{figDetailLe0Eta1}
\vspace{2cm} % Remove few lines of text at bottom of page
\end{figure*}

\begin{figure*}
%\vspace{-5mm}
\centering
\begin{tabular}{ccc}
       \includegraphics[width=.3\textwidth,clip=]{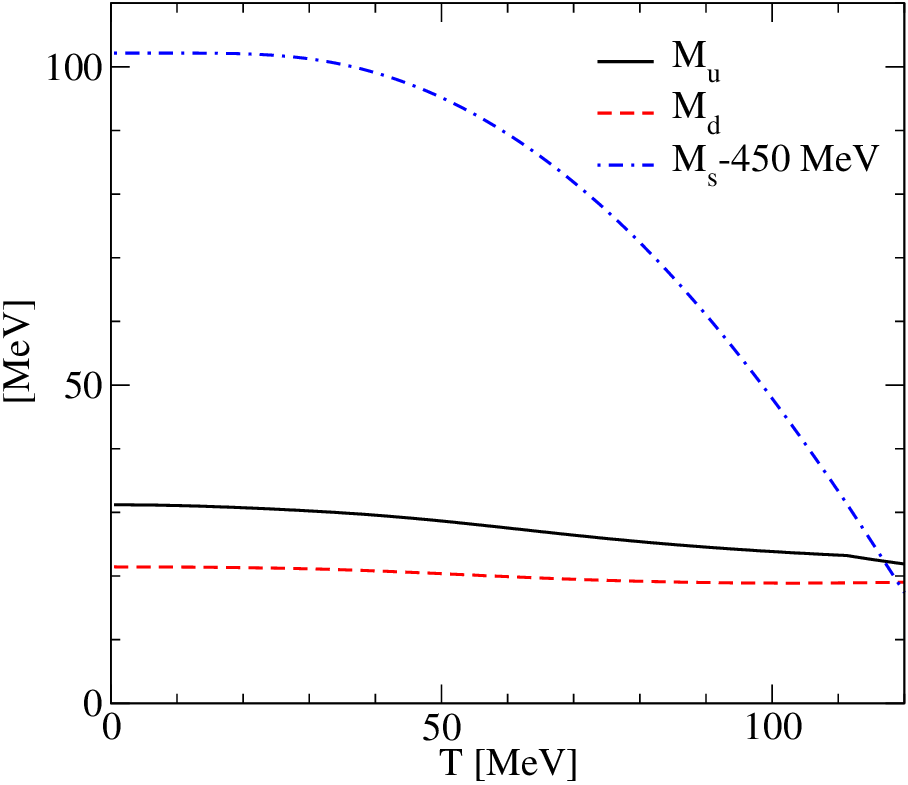} &
       \includegraphics[width=.3\textwidth,clip=]{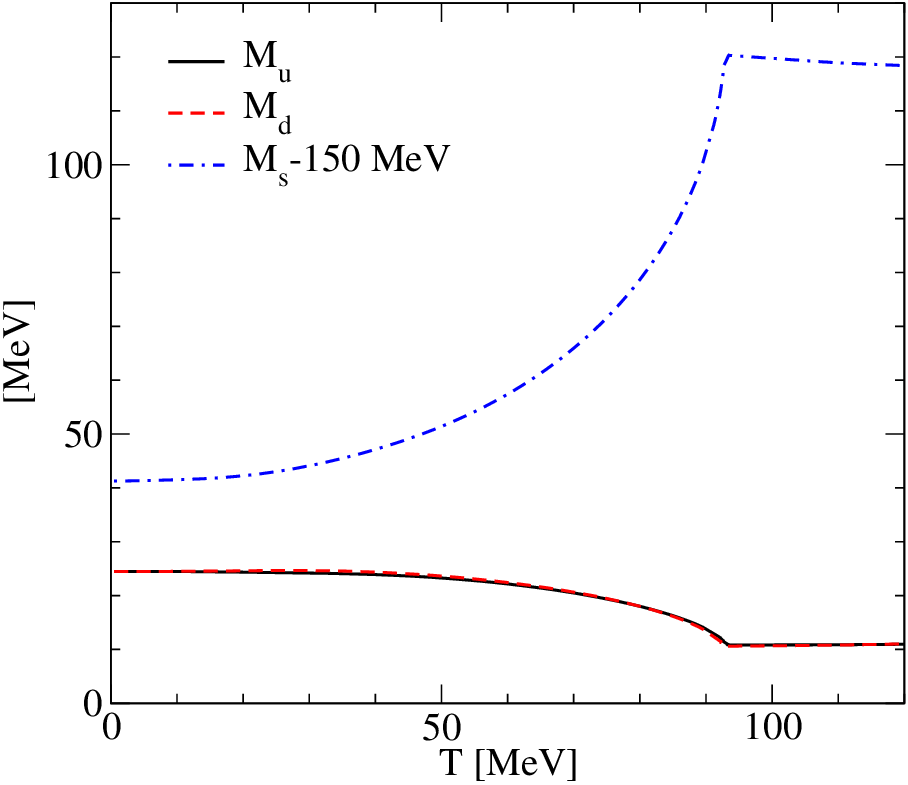} &
       \includegraphics[width=.3\textwidth,clip=]{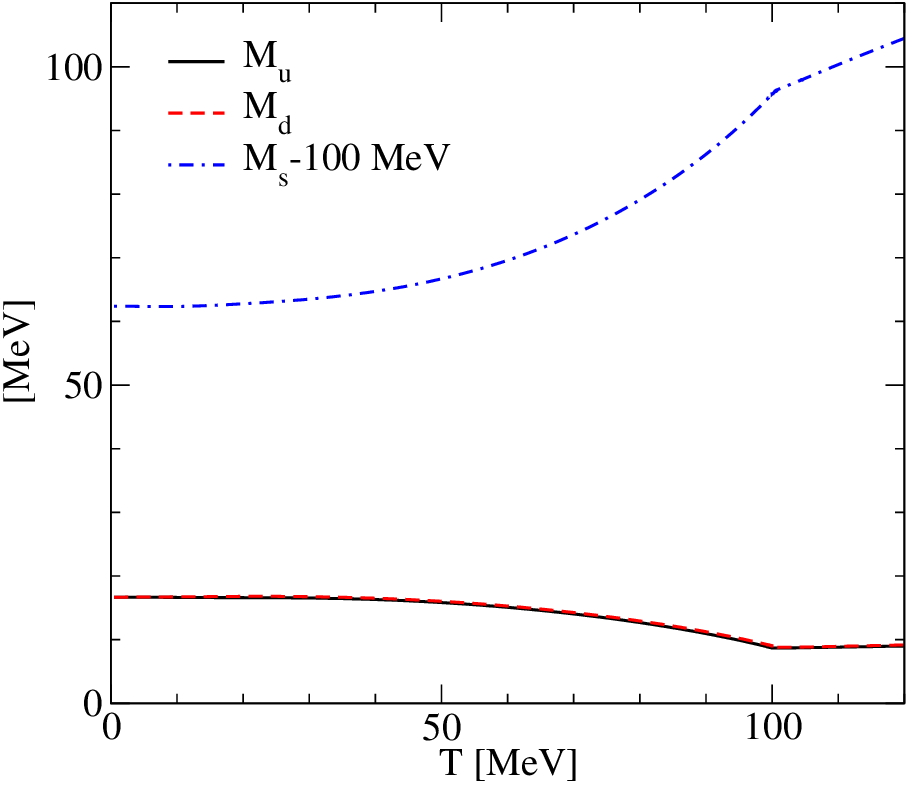} \\
       \includegraphics[width=.3\textwidth,clip=]{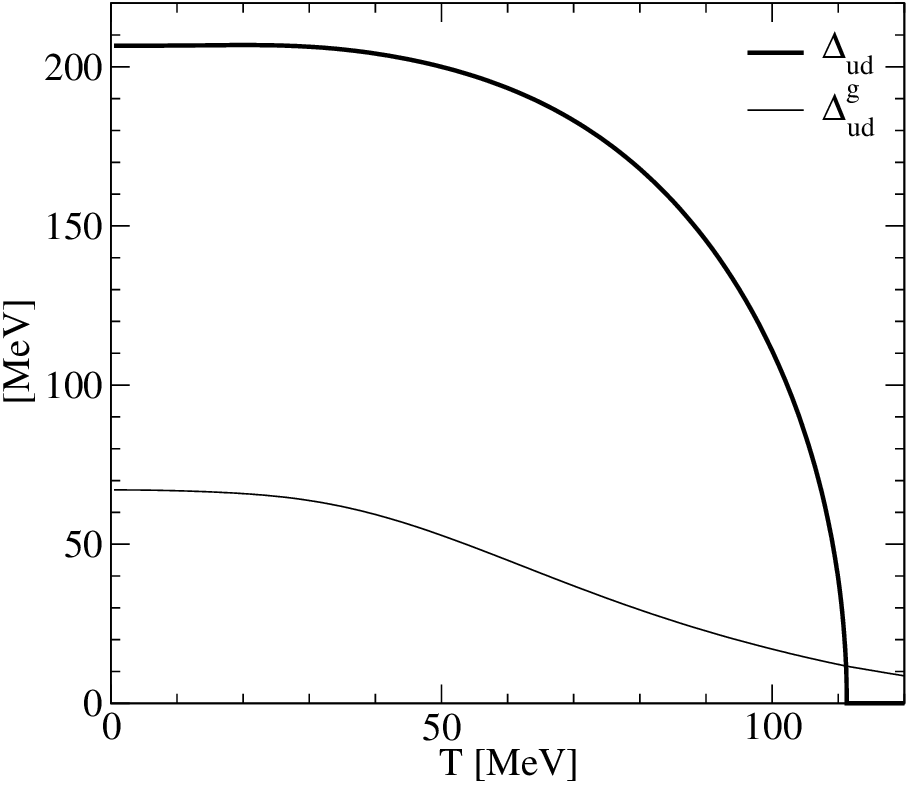} &
       \includegraphics[width=.3\textwidth,clip=]{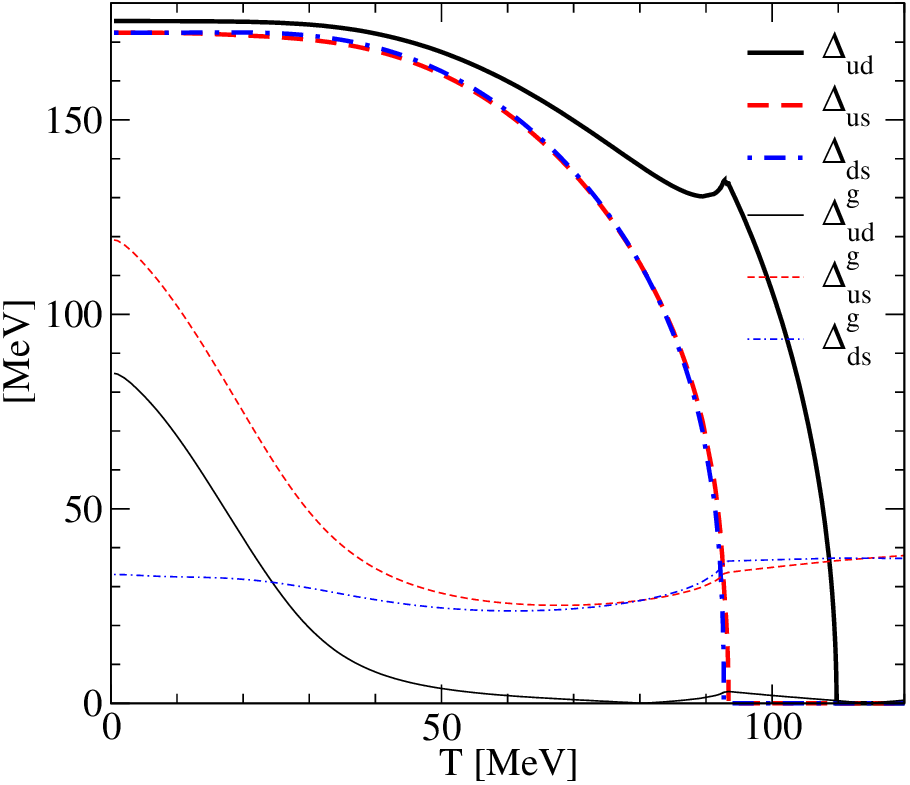} &
       \includegraphics[width=.3\textwidth,clip=]{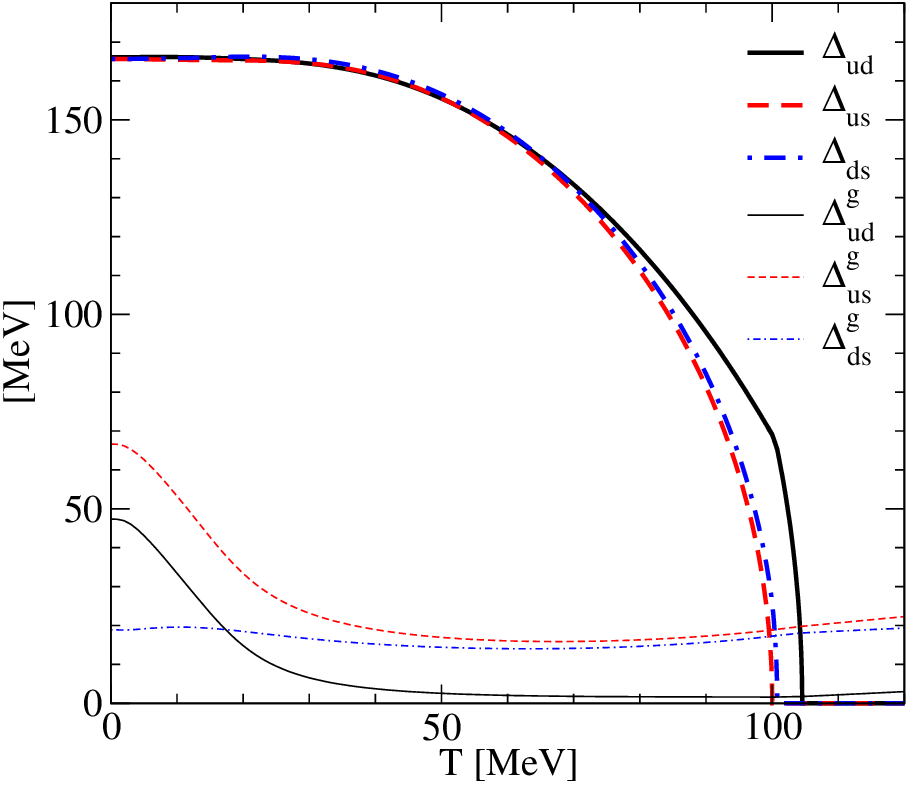} \\
       \includegraphics[width=.3\textwidth,clip=]{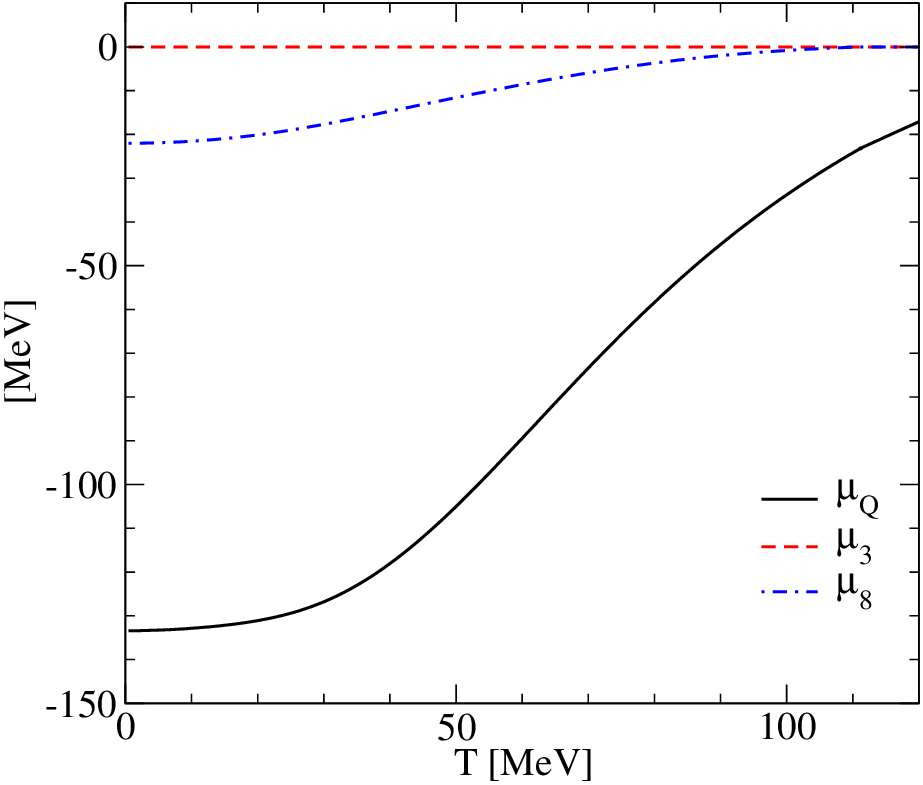} &
       \includegraphics[width=.3\textwidth,clip=]{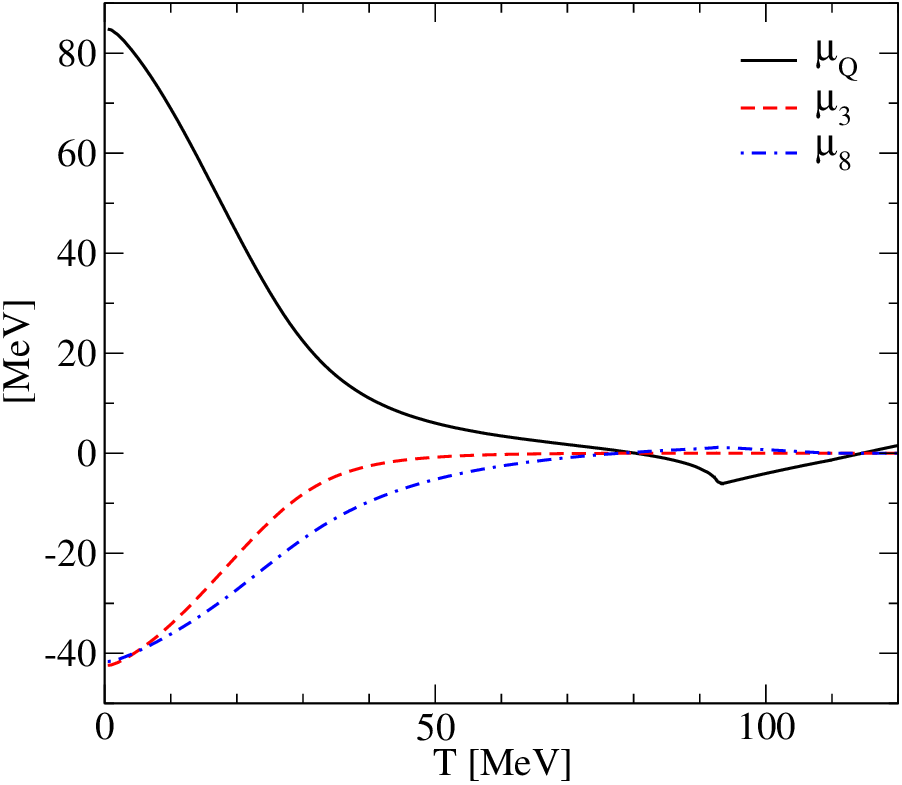} &
       \includegraphics[width=.3\textwidth,clip=]{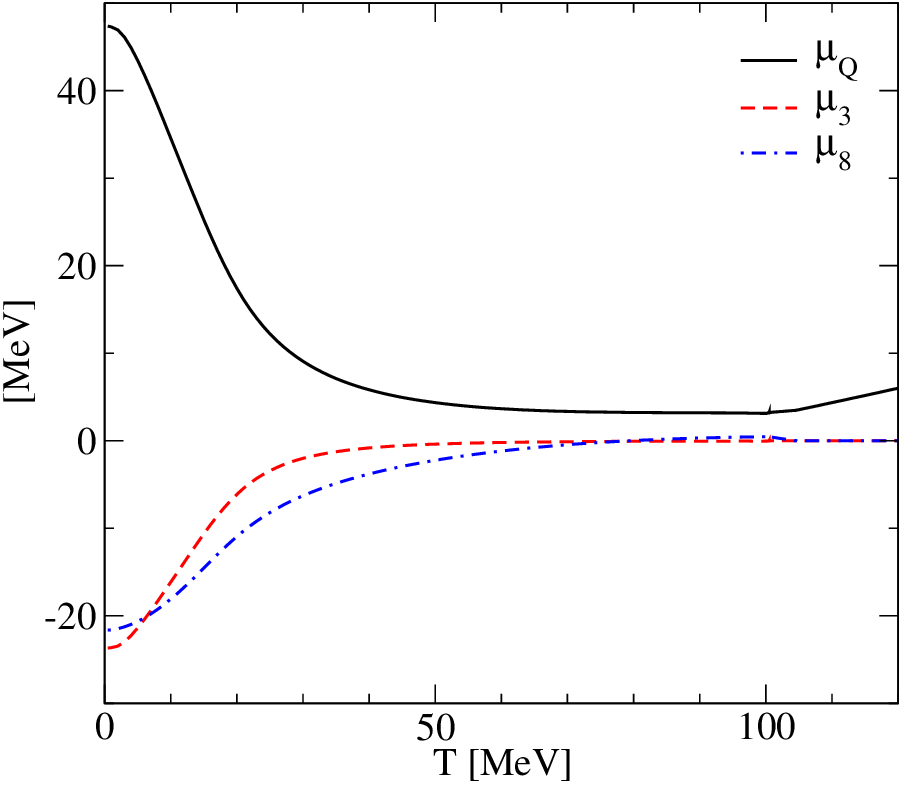} \\
       \includegraphics[width=.3\textwidth,clip=]{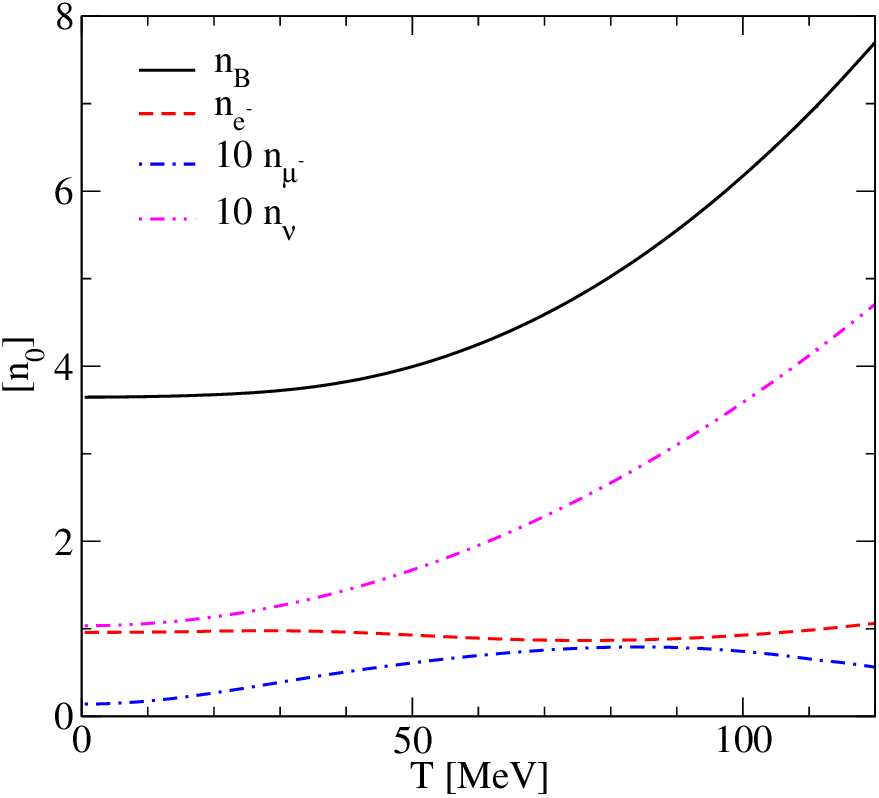} &
       \includegraphics[width=.3\textwidth,clip=]{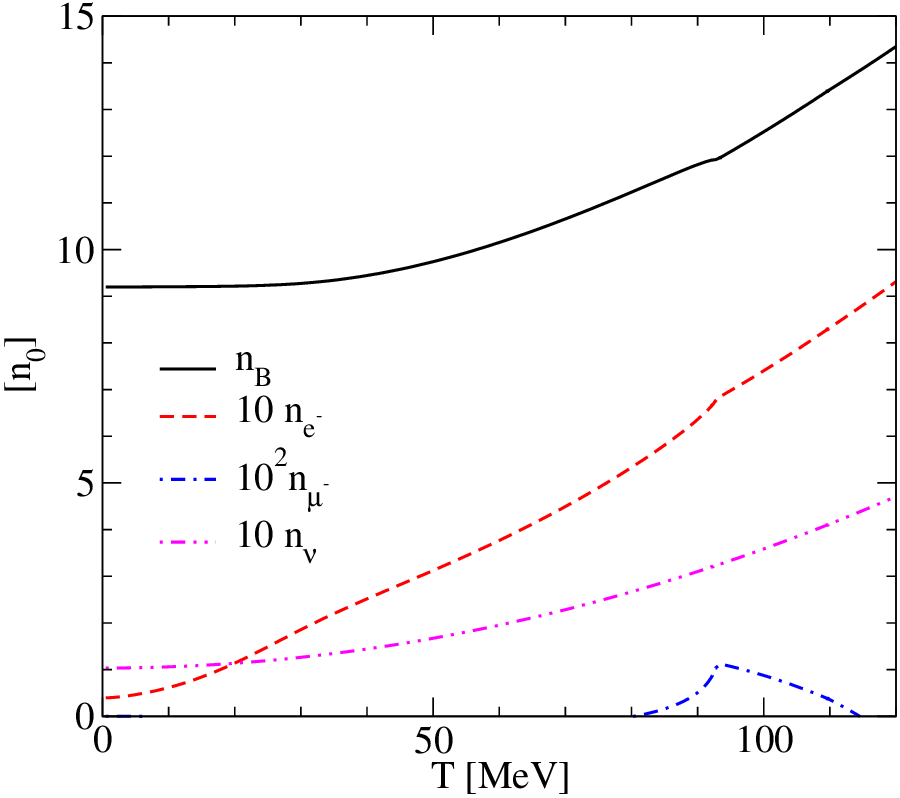} &
       \includegraphics[width=.3\textwidth,clip=]{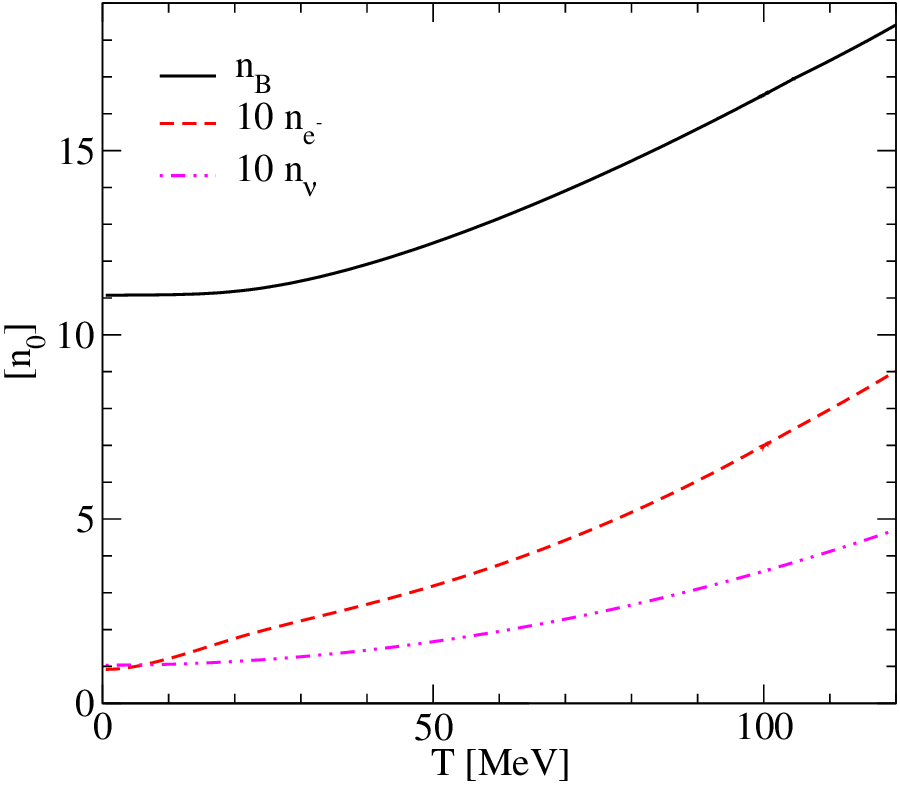}
\end{tabular}
\caption{(Color online)
The constituent quark masses, gaps, charge chemical potentials,
and number densities of charge neutral quark matter in $\beta$-equilibrium
with trapped neutrinos, $\mu_\nu=200$~MeV, at strong coupling, $\eta=1.0$.
Line styles as in \fig{figDetailLe0Eta075}.}
\label{figDetailLe200Eta1}
\vspace{2cm} % Remove few lines of text at bottom of page
\end{figure*}

\clearpage

\begin{figure}
%\vspace{-5mm}
\includegraphics[width=0.45\textwidth,angle=0]{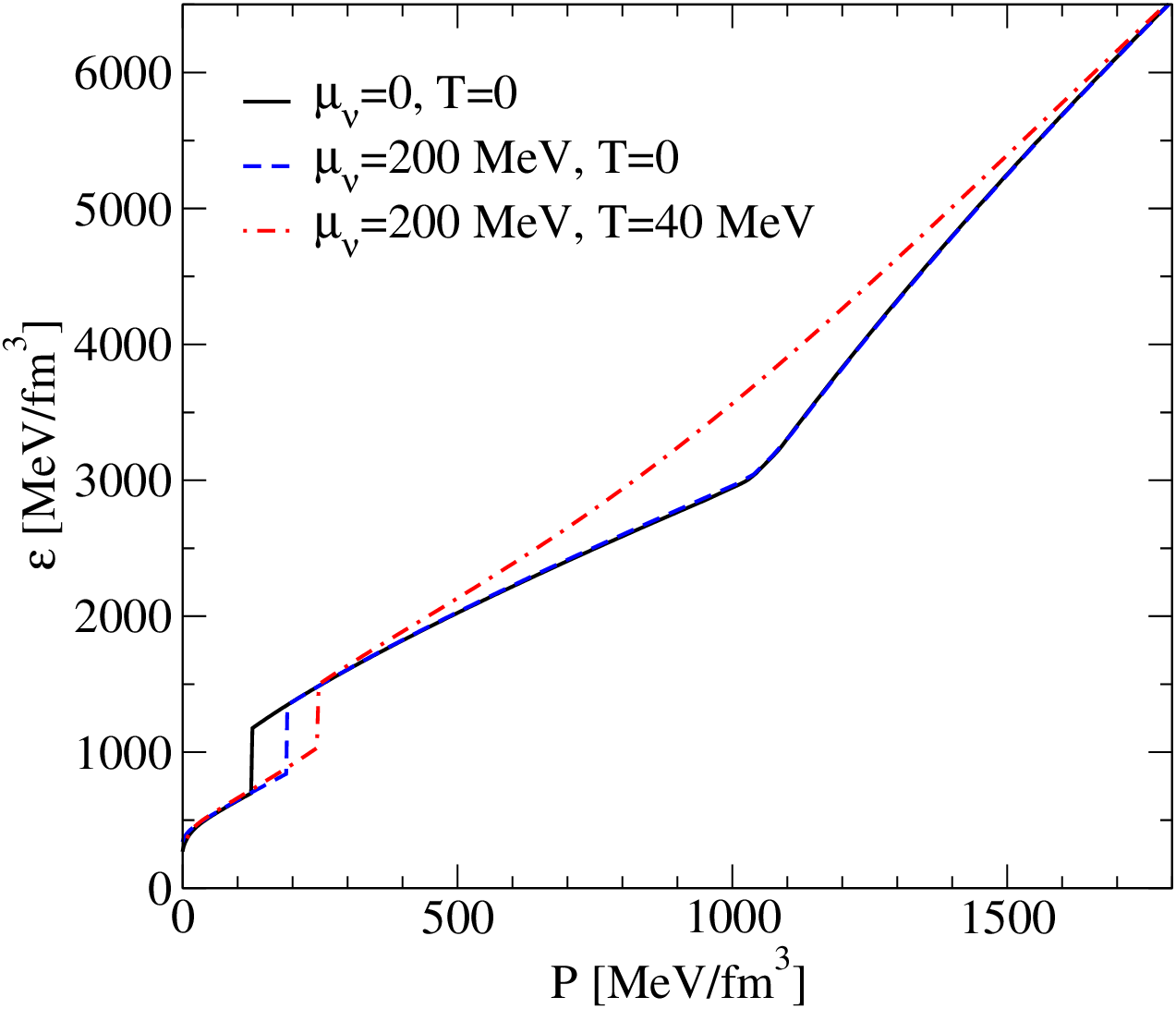}
\caption{(Color online)
        The equation of state for charge neutral quark matter in
        $\beta$-equilibrium at strong coupling, $\eta=1.0$, and
        fixed values of the temperature and electron neutrino chemical
        potential.}
\label{figEosEta1}
\end{figure}

\begin{figure}
%\vspace{-5mm}
\includegraphics[width=0.45\textwidth,angle=0]{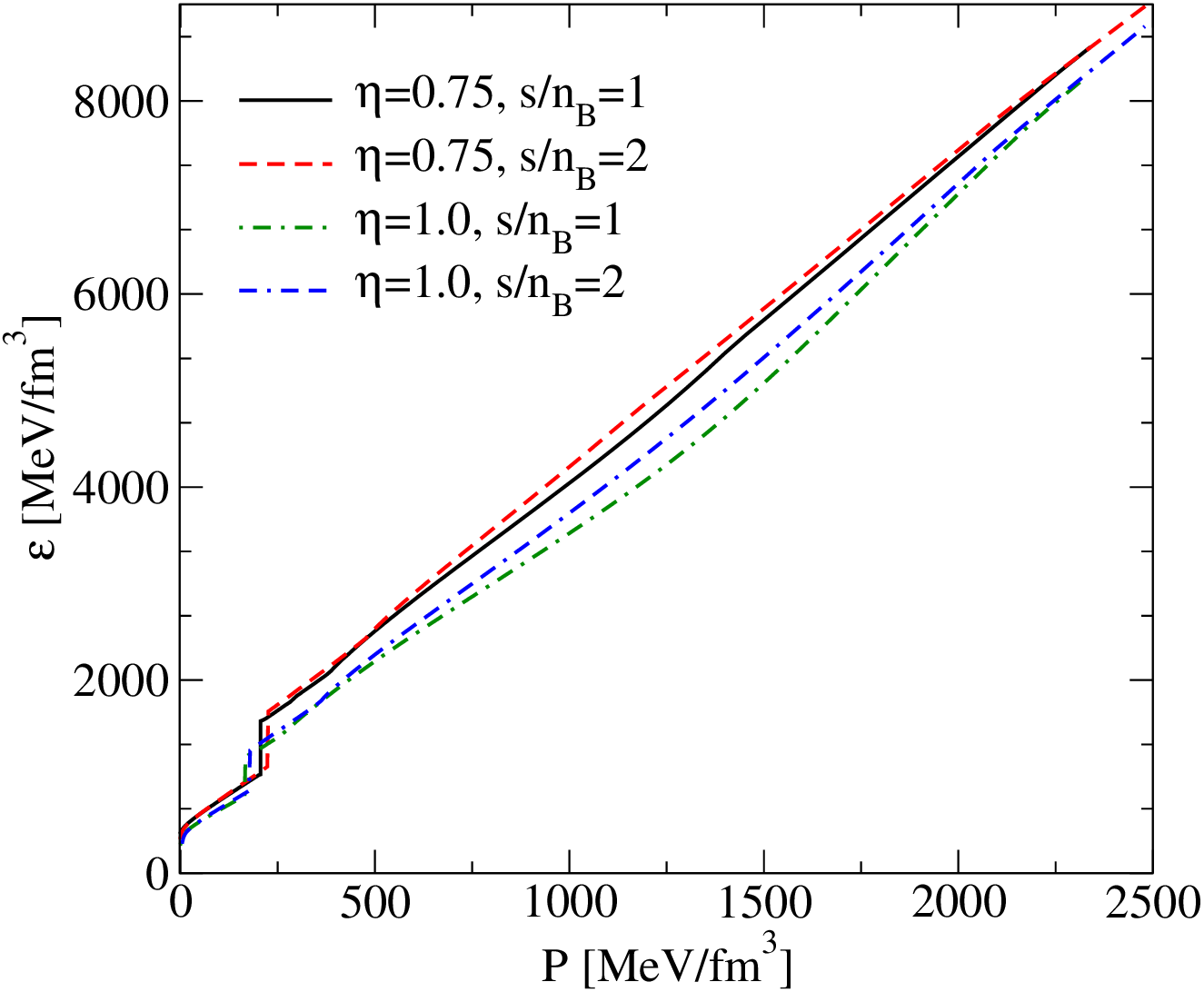}
\caption{(Color online)
        The equation of state for charge neutral quark matter in
        $\beta$-equilibrium for fixed values of the lepton fraction,
        $Y_{L_e}(T,\mu_\nu)=0.4$, and the entropy per baryon, 
        $s(T,\mu_\nu)/n_B(T,\mu_\nu)$.}
\label{figEosYLe04}
\end{figure}

\noindent mate) chiral symmetry is broken by
a first order phase transition into the $\chi$SB phase.

\begin{figure}
%\vspace{-5mm}
\includegraphics[width=0.45\textwidth,angle=0]{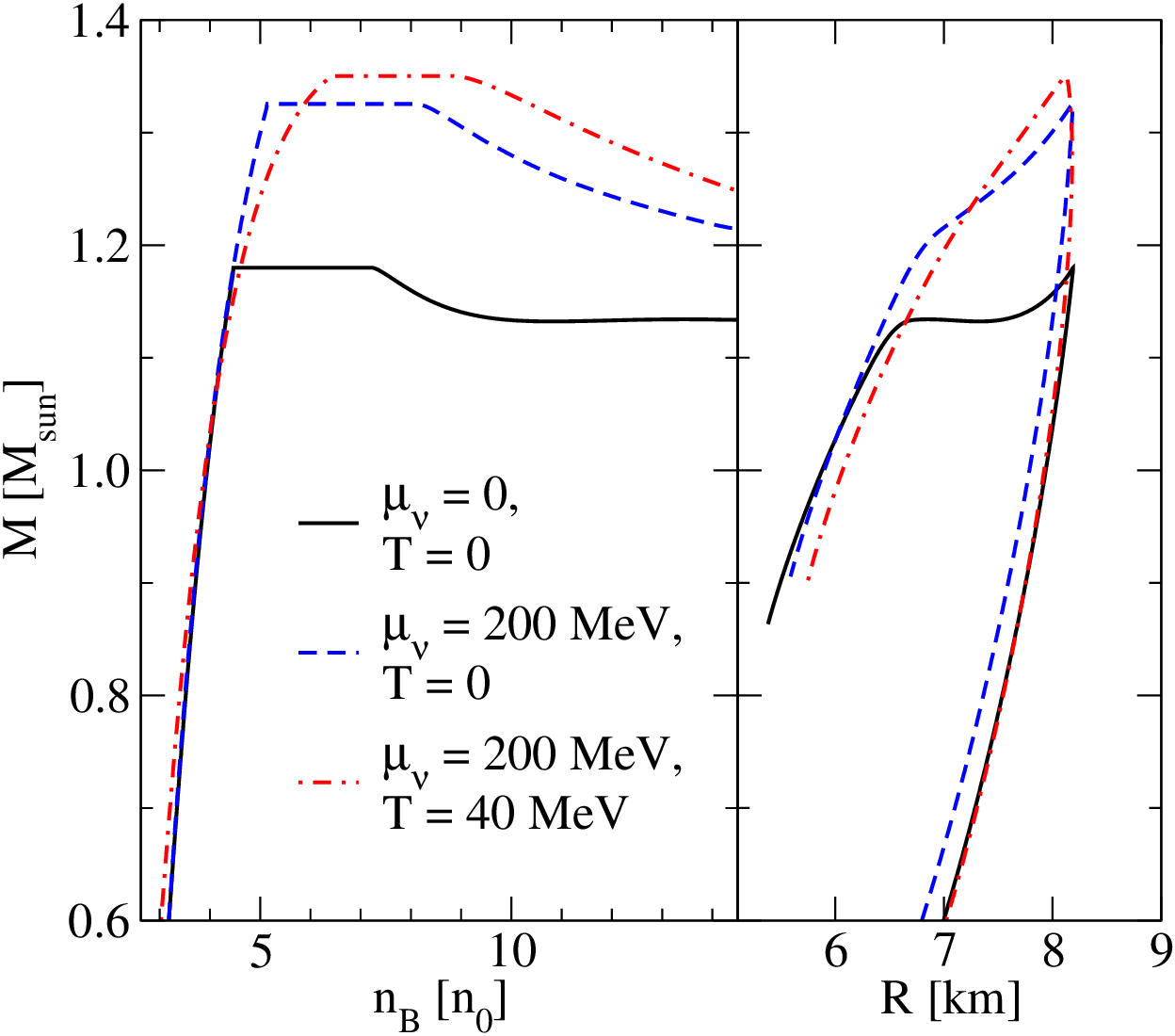}
\caption{(Color online)
Quark star sequences for intermediate coupling, $\eta=0.75$, and fixed
values of the temperature and electron neutrino chemical potential.
The discontinuity at the maximum mass configuration is a consequence
of the appearance of CFL matter in the center of the stars. Stars with a
CFL core are marginally stable and exist if the temperature and neutrino
chemical potential are not too high, see \fig{figPhaseEta075}.}
\label{figSeqEta075}
\end{figure}

\begin{figure}
%\vspace{-5mm}
\includegraphics[width=0.45\textwidth,angle=0]{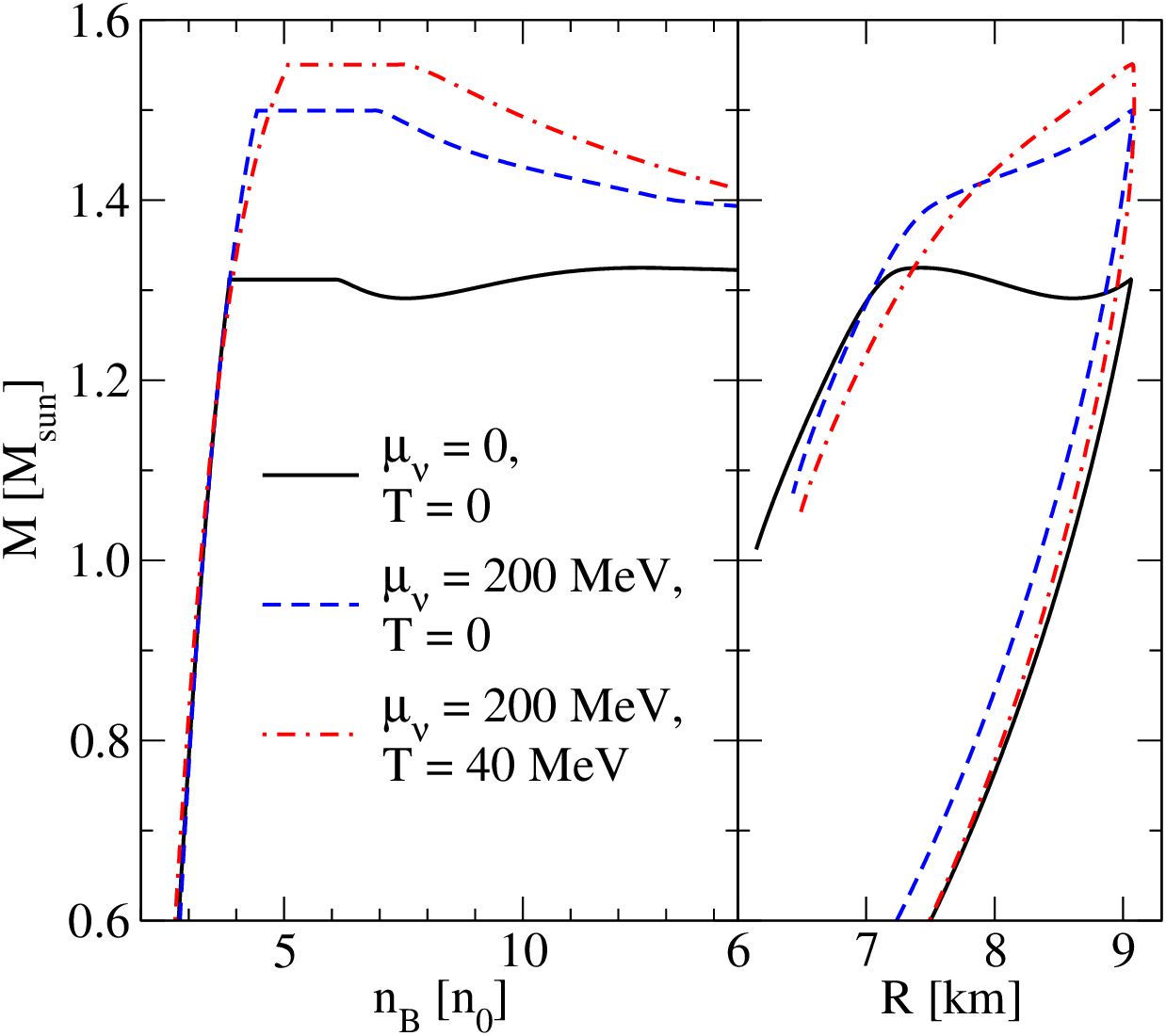}
\caption{(Color online)
Quark star sequences for the strong coupling, $\eta=1.0$, and fixed
values of the temperature and electron neutrino chemical potential. Stable
stars with a CFL core exist if the temperature and neutrino chemical potential 
are not too high, see \fig{figPhaseEta1}.}
\label{figSeqEta1}
\end{figure}

\begin{figure}
%\vspace{-5mm}
\includegraphics[width=0.45\textwidth,angle=0]{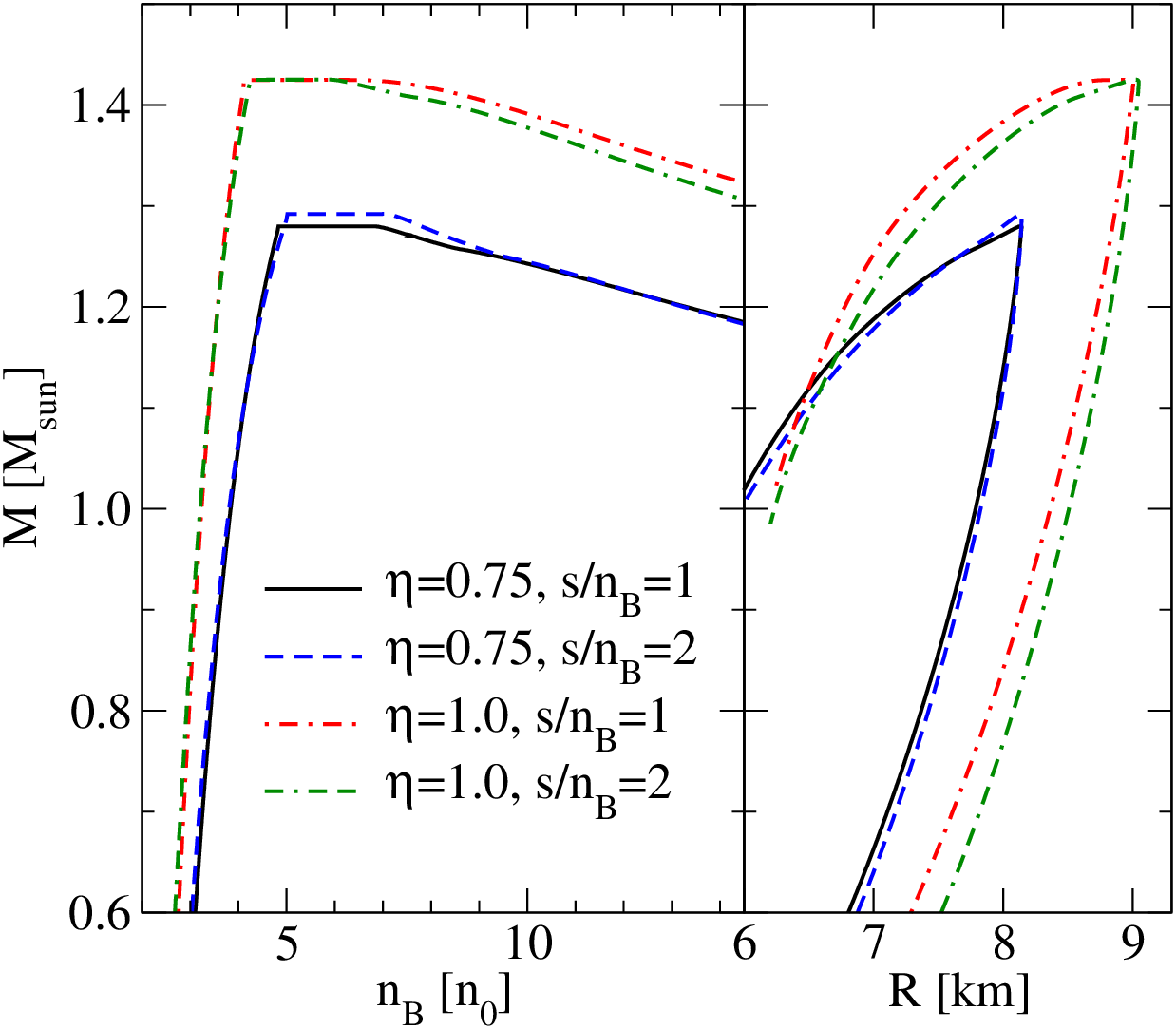}
\caption{(Color online)
Quark star sequences for fixed values of the lepton fraction,
$Y_{L_e}(T,\mu_\nu)=0.4$, and the entropy per baryon, 
$s(T,\mu_\nu)/n_B(T,\mu_\nu)$.
Stars with a CFL core are unstable (marginally unstable for $\eta=1$).}
\label{figSeqYLe04}
\end{figure}

\begin{figure}
%\vspace{-5mm}
\includegraphics[width=0.45\textwidth,angle=0]{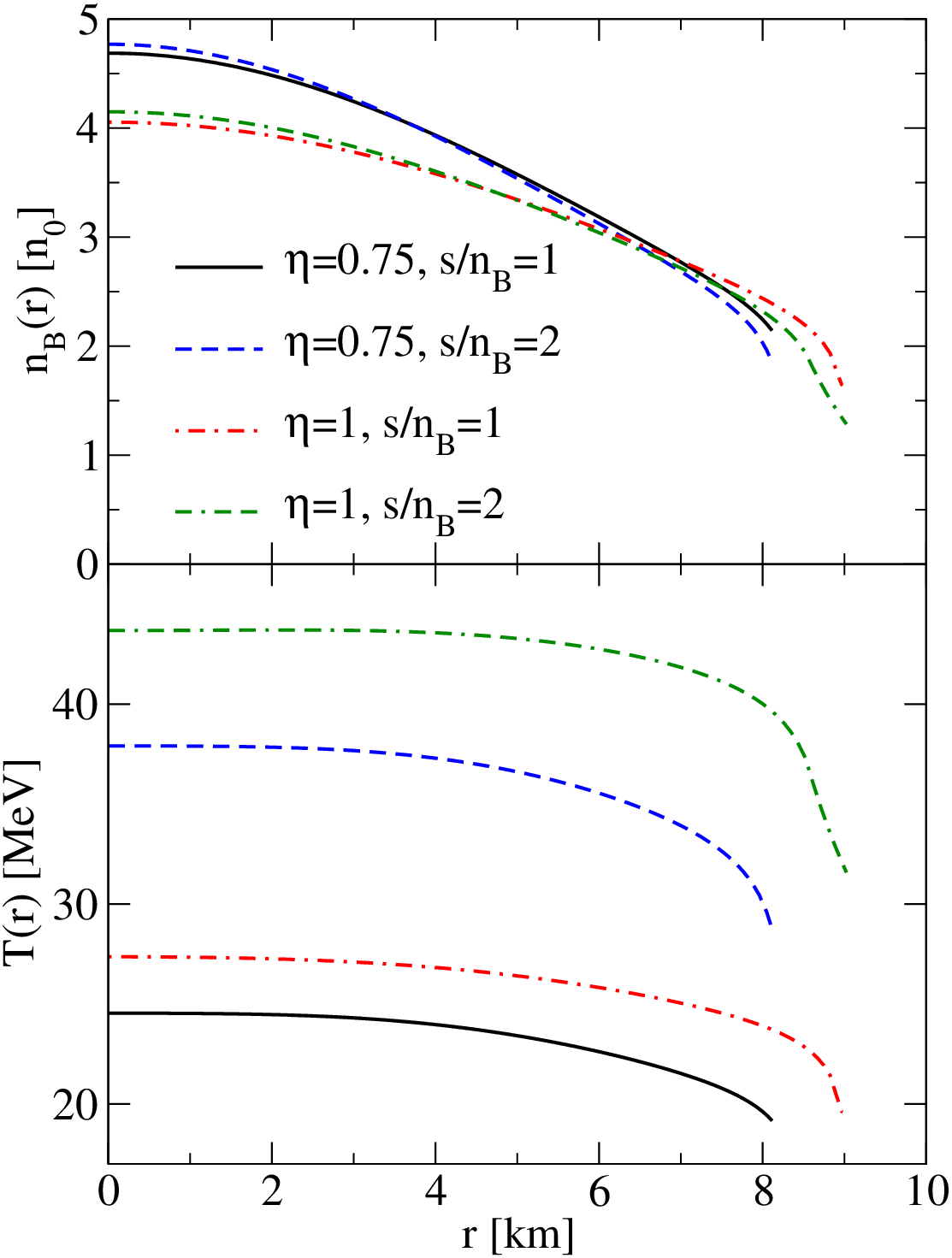}
\caption{(Color online)
The temperature and baryon number density vs. the coordinate radius
of four different quark star solutions, for fixed values of the lepton
fraction, $Y_{L_e}(T,\mu_\nu)=0.4$, and the entropy per baryon, 
$s(T,\mu_\nu)/n_B(T,\mu_\nu)$. The mass of these quark star solutions is
$1.25$~M$_\odot$ ($1.4$~M$_\odot$) for intermediate (strong) coupling.}
\label{figProfYLe04}
\end{figure}

In \fig{figSeqEta075} the quark star sequences for intermediate 
coupling, $\eta=0.75$, and fixed $(\mu_\nu,T)$ values are plotted.
The discontinuity at the maximum mass configuration is a consequence
of the appearance of CFL matter in the center of the stars. Stars with a
CFL core are marginally stable and exist if $T$ and $\mu_\nu$
are not too high, see \fig{figPhaseEta075} in the following Subsection.
Increasing the coupling to $\eta=1.0$ leads to an interesting situation,
shown in \fig{figSeqEta1}. There is a small interval of star masses 
at $M\sim 1.3$~M$_\odot$ for which mass twins occur: pure 2SC quark stars
have stable, high-density mass isomers with a CFL quark core and a smaller 
radius. \fig{figSeqEta1} suggests an interesting scenario: 
upon mass accretion, a 2SC star could undergo a phase transition
to its more compact twin with a CFL core, whereby binding energy
is released.
However, the presence of a hadronic crust could render the twin
configurations unstable and this scenario therefore remains to be
investigated thoroughly.
In \fig{figSeqYLe04} we show the star sequences corresponding to the
EsoS in \fig{figEosYLe04} with fixed entropy per baryon (1 or 2) and
fixed lepton fraction $Y_{L_e}=0.4$. 
While the increase of coupling strength increases the maximum mass and the 
radius of the stars, both values of the entropy per baryon produce rather
similar star sequences. A finite entropy per baryon of $1-2$ correspond
to rather high temperatures that do not allow for stable CFL cores.
In \fig{figProfYLe04} the radial dependence of the temperature and baryon
number density in PNS is illustrated for fixed values of the lepton
fraction and entropy per baryon. These configurations, which resemble
realistic PNS formed in the adiabatic compression of the cores in
massive stars, have approximately constant temperature and neutrino
chemical potential in their interior. Consequently, at the qualitative
level the initial state of a PNS can be modeled by a fixed initial
temperature, $T(r)=T$, and neutrino chemical potential,
$\mu_\nu(r)=\mu_\nu$. This result justifies the alternative neutrino 
trapping/untrapping scenario considered in this paper.

%%%%%%%%%%%%%%%%%%%%%%%%%%%%%%%%%%%%%%%%%%%%%%%%%%%%%%%%%%%%%%%%%%%%%%%%%%%%%

\subsection{Phase diagrams}

\begin{figure*}[ht]
%\vspace{-5mm}
\centering
\begin{tabular}{cc}
        \includegraphics[width=.45\textwidth]{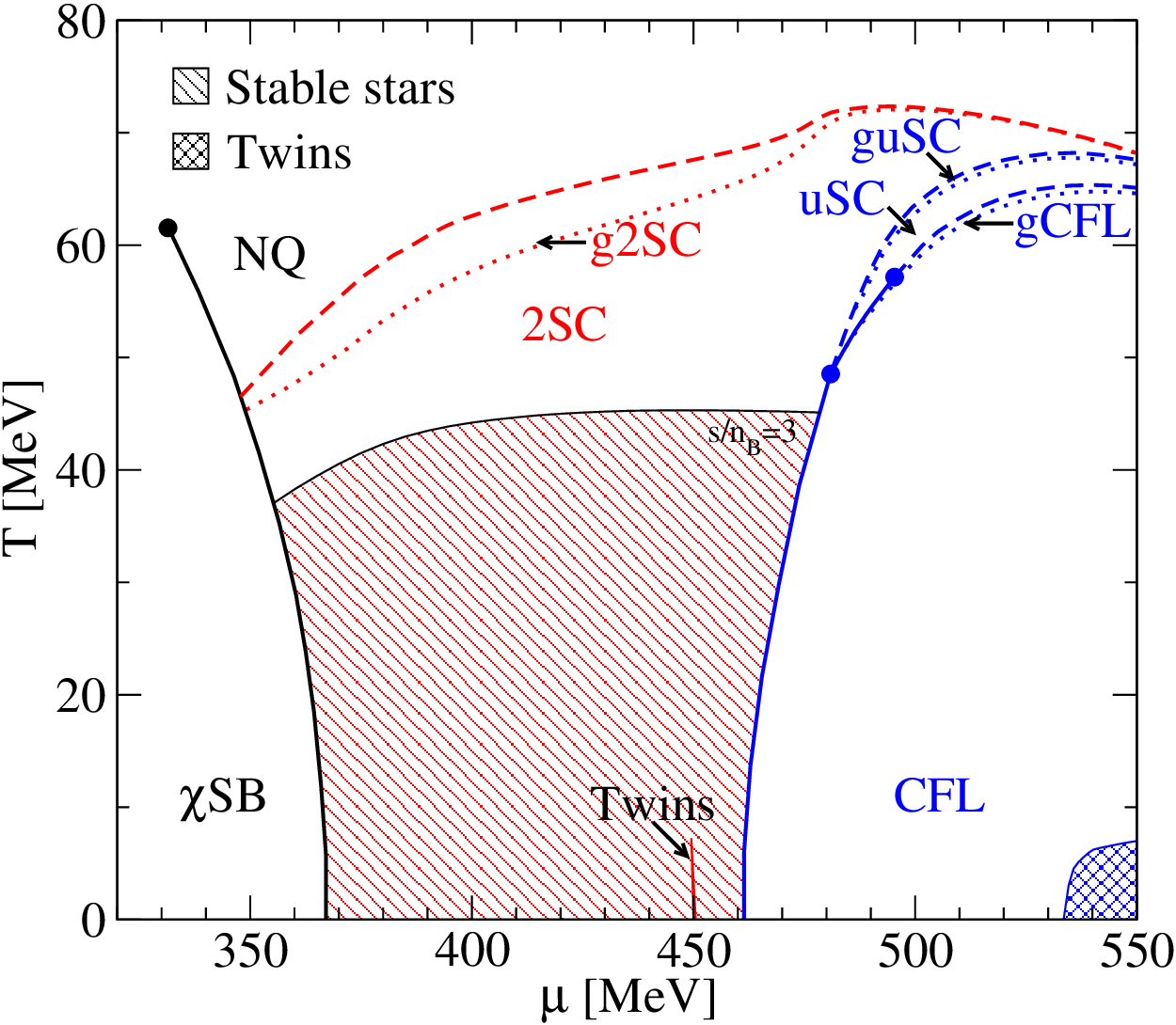} &
        \includegraphics[width=.45\textwidth]{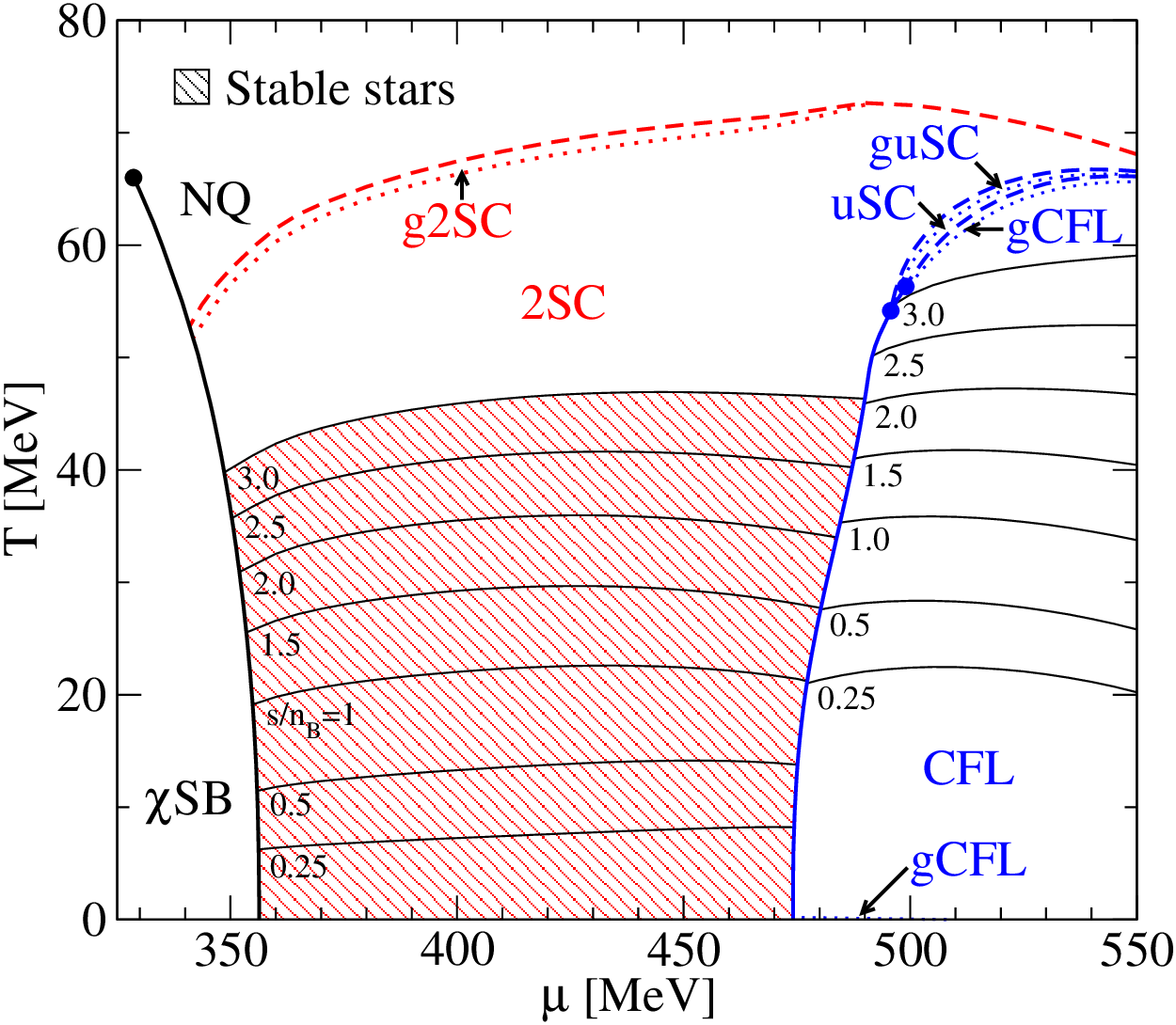}
\end{tabular}
\caption{(Color online)
Phase diagrams of charge neutral quark matter in $\beta$-equilibrium
at intermediate coupling, $\eta=0.75$, for fixed values of the electron
neutrino chemical potential, $\mu_{\nu}=0$ (left-hand side) and 
$\mu_{\nu}=200$~MeV (right-hand side).
First-order phase transition boundaries are indicated by bold solid lines,
while bold dashed lines represent second-order phase boundaries. The dotted
lines indicate gapless phase boundaries and the thin solid lines are level
curves of constant entropy per baryon. Hatched regions represent stable
compact star solutions, with central quark number chemical potential 
$\mu(r=0)=\mu$ and temperature $T(r=0)=T$. The cross-hatched regions 
correspond to baryon number twins\ie for these values of $\mu(r=0)$ and 
$T(r=0)$ there exist stable 2SC stars and 2SC-CFL stars with equal baryon 
number.}
\label{figPhaseEta075}
\end{figure*}

Next we present for the first time the central temperatures and quark number
chemical potentials of stable quark star solutions in the phase diagrams of
quark matter, with and without neutrino trapping.
In \figs{figPhaseEta075}{figPhaseEta1} the phase diagrams of charge neutral
quark matter in $\beta$-equilibrium are shown for intermediate, $\eta=0.75$,
and strong, $\eta=1$, coupling and for two different values of the electron
neutrino chemical potential, $\mu_{\nu}=0$ and $200$~MeV.

\begin{figure*}[htb]
%\vspace{-5mm}
\centering
\begin{tabular}{cc}
        \includegraphics[width=.45\textwidth]{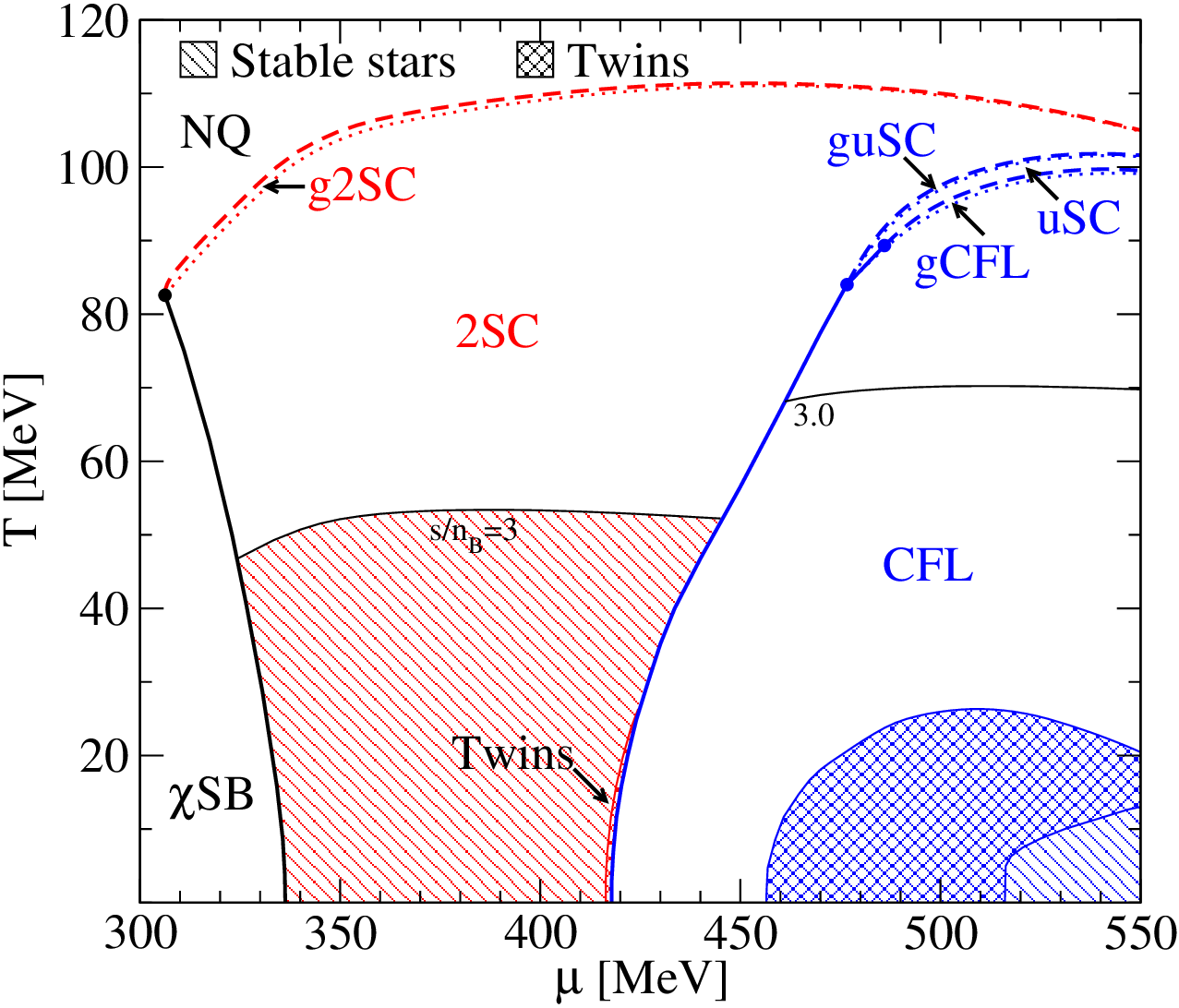} &
        \includegraphics[width=.45\textwidth]{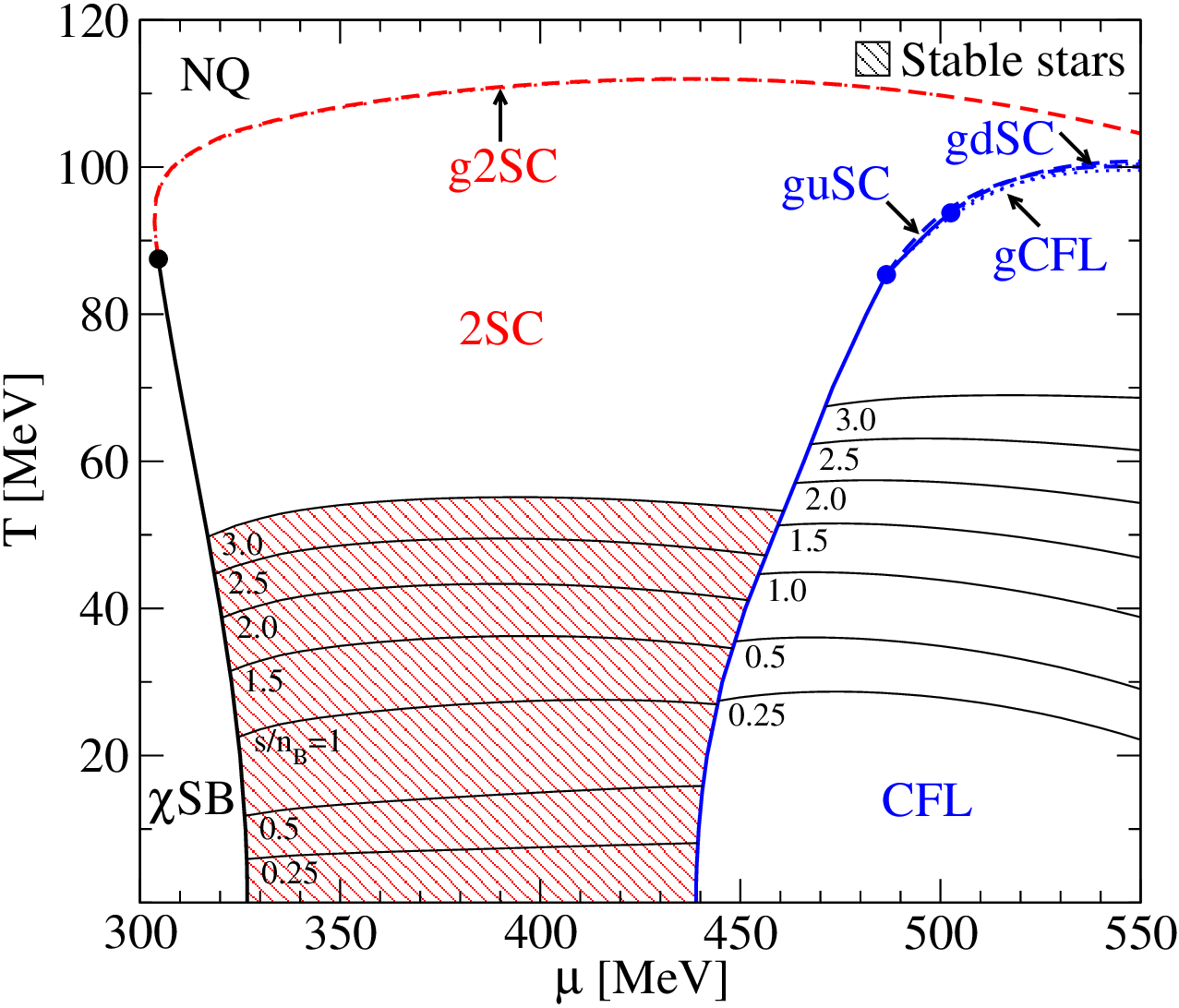}
\end{tabular}
\caption{(Color online)
Phase diagrams of charge neutral quark matter in $\beta$-equilibrium
at strong coupling, $\eta=1.0$, for fixed values of the electron neutrino
chemical potential, $\mu_{\nu}=0$ (left-hand side) and $\mu_{\nu}=200$~MeV
(right-hand side). Line styles as in \fig{figPhaseEta075}.}
\label{figPhaseEta1}
\end{figure*} 

\clearpage

% This text appears in the caption, no reason to duplicate it here.
%%%%%%%%%%%%%%%%%%%%%%%%%%%%%%%%%%%%%%%%%%%%%%%%%%%%%%%%%%%%%%%%%%%%
%First (second) order phase transitions are denoted with bold solid (dashed)
%lines, while dotted lines denote gapless phase boundaries. The thin solid
%lines are level curves of constant entropy per baryon. Hatched regions
%represent stable compact star solutions, with central quark number chemical
%potential $\mu(r=0)=\mu$ and temperature $T(r=0)=T$. The cross-hatched regions 
%correspond to baryon number twins\ie for these values of $\mu(r=0)$ and 
%$T(r=0)$ there exist stable 2SC stars and 2SC-CFL stars with equal total 
%baryon number.

As an upper limit estimate for the initial core temperatures of quark stars
we show the lines corresponding to an entropy per baryon of 3, which in the
2SC phase leads to a maximum temperature of $\sim 40$~MeV ($\sim 50$~MeV) for
intermediate (strong) coupling.
For sufficiently high neutrino chemical potentials, quark stars with a CFL
core are unstable. In particular, by comparing the left-hand ($\mu_{\nu}=0$)
and right-hand ($\mu_{\nu}=200$~MeV) panels of \fig{figPhaseEta1} it is clear
that stars with a CFL core are rendered unstable by the finite neutrino
chemical potential.
For a typical value of $\mu_{\nu}=200$~MeV, which should be realized in the
evolution of a PNS core \cite{Burrows:1986me,Prakash:1996xs},
we find no stable CFL cores. Note that the discussion of star temperatures
exceeding the neutrino opacity temperature, $T_c \sim 1$ MeV, makes sense
only during the PNS era\ie when neutrinos are trapped and
the neutrino chemical potential is finite.
Consequently, the region of stable CFL core stars shown in the panel on the
left-hand side of \fig{figPhaseEta1} may not be realized at all during stages
of hot PNS (quark star) evolution. 
We will return to this issue in the next Subsection.

\begin{figure}
%\vspace{-5mm}
\includegraphics[width=0.45\textwidth,angle=0]{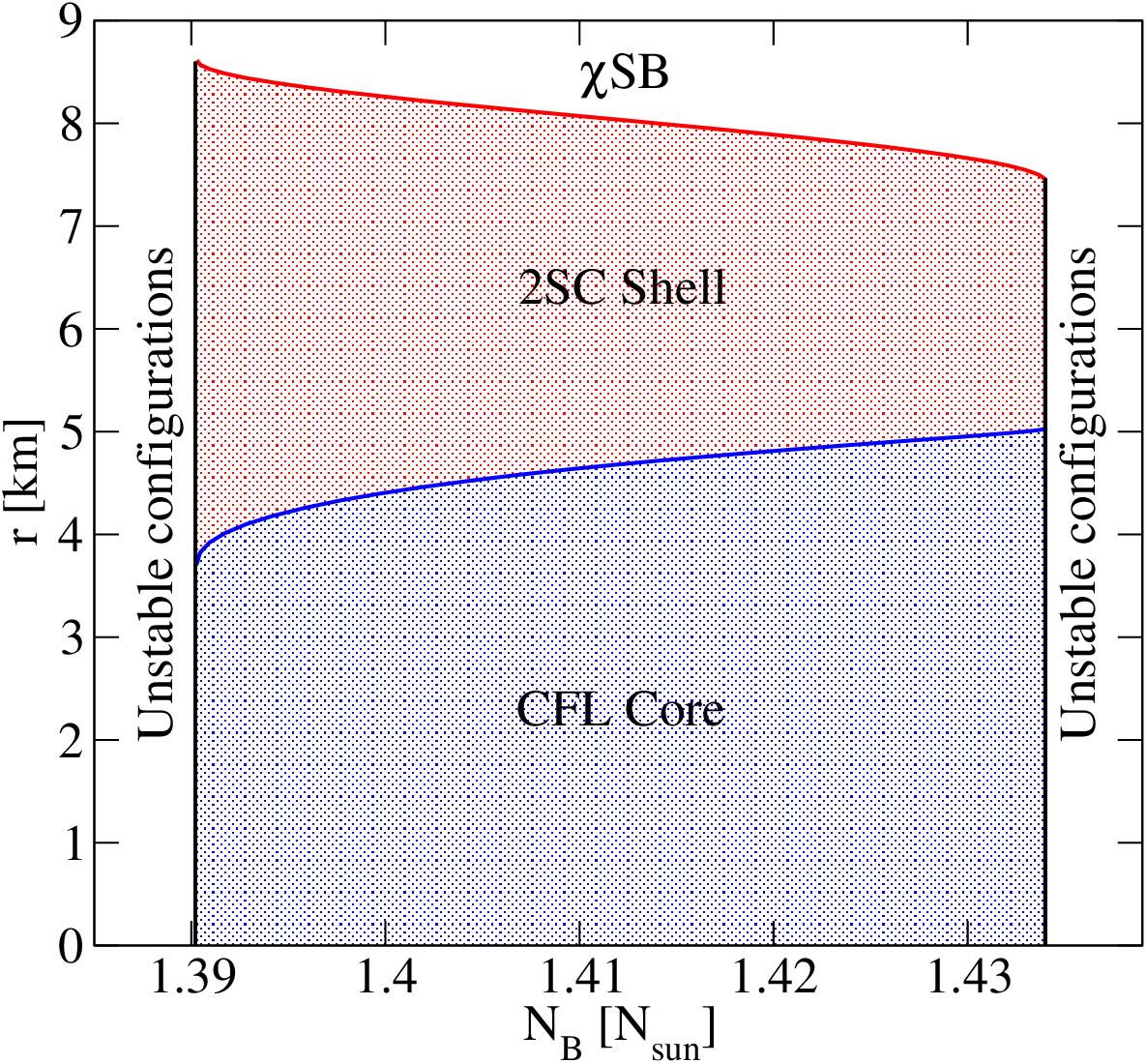}
\caption{(Color online)
Radial phase structure of final states with CFL cores at
strong coupling, $\eta=1.0$. The $T=0$, $\mu_\nu=0$ compact star sequence
in \fig{figSeqEta1} has a branch of stable stars with CFL cores,
this Figure shows the radial phase structure of these stars.}
\label{figStructureT0Eta1}
\end{figure}

For temperatures below $T_c $, we find a range of baryon numbers
$N_B=1.39 - 1.43(5)$~N$_\odot$, for which quark stars with a CFL core
and a 2SC shell are stable. The phase structure of these stars is
illustrated in \fig{figStructureT0Eta1}.
Gapless phases of superconducting quark matter exist for both coupling strengths,
at temperatures well above those of relevance for the cores of PNS,
see \figs{figPhaseEta075}{figPhaseEta1}. The gapless phases can therefore
be neglected in the present discussion of PNS evolution
\footnote{Note that a gapless CFL phase exist at low temperature for 
intermediate coupling and $\mu_\nu=200$~MeV. This is irrelevant for 
the evolution of PNS for two reasons: (1) at very low $T$ neutrinos
are not trapped and (2) the corresponding quark star configurations
are unstable. In addition, low-temperature gapless phases suffer from
a chromomagnetic instability and are typically unphysical.}.

%%%%%%%%%%%%%%%%%%%%%%%%%%%%%%%%%%%%%%%%%%%%%%%%%%%%%%%%%%%%%%%%%%%%%%%%%%%%%

\subsection{Cooling evolution and untrapping transition}

\begin{figure}
%\vspace{-5mm}
\includegraphics[width=0.45\textwidth,angle=0]{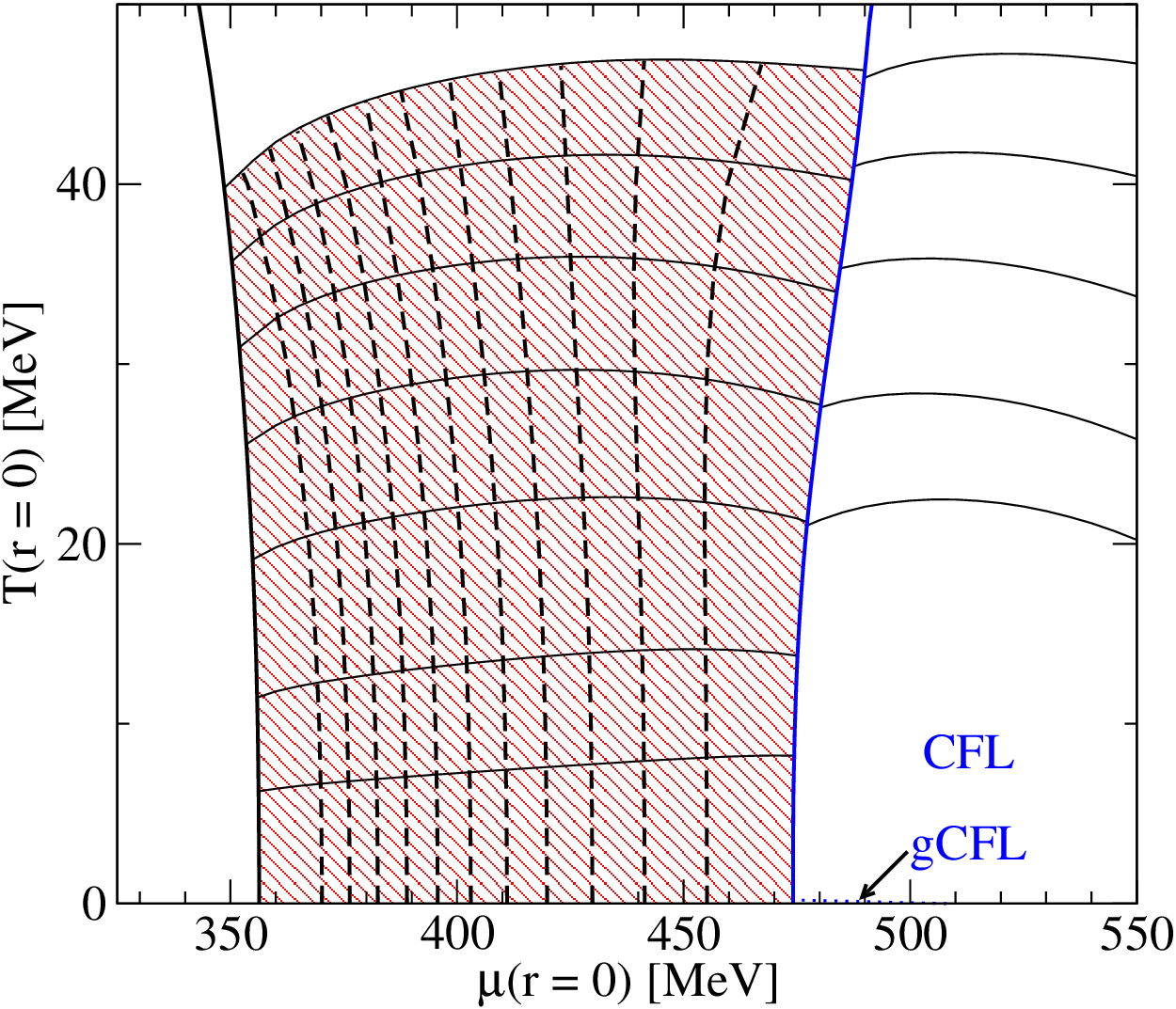}
\caption{(Color online)
Cooling of PNS with trapped neutrinos, $\mu_\nu=200$~MeV, at intermediate
coupling, $\eta=0.75$.
The dashed lines indicate the central temperature and quark number chemical 
potential of configurations with fixed baryon number, 
$N=0.2,~0.3,~0.4,~\ldots, 1.2$~N$_\odot$.
Solid lines indicate phase boundaries and level curves of constant entropy per 
baryon, see \fig{figPhaseEta075}.}
\label{figCoolingLe200Eta075}
\end{figure}

\begin{figure}
%\vspace{-5mm}
\includegraphics[width=0.45\textwidth,angle=0]{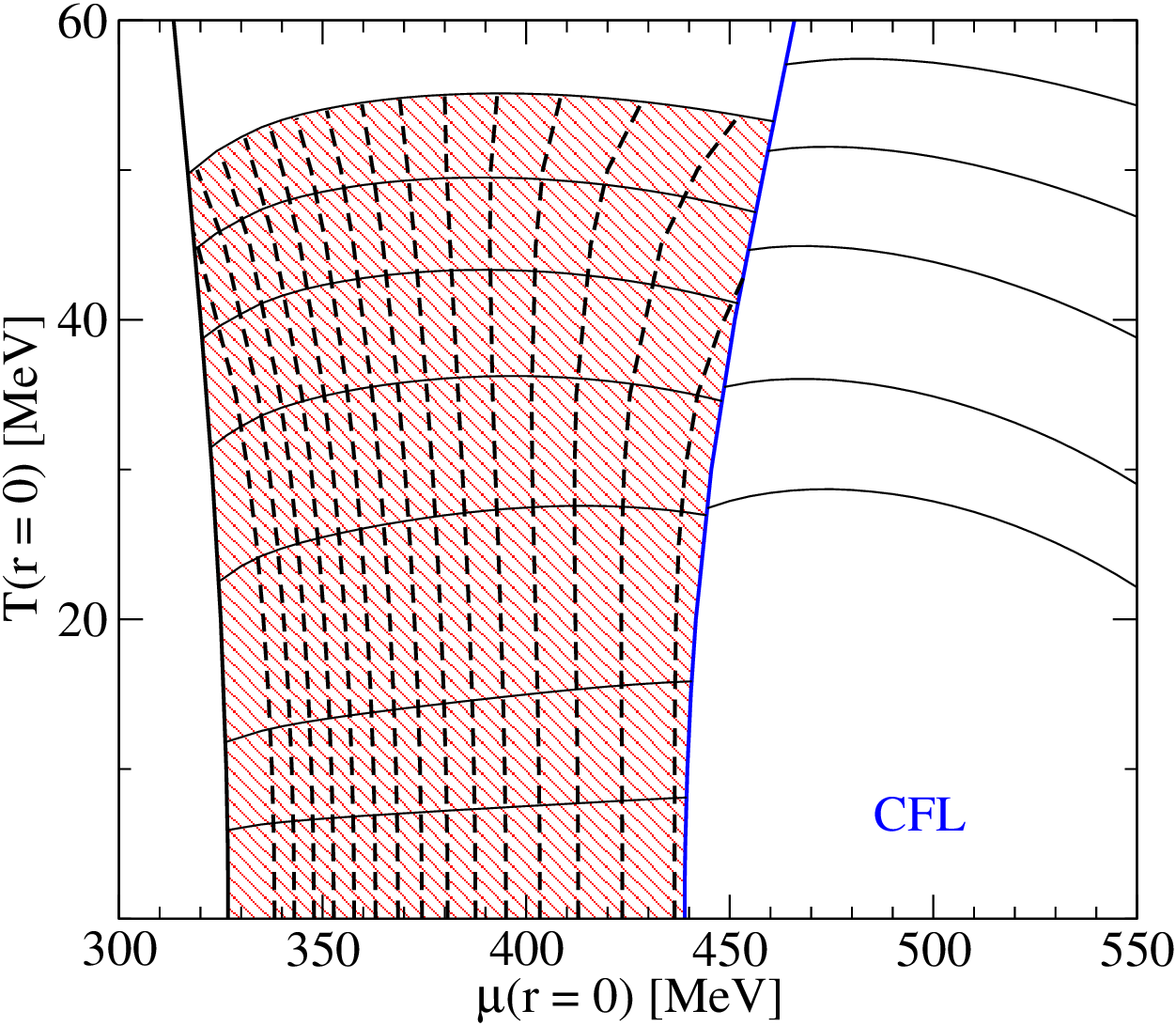}
\caption{(Color online)
Cooling of PNS with trapped neutrinos, $\mu_\nu=200$~MeV, at strong coupling, 
$\eta=1.0$.
The dashed lines indicate the central temperature and quark number chemical 
potential of configurations with fixed baryon number, 
$N=0.2,~0.3,~0.4,~\ldots, 1.6$~N$_\odot$.
Solid lines indicate phase boundaries and level curves of constant entropy per 
baryon, see \fig{figPhaseEta1}.}
\label{figCoolingLe200Eta1}
\end{figure}

In this Subsection we consider a scenario of quark star cooling where
the baryon number of the stars is conserved\ie we assume that there are
neither accretion nor mass loss.
In \figs{figCoolingLe200Eta075}{figCoolingLe200Eta1}
we show the cooling evolution of PNS configurations in the phase diagrams
of quark matter with trapped neutrinos, $\mu_\nu=200$~MeV, for intermediate
and strong coupling.
The dashed lines indicate the central temperature and quark number chemical 
potential of configurations with fixed baryon number. 
The vertical solid lines indicate phase boundaries of the 2SC phase, while 
the horizontal lines denote curves of constant entropy per baryon.
We observe that, as long as the neutrino chemical potential remains fixed, 
the cooling trajectories that start from stable configurations remain inside
the region of stability. The reverse, however, does not hold. Upon heating, a
configuration close to the CFL phase border may become unstable at 
high temperature. See\eg the trajectory for $N=1.6$~N$_\odot$ in 
\fig{figCoolingLe200Eta1}. 

\begin{figure}
%\vspace{-5mm}
\includegraphics[width=0.45\textwidth,angle=0]{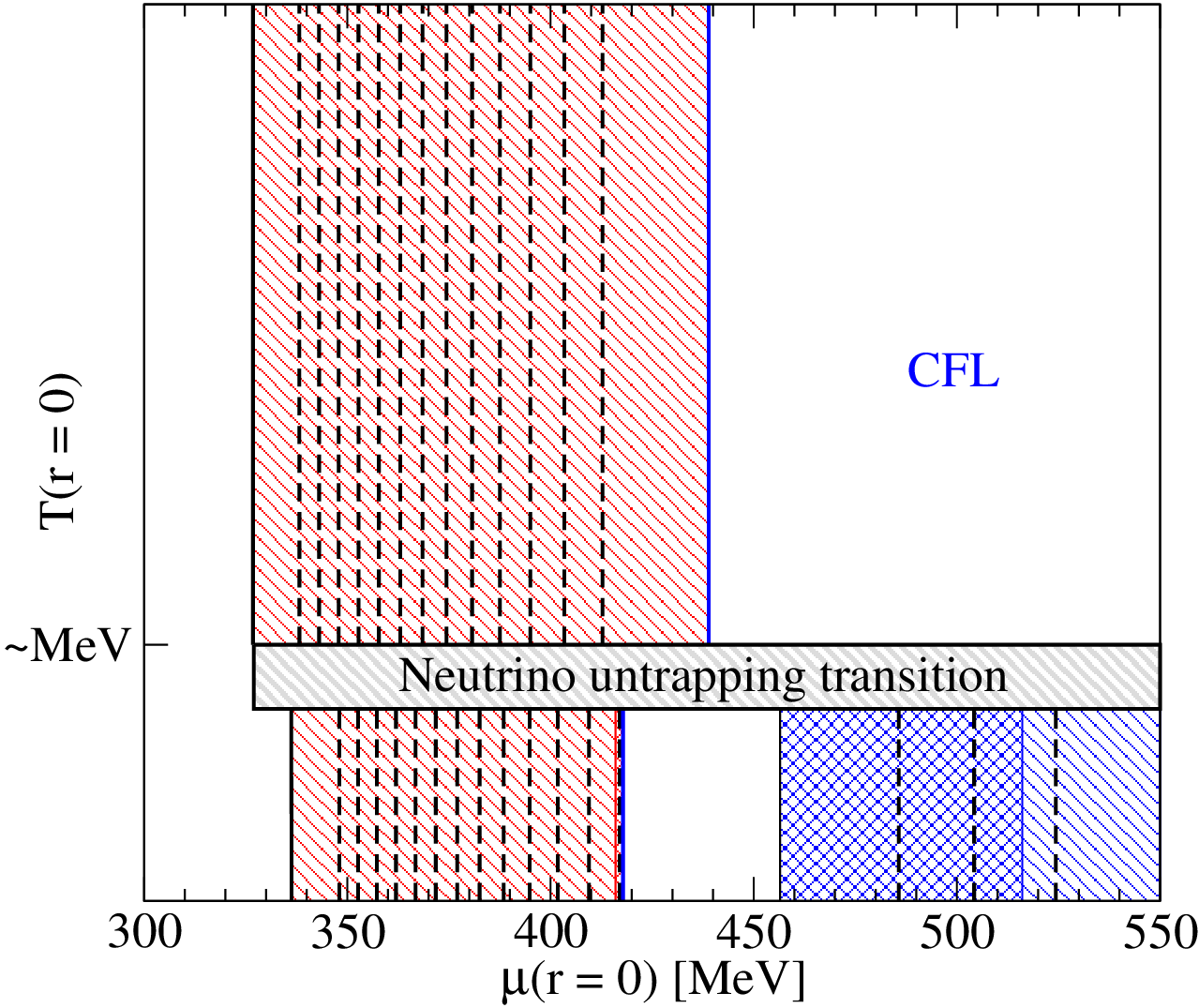}
\caption{(Color online)
Neutrino untrapping transition at strong coupling, $\eta=1.0$.
The dashed lines in the 2SC phase (See \fig{figPhaseEta1}) indicate the 
central temperature and quark number chemical
potential of configurations with fixed baryon number, 
$N=0.2,~0.3,~0.4,~\ldots, 1.4$~N$_\odot$.
Above the critical temperature, $T_c\sim 1$~MeV, of the neutrino untrapping 
transition the neutrino mean free path is shorter than the size of the stars 
and $\mu_\nu$ is therefore fixed at $200$~MeV.
This corresponds to cold PNS with trapped neutrinos. At $T_c$ the mean free
path of the neutrinos is similar to the size of the stars and 
$\mu_\nu \rightarrow 0$.
The central baryon number chemical potential of configurations with fixed 
baryon number increases in the untrapping transition.
When the neutrinos have escaped there are stable 2SC stars for 
$N\lesssim 1.41$~N$_\odot$ and stable stars with a CFL core for 
$1.39$~N$_\odot \lesssim N \lesssim 1.43$~N$_\odot$.
The dashed lines in the CFL phase indicate the central temperature and quark 
number chemical potential of configurations with fixed baryon number, 
$N=1.4,~1.41,~1.42$~N$_\odot$. 
The configurations with $N=1.4$ and $1.41$~N$_\odot$ are CFL baryon number 
twins\ie for these baryon numbers there exist also pure 2SC stars. 
See \fig{figPhaseEta1} and \fig{figUntrapEta1} for further information.}
\label{figCoolingLe0and200Eta1}
\end{figure}

Another instability occurs during the cooling evolution when a star reaches 
the neutrino opacity temperature, $T_c\sim 1$ MeV, and neutrinos start leaving 
the system. 
The microscopic processes behind this neutrino untrapping transition are
elastic and inelastic neutrino-quark collisions, which lead to a neutrino
mean free path that exceeds the radius of the star as the temperature
decreases to $T_c$. This characterizes the transition from the neutrino-diffusion
regime with finite $\mu_\nu$ to the free-streaming regime with $\mu_\nu=0$,
which takes place within the transport timescale of a few milliseconds
\cite{Berdermann:2006rk}.
Since this timescale is well above the strong and weak interaction timescales,
which establish the local thermal and chemical equilibrium, the
untrapping transition can be considered as a quasi-static process that
connects the initial state with finite neutrino chemical potential and
the final state with vanishing $\mu_\nu$ by a sequence of equilibrium
star configurations.
This justifies the estimate of the energy release due to neutrino untrapping
from the mass defect between initial and final configurations.
Cooling trajectories that are close to the CFL phase border above
$T_c$ continue within the CFL-twin region below it\ie conservation of baryon
number requires that the phase structure of these stars change near $T_c$.
Leaving aside the details of the untrapping transition itself, we illustrate 
this situation in \figs{figCoolingLe0and200Eta1}{figCoolingLe0and200Eta075} 
for strong and intermediate coupling, respectively. 
In these figures, the phase diagram above (below) $T_c$ represents quark matter
with trapped (untrapped) neutrinos.
We now discuss the strong coupling case illustrated in 
\fig{figCoolingLe0and200Eta1} in more detail. 
At $T_c$ the mean free path of the neutrinos is similar to the size of the 
stars and 
$\mu_\nu \rightarrow 0$.
The central quark (or baryon) number chemical potential of configurations with
fixed baryon number increases in the untrapping transition.
When the neutrinos have escaped there are stable 2SC stars for 
$N\lesssim 1.41$~N$_\odot$ and stable stars with a CFL core for 
$1.39$~N$_\odot \lesssim N \lesssim 1.43$~N$_\odot$.
The dashed lines in the CFL phase indicate the central temperature and quark 
number chemical potential of configurations with fixed baryon number, 
$N=1.4,~1.41,~1.42$~N$_\odot$. 
The configurations with $N=1.4$ and $1.41$~N$_\odot$ are CFL baryon number 
twins\ie for these baryon numbers there exist also pure 2SC stars. 

\begin{figure}
%\vspace{-5mm}
\includegraphics[width=0.45\textwidth,angle=0]{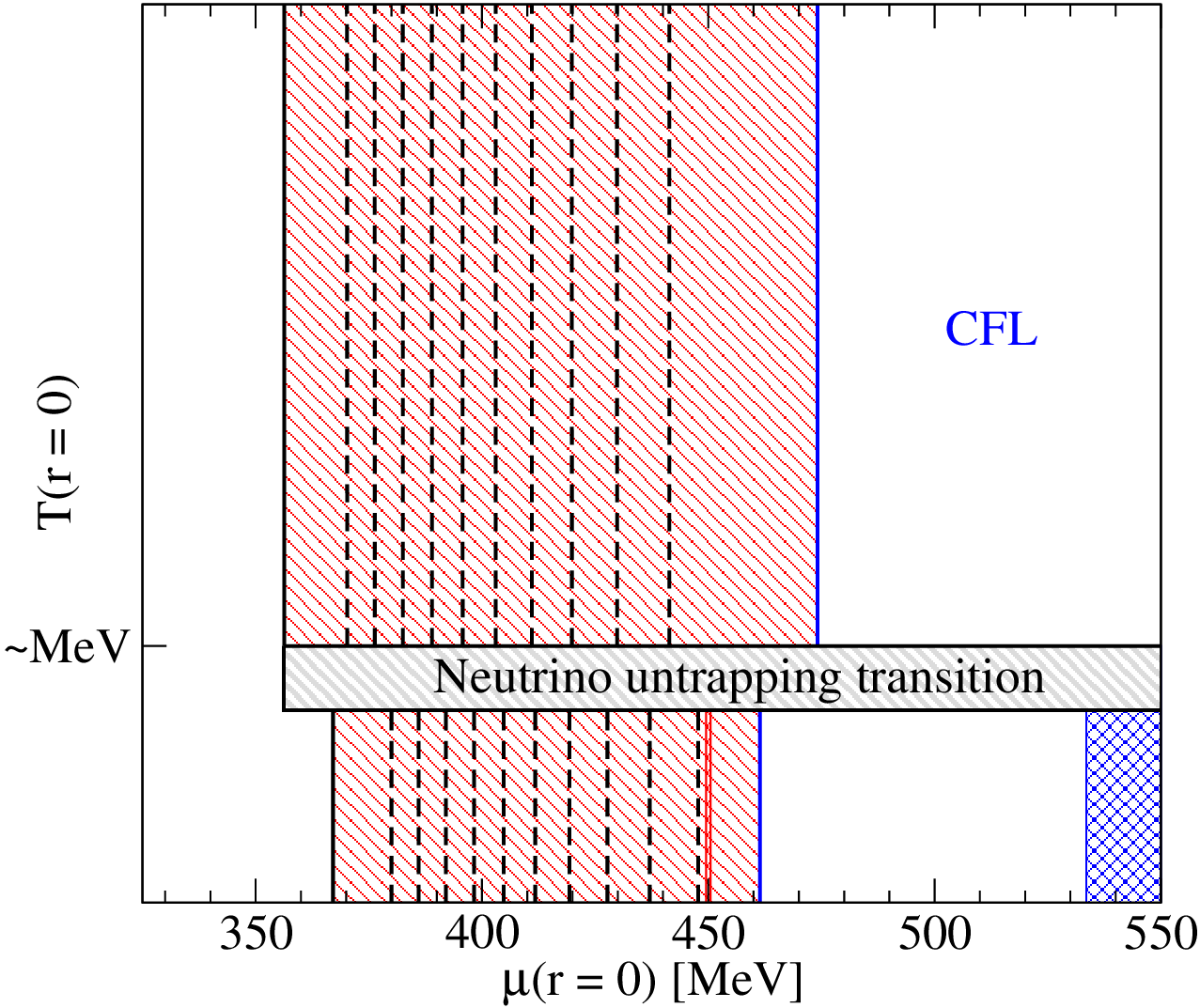}
\caption{(Color online)
Neutrino untrapping transition at intermediate coupling, $\eta=0.75$.
The dashed lines indicate the central temperature and quark number chemical
potential of configurations with fixed baryon number, 
$N=0.2,~0.3,~0.4,~\ldots, 1.1$~N$_\odot$.
CFL baryon number twins are marginally stable and exist only for a narrow 
interval of baryon numbers. They are therefore omitted. 
For further information, see \fig{figPhaseEta075} and 
\fig{figCoolingLe0and200Eta1}.}
\label{figCoolingLe0and200Eta075}
\end{figure}

For intermediate coupling, shown in \fig{figCoolingLe0and200Eta075}, there
is only a narrow interval of baryon numbers for which a transition to 
marginally stable CFL baryon number twins is possible during untrapping. 
These baryon numbers are not shown in that Figure.
 
%%%%%%%%%%%%%%%%%%%%%%%%%%%%%%%%%%%%%%%%%%%%%%%%%%%%%%%%%%%%%%%%%%%%%%%%%%%%%

\subsection{Mass defects due to cooling and neutrino untrapping}

\begin{figure}
%\vspace{-5mm}
\includegraphics[width=0.45\textwidth,angle=0]{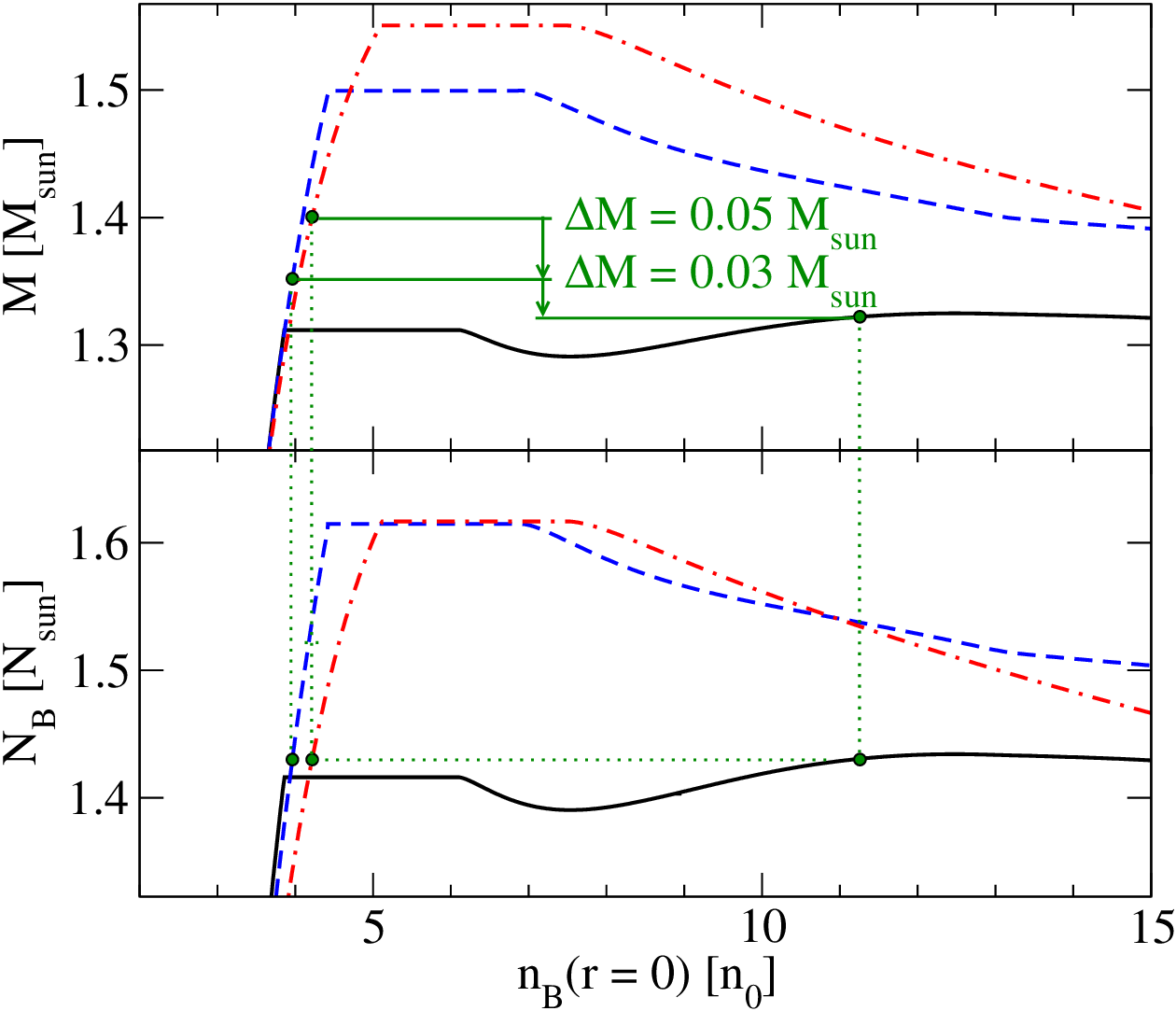}
\caption{(Color online)
The mass defect due to cooling and neutrino untrapping at strong
coupling, $\eta=1.0$. The dash-dotted line represents the sequence of hot
($T=40$~MeV) PNS with trapped neutrinos ($\mu_\nu=200$~MeV). 
 The PNS cool rapidly by neutrino emission. 
A few seconds after birth the stars are cold on the nuclear scale ($T\sim$~MeV)
and further cooling has no direct impact on their structure. 
However, the neutrino mean-free path is sensitive to
the temperature. Below some critical temperature ($\sim$MeV) the neutrinos that
are trapped in the PNS escape and this has an effect on the structure. 
The dashed (solid) line represent the sequence of cold stars with trapped 
(untrapped) neutrinos\ie $T=0$ and $\mu_\nu=200$~MeV ($\mu_\nu=0$).
 When a hot isothermal $1.40$~M$_\odot$ PNS with baryon number 
$N=1.43$~N$_\odot$ cools to low temperature by neutrino emission, the mass 
decreases to $1.35$~M$_\odot$.
At low temperature the neutrinos escape from the star and the mass decreases
to $1.32$~M$_\odot$. The relative mass defects are $3.5$\% due to cooling,
and $2.1$\% due to neutrino untrapping.}
\label{figUntrapEta1}
\end{figure}

The cooling evolution of a hot PNS from $T\sim 40$~MeV to $T_c\sim 1$~MeV
and the subsequent neutrino untrapping transition entail significant changes
to the EoS and the structure of the star.
This results in a decrease of the gravitational mass. The mass defect\ie the
difference of the gravitational masses before and after the cooling and/or
untrapping transition, corresponds to the energy that is released from the star,
predominantly by neutrino emission. A fraction of the neutrinos might, however,
be converted into photons that give rise to a gamma- or X-ray burst.
The PNS cool rapidly by neutrino emission. 
A few seconds after birth the stars are cold on the nuclear scale 
($T\sim 1$~MeV) and further cooling has no direct impact on their structure. 
However, the neutrino mean-free path is sensitive to
the temperature. Below some critical temperature ($\sim$MeV) the neutrinos that
are trapped in the PNS escape and this has an effect on the structure. 
In \figs{figUntrapEta1}{figUntrapTwinEta1} the energy release is estimated
from the mass defects due to cooling and neutrino untrapping for strong
coupling and two different values of the baryon number.

\begin{figure}
%\vspace{-5mm}
\includegraphics[width=0.45\textwidth,angle=0]{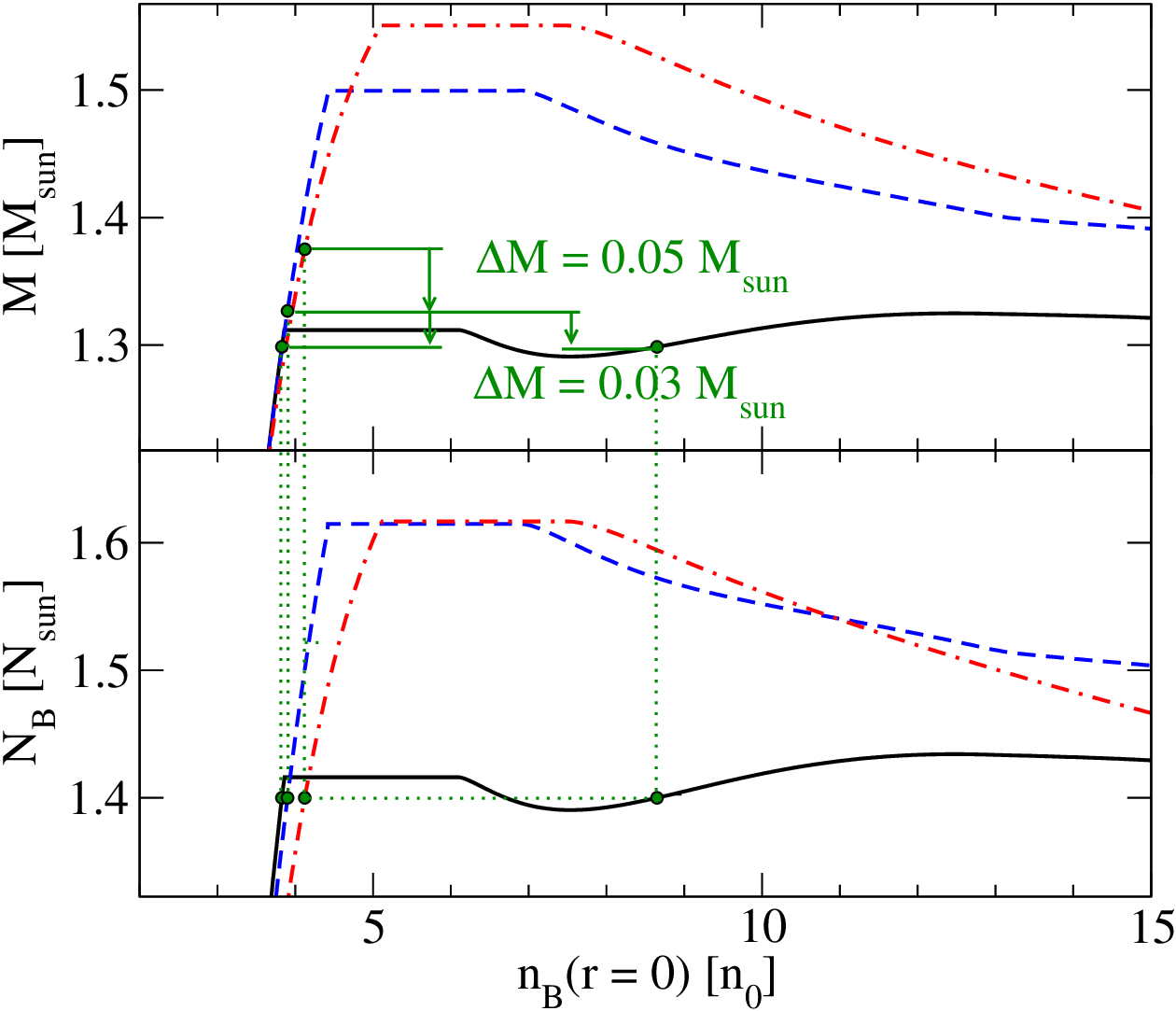}
\caption{(Color online)
The mass defect due to cooling and neutrino untrapping at strong
coupling, $\eta=1.0$. Line styles as in~\fig{figUntrapEta1}.
When a hot isothermal PNS with baryon number $N=1.4$~N$_\odot$ cools to
low temperature by neutrino emission, the mass decreases from $1.38$~M$_\odot$
to $1.33$~M$_\odot$. 
At low temperature the neutrinos escape from the star and the mass decreases 
to $1.30$~M$_\odot$. 
The final state is either a homogenous 2SC star or a more tightly bound 2SC 
star with a CFL core.
The mass difference between these two final states is $0.01$\%.}
\label{figUntrapTwinEta1}
\end{figure}

\begin{figure}
%\vspace{-5mm}
\includegraphics[width=0.45\textwidth,angle=0]{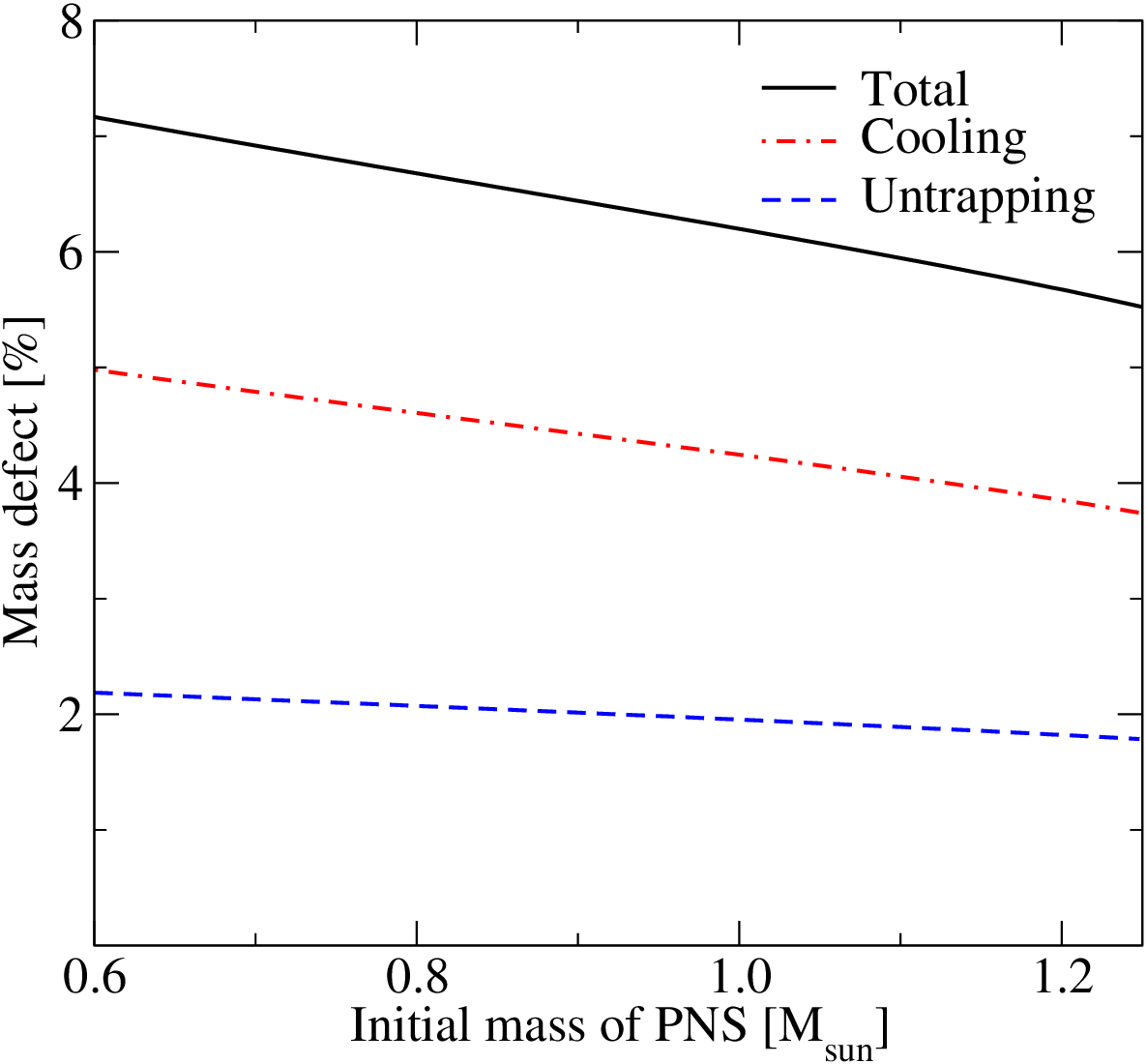}
\caption{(Color online)
        The mass defects due to cooling ($T$: $40$~MeV $\rightarrow$ MeV)
        and neutrino untrapping ($\mu_\nu$: $200$~MeV $\rightarrow 0$) at
        intermediate coupling, $\eta=0.75$.}
\label{figUntrap2Eta075}
\end{figure}

\begin{figure}
%\vspace{-5mm}
\includegraphics[width=0.45\textwidth,angle=0]{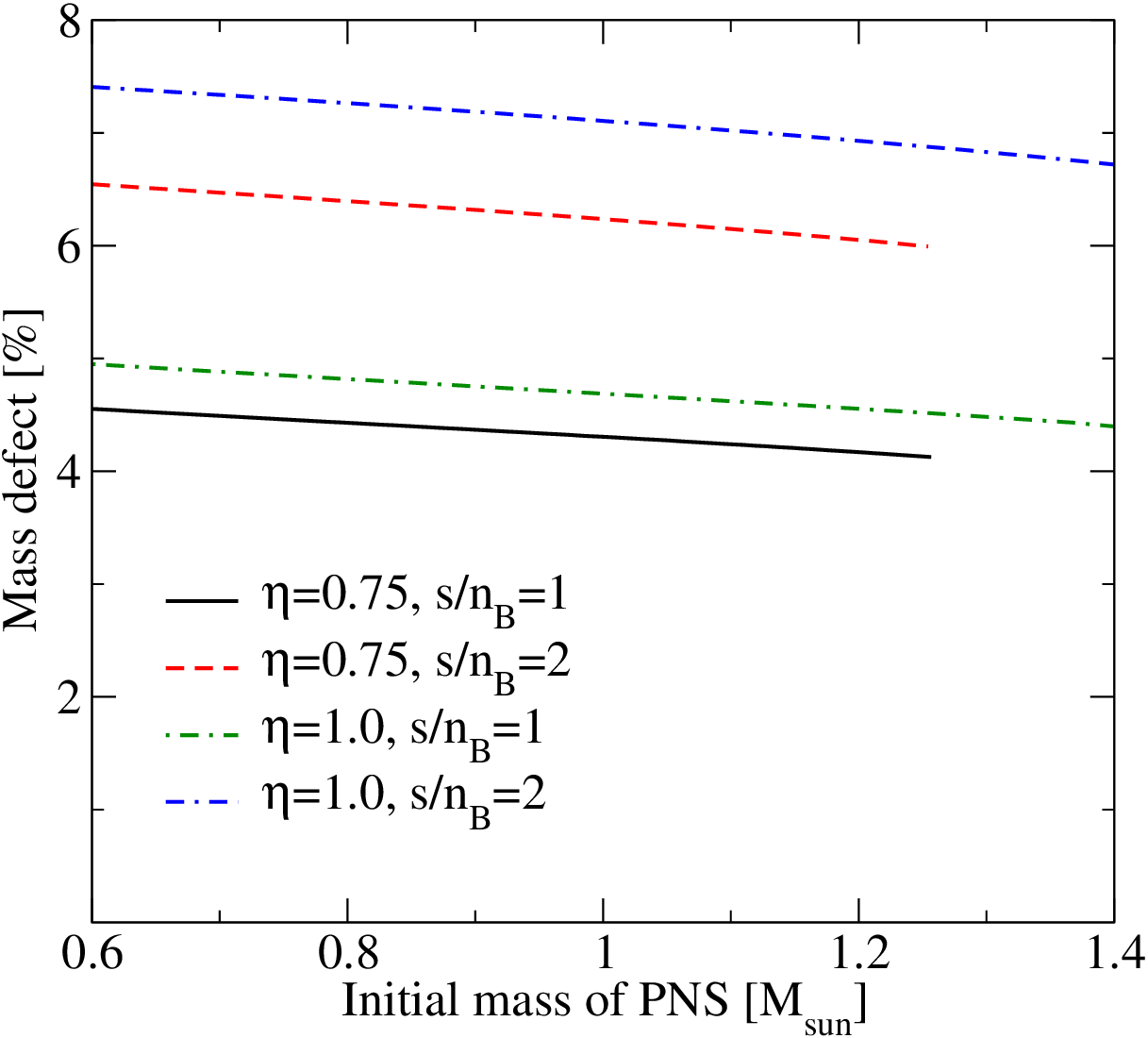}
\caption{(Color online)
        The total mass defects due to cooling ($s(T,\mu_\nu)/n_B(T,\mu_\nu)$: 
        $1$~or~$2$ $\rightarrow 0$)
        and neutrino untrapping ($Y_{Le}(T,\mu_\nu)$: $0.4$ $\rightarrow 0$) at
        intermediate, $\eta=0.75$, and strong coupling, $\eta=1.0$.}
\label{figUntrapYLe04}
\end{figure}

In \figs{figUntrap2Eta075}{figUntrap2Eta1} the energy release due to cooling
and neutrino untrapping is plotted vs. the initial mass of the PNS, for
intermediate and strong coupling, respectively.
\fig{figUntrapYLe04} shows the net energy release when PNS cool from
a given entropy per baryon of 1 or 2 to zero and the neutrinos are
untrapped, $Y_{L_e}=0.4\rightarrow 0$.

\begin{figure}
%\vspace{-5mm}
\includegraphics[width=0.45\textwidth,angle=0]{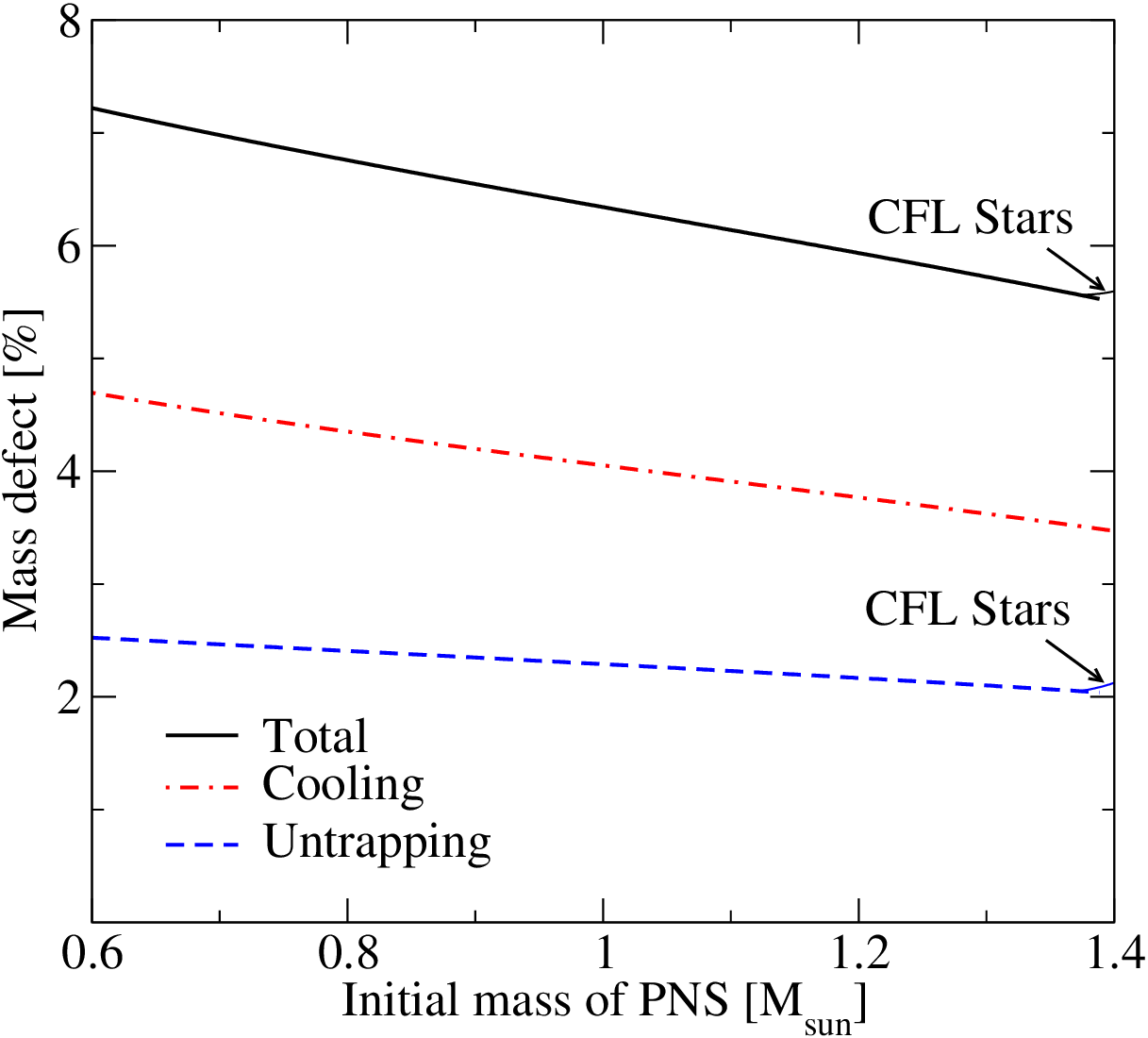}
\caption{(Color online)
        The mass defects due to cooling ($T$: $40$~MeV $\rightarrow$ MeV)
        and neutrino untrapping ($\mu_\nu$: $200$~MeV $\rightarrow 0$) at
        strong coupling, $\eta=1.0$.}
\label{figUntrap2Eta1}
\end{figure}

%%%%%%%%%%%%%%%%%%%%%%%%%%%%%%%%%%%%%%%%%%%%%%%%%%%%%%%%%%%%%%%%

\section{Conclusions}
We have studied the effect of finite neutrino chemical potentials on the 
phase diagram and the finite-temperature EoS of three-flavor quark matter
with self-consistently determined quark masses and pairing gaps. 
We confirm the results of R\"uster {\it et al.} \cite{Ruster:2005ib} 
that the 
phase transition to strange quark matter, such as the CFL phase is
shifted to higher densities when neutrinos are trapped in the system, thus
making it unlikely that strange matter exist in PNS cores. 
We find that in the presence of trapped neutrinos, the isospin mismatch
induced by the $\beta$-equilibrium conditions is reduced and the 2SC phase
becomes more favorable. In particular, the onset of the 2SC phase is
shifted to lower densities in the presence of trapped neutrinos.
This result seems to be robust, as it has been found also for inhomogeneous 
phases of the LOFF type \cite{Laporta:2005be}. 
The investigation of the neutrino trapping effect on the LOFF phases in the 
phase diagram with self-consistently determined strange quark masses is an 
interesting task for future work.

A new result of this study is the systematic analysis of the regions
in the quark matter phase diagram that are realised in stable quark
star cores. This analysis is carried out for two different strengths of
the coupling in the diquark channel and for two different values of the
neutrino chemical potential\ie with and without neutrino trapping.
A remarkable finding is that for the initial evolution of a PNS, when
the neutrinos are trapped, all configurations with a CFL core are unstable. 
In absence of trapped neutrinos there are small, isolated regions in
the temperature-density plane where stable configurations with a CFL
core exists. A fraction of these 2SC-CFL stars have baryon number
twins in the 2SC state. 
We suspect, however, that the configurations with a CFL core would be
unstable when the influence of a hadronic crust is taken into account.
See\eg Ref. \cite{Klahn:2006iw} for an analysis of such hybrid star
configurations. All 2SC quark star solutions considered in this work
are stable, while the phase border in-between the 2SC and CFL phases
marks the endpoint of stability for PNS sequences in the plane of
central temperature and central quark number chemical potential.
 
The maximum temperature of PNS are estimated to about $50$~MeV for
an entropy per baryon of 3, which is an upper limit for the entropy
generated in the adiabatic compression of the core during collapse
of a massive star.
For this value of the entropy per baryon, the temperature in the CFL
phase could reach about $65$~MeV, but the corresponding PNS configurations
are unstable.

When hot PNS cool by neutrino emission and reaches the neutrino opacity
temperature, $T_c\sim 1$~MeV\ie the temperature where the neutrino mean
free path becomes comparable to the size of the quark star core, the
neutrinos created by electron capture in the collapse are untrapped.
By calculating and comparing the compact star sequences for different
neutrino untrapping scenarios we find that the energy release due to
neutrino untrapping (and/or cooling) is of the order of 100 bethe. % (= $10^{53}$~erg).
This process could therefore be important for the inner engine of 
supernov{\ae} and gamma-ray bursts. 

The discussion of superconducting quark matter phases in the present
work concerns the stability of PNS configurations with a CFL core and
a neutrino heating mechanism for the outer core of newborn PNS due to
the fact that the opacity temperature in (superconducting) quark matter
exceeds that in nuclear matter.
A detailed study of the neutrino untrapping transition in hybrid stars
is beyond scope of the present work and will be given elsewhere.
   
%%%%%%%%%%%%%%%%%%%%%%%%%%%%%%%%%%%%%%%%%%%%%%%%%%%%%%%%%%%%%%%%

\section{Acknowledgements}

F.S. acknowledges support from the Swedish Graduate School of Space
Technology and the Royal Swedish Academy of Sciences. D.B. is grateful
for support from the Polish Ministry of Science and Higher Education.

\vspace{1cm}

%%%%%%%%%%%%%%%%%%%%%%%%%%%%%%%%%%%%%%%%%%%%%%%%%%%%%%%%%%%%%%%%

%%%%%%%%%%%%%%%%%%%%%%%%%%%%%%%%%%%%%%%%%%%%%%%%%%%%%%%%%%%%%%%%

\end{document}